\documentclass[%
aip,
pof,
amsmath,amssymb,
preprint
]{revtex4-2}

\usepackage{graphicx}% Include figure files
\usepackage{dcolumn}% Align table columns on decimal point

\usepackage[utf8]{inputenc}
\usepackage[T1]{fontenc}
\usepackage{mathptmx}
\usepackage{etoolbox}

\usepackage{siunitx}
\usepackage{amssymb}
\usepackage{amsmath,amssymb,amsfonts,amsthm,bm}
\usepackage{hyperref}
\usepackage[lofdepth,lotdepth]{subfig}

\usepackage{array}
\usepackage{colortbl}
\usepackage{epstopdf}

\makeatletter
\def\@email#1#2{%
	\endgroup
	\patchcmd{\titleblock@produce}
	{\frontmatter@RRAPformat}
	{\frontmatter@RRAPformat{\produce@RRAP{*#1\href{mailto:#2}{#2}}}\frontmatter@RRAPformat}
	{}{}
}%
\makeatother
\begin{document}
	
	%\preprint{AIP/123-QED}
	
	\title[]{{Small-Scale air turbulence structure, microphysical time scales and  local supersaturation balance at a warm Cloud Top Boundary}}
	% Force line breaks with \\
	\author{Ludovico Foss\`{a}}
	\altaffiliation[Present address: ]{Department of Mechanical engineering, the University of Sheffield, the United Kingdom.}%Lines break automatically or can be forced with \\
	\author{Shahbozbek Abdunabiev}%
	\author{Mina Golshan}
	\author{Daniela Tordella}
	\email{daniela.tordella@polito.it}
	\affiliation{Dipartimento di Scienza Applicata e Tecnologia, Politecnico di Torino, Italy}%\\This line break forced with \textbackslash\textbackslash}%
	
	%\author{C. Author}
	% \homepage{http://www.Second.institution.edu/~Charlie.Author.}
	%\affiliation{%Second institution and/or address%\\This line break forced% with \\}%
	
	\date{\today}% It is always \today, today,
	%  but any date may be explicitly specified
	
	\begin{abstract}
		%CHECKED MARGUERITE JONES
		
		{Recent results have shown that there is an acceleration in the spread of the size distribution of droplet populations in the region bordering the cloud and undersaturated ambient.
			We have analyzed the supersaturation balance in this region, which is typically a highly intermittent shearless turbulent mixing layer, under a condition where there is no mean updraft. We have investigated the evolution of the cloud-clear air interface and of the droplets therein via direct numerical simulations.
			We have compared horizontal averages of the phase relaxation, evaporation, reaction and condensation times within the cloud-clear air interface for the size distributions of the initial monodisperse and polydisperse droplets.
			For the monodisperse population, a clustering of the values of the reaction, phase and evaporation times, that is around 20-30 seconds, is observed in the central area of the mixing layer, just before the location where the maximum value of the supersaturation turbulent flux occurs.
			This clustering of values is similar for the polydisperse population but also includes the condensation time.
			The mismatch between the time derivative of the supersaturation and the condensation term  in the interfacial mixing layer is correlated with the planar covariance of the horizontal longitudinal velocity derivatives of the carrier air flow and the supersaturation field, thus suggesting that a quasi-linear  relationship may exist between these quantities. }
	\end{abstract}
	
	\maketitle
	
	%\begin{quotation}
	%ADD \citep{Sidin2009,Onishi2009,Jeffery2001} POF
	%\end{quotation}
	
	\section{Introduction}
	
	The large-scale dynamics of warm atmospheric clouds is closely coupled with small-scale phenomena. Lukewarm clouds are a stage of a complex interplay between competing turbulent and microphysical processes, which determine their evolution over time. %, and have been observed to readily precipitate after a few minutes from their formation. 
	However, many physical processes that are relevant for cloud dynamics have not yet been completely unravelled, and still constitute a matter of debate in the cloud physics and turbulence communities. In recent years, a great deal of attention has been paid to the effect of turbulent mixing at cloud boundaries as well as its impact on droplet condensation (evaporation) and collision. The interfacial mixing of cloud and clear air has often been identified as the main cause of the observed broadening of droplet size distributions and the rapid onset of precipitation. 
	
	Warner (1969)\cite{Warner1969} suggested the importance of mixing at a growing cloud top in unstable environments. Latham and Reed (1977) \cite{Latham1977} and Baker, Corbin and Latham (1980)\cite{Baker1980} were the first to recognize a difference between homogeneous and inhomogeneous mixing, where the microphysics time scale can be either longer or shorter than the time scale of the turbulent motions. The ratio between the turbulent time scale and the microphysical time scale is represented by the Damk\"{o}hler number,  $\textrm{Da}$. The same turbulent flow encompasses a wide range of $\textrm{Da}$ along the energy cascade \citep{Kumar2013}. %, and the two regimes are expected to coexist on different spatial scales. 
	Several time scales have been used to define the Damk\"{o}hler number \citep{Lu2018}, and to parameterize the impact of the entrainment and mixing of clear air at cloud boundaries. These scales include evaporation, phase relaxation and reaction time scales. The fundamental variable that drives the condensation (evaporation) of a droplet is supersaturation $S=RH-1$, where $RH$ is the relative humidity \citep{Squires1952}. $S$ varies over both time and space, and is determined by the local, instantaneous concentration of water vapor $\rho_v$ and temperature $T$ through the Clausius-Clapeyron equation. %Apart from driving the condensation (evaporation) process, supersaturation is also affected by the absorption (depletion) of $\rho_v$ and the release of latent heat at a droplet surface. 
	However, supersaturation $S$ has often been described as a somewhat global property of a cloud parcel \citep{RogersYau1996}, and the local value of the vertical velocity and the microphysical properties are generally taken into account for its estimation. The supersaturation balance within a cloud is often described by means of a production-condensation model of the type proposed by \cite{Twomey1959}, where the mean updraft velocity $w$ and the mean radius of the droplet population $\overline{R}$
	%- defined as the first-order moment of the droplet size distribution $N\overline{R}$ - of the droplet population 
	are the main contributors to the time derivative of $S$
	\begin{equation}\label{Twomey_introduction}
		\frac{dS}{dt} %Q_1w-Q_2\frac{dLWC}{dt}
		\cong c_1w-\frac{S}{\tau_{phase}} 
	\end{equation}
	{where $\tau_{phase} = (c_2 n_d \overline{R})^{-1}$ is the phase relaxation timescale \citep{Politovich-Cooper1988, Devenish2012}, $n_d$ is the droplet number density and $c_1, c_2$ are coefficients that depend, albeit only slightly,  on the temperature, $c_1$ and temperature/pressure, $c_2$,  \cite{Grabowski-Wang2012}. Cooper (1989)\cite{Cooper1989} described a theoretical framework in which the variability of $S$, and the subsequent broadening of the droplet size distribution are determined by the value of the integral radius as well as by the covariance of the integral radius and the vertical velocity fluctuation.} 
	
	Sardina et al. 2015  generalized Twomey's model to a scalar transport equation that they used in their direct numerical simulation (DNS) study of cloud cores \cite{Sardina2015}. They showed that the contribution of the diffusive effect is negligible for large Reynolds numbers. Chandrakar et al. (2016)\cite{Chandrakar2016}, in a laboratory experiment, used the stochastic condensation model of \cite{Sardina2015} to investigate the effects of an aerosol concentration on the broadening of the droplet size distribution. They argued that supersaturation fluctuations determine diverse growth conditions inside cloud cores with low aerosol (and droplet) concentrations. \cite{Siebert2017}, who studied the impact of turbulent temperature and water vapor density fluctuations on supersaturation by performing in-situ measurements of shallow-cumulus clouds, suggested the same. Their data show a reduction in the standard deviation of supersaturation inside cloud cores compared to regions where few or no droplets are located. They used the phase relaxation time $\tau_{phase}$ as the microphysical time scale in the Damk\"{o}hler number. Prabhakaran (2020)\cite{Prabhakaran2020} used the stochastic condensation approach to study the activation of dry-sodium chloride aerosols, as well as droplet nucleation and growth via laboratory experiments. They used a climate chamber where statistically steady-state Rayleigh-B\'{e}nard turbulence had been generated. They claimed that their results can be extended to a context in which the effects of entrainment and mixing are important, and that, in this case, droplet activation is governed by a fluctuation-dominated regime, even though such a region is subsaturated on the whole.
	
	However, many DNS studies have focused on both statistically steady-state and transient shearless mixing layers located at a vertical droplet-laden, cloud-clear air interface \citep{Kumar2014,Gao2018}.  Kumar et al.(2018)\cite{Kumar2018} investigated the effects of the range of the energy cascade on the relative dispersion of a droplet population, which was observed to increase for larger initial values of the domain size-based $\textrm{Da}$. Miller and Bellan (2000)\cite{Miller2000} performed direct numerical simulations (DNS) of a droplet-laden shear layer that featured a two-way interphase coupling and a Lagrangian tracking system for the droplets. Onishi, Takahashi and Komori (2009)\cite{Onishi2009} studied the influence of gravity on droplet collision and coalescence. Sidin, Ijzermans and Reeds (2009)\cite{Sidin2009} used a synthetic turbulent field to investigate the impact of both large and small-scale turbulent eddies on droplet condensation and evaporation. Their DNS model did not take into account the effects of condensation and evaporation. {Recently, Golshan et al. (2021)\cite{Golshan2021} have recently performed direct numerical simulations of a horizontal, droplet-laden, interfacial shearless mixing layer subject to unstable stratification. They observed a remarkable acceleration in the dynamics of the droplet population in the mixing layer, in particular in the temporal evolution of the droplet collision/evaporation rates and in their  spectrum broadening. These findings were linked to the large intermittency of the  small-scale turbulence, which is driven by the anisotropy of the carrier flow shearless layer  and by the active scalars  transported there.
		
		The aim of the present work is twofold: first, to compute and compare the various microphysical time scales in the cloudy - clear air interfacial layer  so far proposed in the literature, second, to infer a possible  source term for Twomey's equation (\ref{Twomey_introduction}) that accounts for the small-scale statistics of the carrier flow at a cloud-top boundary where the updraft is null. We have used the dataset computed from the aforementioned direct numerical simulation campaign performed by \cite{Golshan2021}. We have adopted a high-resolution pseudospectral method that allows us to observe the temporal evolution of the supersaturation fluctuations and the velocity derivative statistics across the horizontal turbulent shearless mixing layer (for the gas phase dynamics, see \citep{Tordella2011})}. 
	
	The physical model of a shearless cloud-clear air interface, together with  the relevant governing equations of the direct numerical simulations, are presented in section \ref{SECTION_PHYS_SYSTEM}. %Section \ref{SECTION_EXPERIMENTS} provides a brief summary of the numerical experiments that were performed. 
	The obtained results are discussed in section \ref{SECTION_RESULTS}, and the conclusions are drawn in section \ref{SECTION_CONCLUSIONS}.
	
	\section{Physical problem and mathematical framework} \label{SECTION_PHYS_SYSTEM}
	\subsection{Physical model and governing equations}
\begin{figure}[bht!]
		\centering
		\subfloat[a][]{\includegraphics[width=0.4\linewidth]{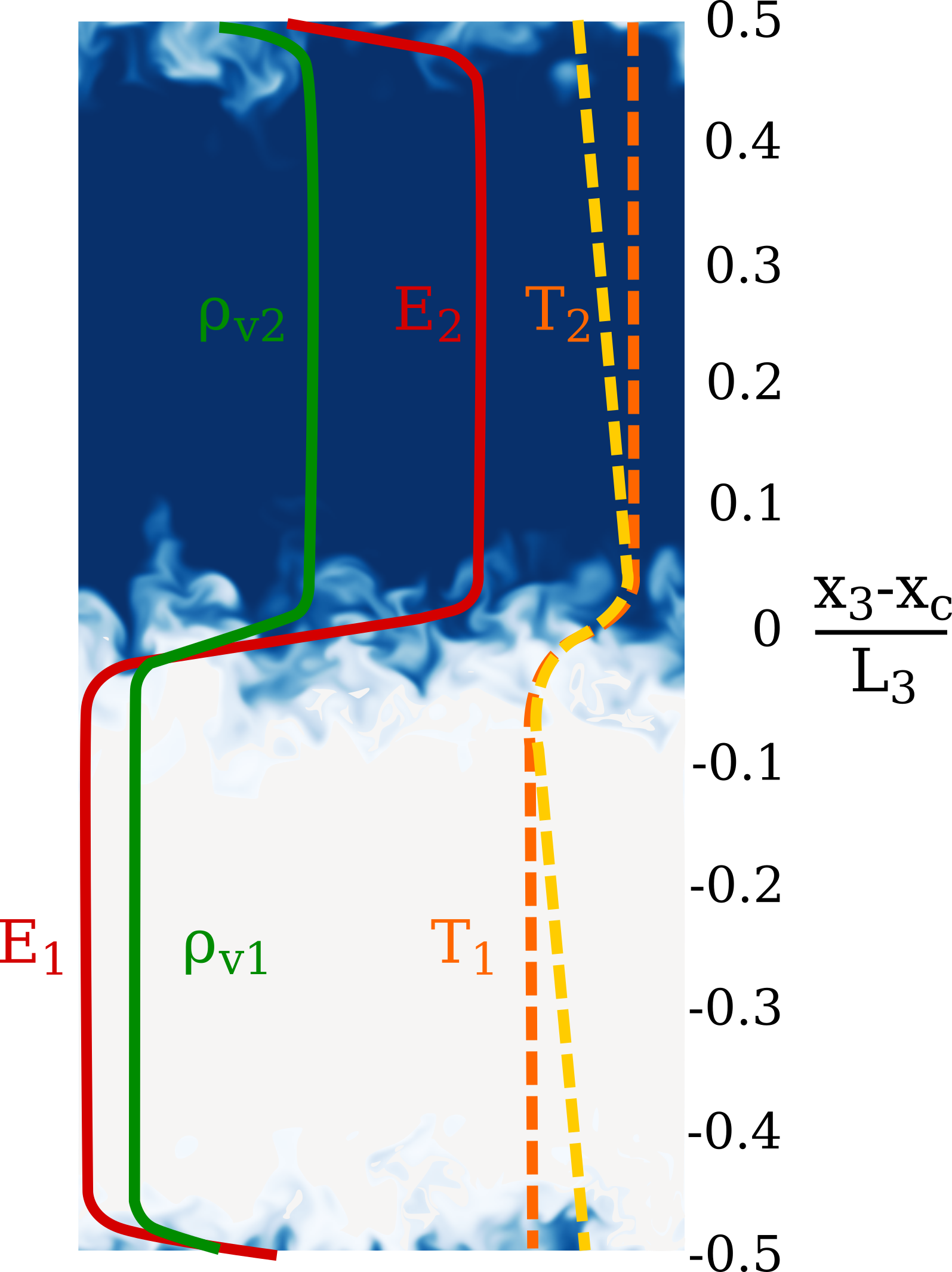}\label{fig:comp_domain_a}}
		\hspace{0.05\linewidth}
		\subfloat[b][]{\includegraphics[width=0.5\linewidth]{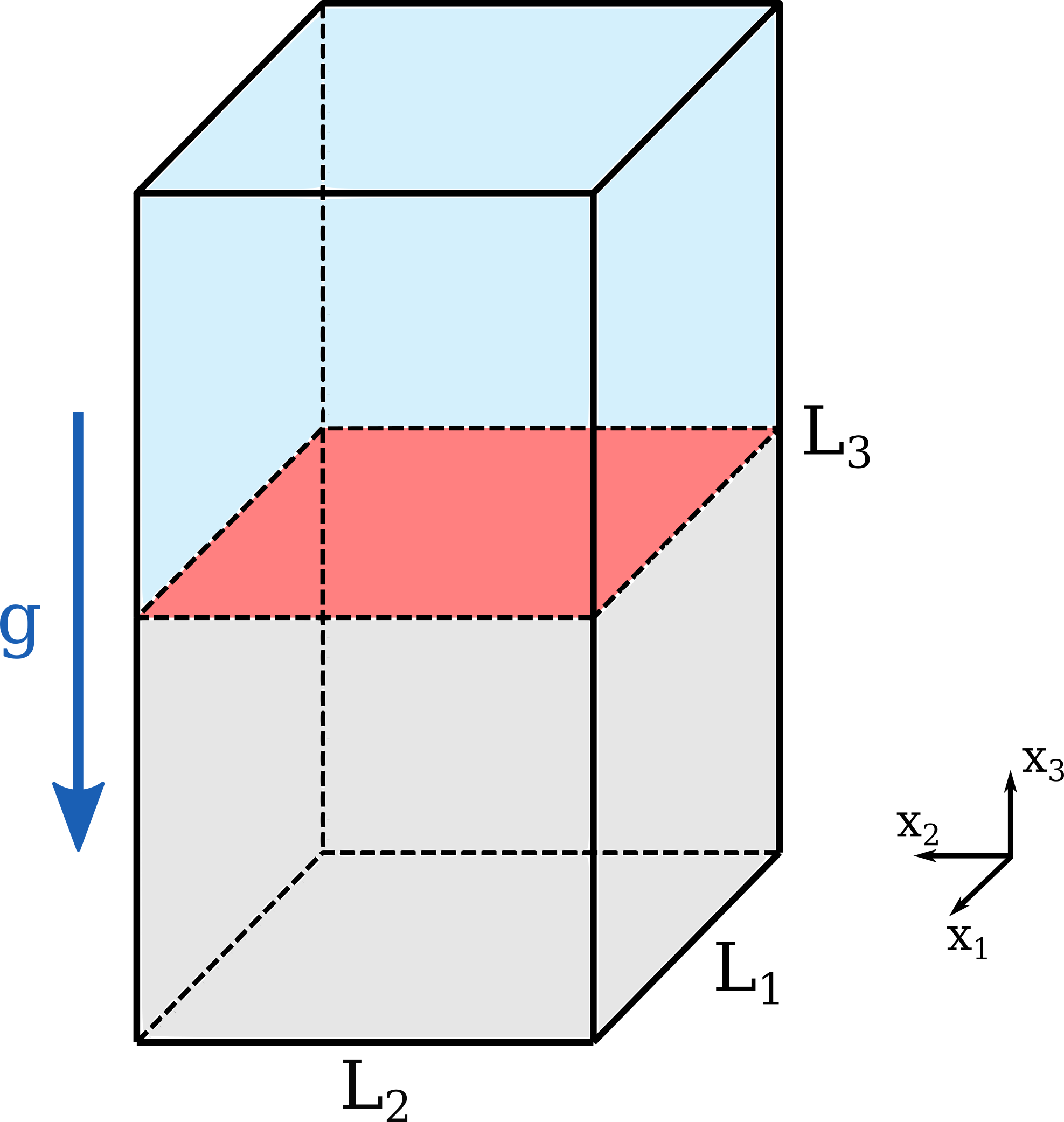}\label{fig:comp_domain_scheme}}
		\caption{Panel (a): initial distributions of kinetic energy $E$ (solid dark red line), water vapor density $\rho_v$ (solid green line) and temperature $T$ (orange and yellow dash-dotted lines) in the vertical direction across the interfacial mixing layer. The mixing layer is located at $x_c=L_{12}$ (see Table \ref{tab:simulation_parameters}). The temperature fluctuation component $T'(x_3)$ of equation (\ref{temperature_continuous}) is plotted with yellow the dash-dotted line, while the non-periodic physical temperature $T$ is plotted with the orange dash-dotted line. Subscripts 1 and 2 refer to cloud and clear-air conditions (see Table \ref{tab:physical_parameters}). 
			Panel (b): The computational domain is a parallelepiped composed of two adjacent cubes. The total height $L_3$ is twice $L_{1,2}$. Subscripts 1 and 2 refer to the horizontal directions parallel to the mixing layer, whereas subscript 3 indicates the vertical direction. %In order to evaluate the evolution of the flow and droplet population properties across the interface, planar statistics are computed for each discrete value of the vertical coordinate $x_3$.
		}
		\label{fig:comp_domain}
	\end{figure}

	The aim here has been to study the transient decay of a top cloud-clear air interface by performing direct numerical simulations of a turbulent shearless mixing layer. This idealized interfacial layer separates two regions. A warmer, droplet-laden cloud region is located at the bottom half of the domain and it is rich in water vapor and kinetic energy. A clear air less energetic area lies in the top half of the computational domain, which is a parallelepiped made up of two adjacent cubes, see Fig. \ref{fig:comp_domain}, panels a and b.
	
	{A turbulent layer without mean shear is a reasonable model of turbulence at the boundary between atmospheric clouds and the surrounding undersaturated air. This flow is considered simple because it is free of the complications associated with the production of turbulence due to the mean flow. However, in reality it is home to dynamic aspects that are not obvious and have not yet been fully described or understood. %, and not conveniently exploited for the control of turbulent motions. 
		We briefly list some of them hereafter. To form a shear-free turbulent layer, it is sufficient that two contiguous non-sheared regions with a different integral scale and  the same kinetic initial energy interact. This, in time, can generate a shear-free layer that hosts a gradient of kinetic energy\cite{Tordella2012}. All the shear-free turbulent  layers are in-homogeneous, thus anisotropic, and also intermittent at the small scale level. Anisotropy appears in the main diagonal of the velocity fluctuation gradient, which is characterised by a substantial absence of significant off-diagonal terms\citep{Tordella_2008, Tordella2011}. The growth or reduction of the thickness of the layer is controlled to a great extent by the concomitant action of the local kinetic energy and spatial macroscale gradients. If  these gradients have opposite signs across the layer, the thickening of the layer decelerates, and vice versa, if the signs are concordant\cite{Tordella_2006}. If the layers are stratified in density, substrates are formed. In the case of stable stratification, the energy collapses below the two formation region levels. Flow transport across the layer is blocked. In the case of unstable stratification, the sublayer hosts an accumulation of energy, and transport is enhanced\cite{Gallana_2022}.
		\\
		Moreover, it should be noted that for the case where the most energetic portion beside the layer (cloud region) hosts both supersaturated water vapor and water droplets, recent results have shown that i) the small-scale intermittency of the air flow in the mixing layer is higly correlated with the drop collision rate of both thr monodisperse and polydisperse drop size distributions, ii)  a more intense widening of the drop population size spectrum is observed in the interfacial region with respect to what happens inside the homogeneous cloud region. These results have prompted our interest in exploring the correlation between supersaturation fluctuations and the small-scale intermittency of air flow turbulence. A relationship  has here been hypothesized to be responsible for the so-called bottleneck problem associated with the interaction of the evaporation-condensation-coalescence processes present in the formation of cumulus rain.}
	
	%\ref{fig:comp_domain_scheme}. 
	Boussinesq Navier-Stokes equations provide the Eulerian description of the incompressible, stratified, velocity fluctuation, $u_i$, along with active scalar transport equations for temparature, $T$, and water vapor density
	$\rho_v$ \citep{Vaillancourt2001,Andrejczuk2004,Kumar2013,Kumar2014,Gotzfried2017,Kumar2018,Gao2018} 
	\begin{subequations}\label{boussinesq_NS}
		\begin{align}
			\displaystyle \frac{\partial u_j}{\partial x_j}= &\, 0 \label{mass}\\
			\displaystyle \frac{\partial u_i}{\partial t}+u_j\frac{\partial u_i}{\partial x_j}=& -\frac{1}{\rho_0}\frac{\partial p}{\partial x_i}+\nu\frac{\partial^2 u_i}{\partial x_j^2}+\mathcal{B}\delta_{3i} \label{momentum}\\
			\displaystyle \frac{\partial T}{\partial t}+u_j\frac{\partial T}{\partial x_j}=& \kappa\frac{\partial^2 T}{\partial x_j^2}+\frac{\mathcal{L}C_d}{\rho_0c_p} \label{energy}\\
			\displaystyle \frac{\partial \rho_v}{\partial t}+u_j\frac{\partial \rho_v}{\partial x_j}=& \kappa_v\frac{\partial^2 \rho_v}{\partial x_j^2}-C_d \label{vapor}
		\end{align}
	\end{subequations}
	where  $\rho_0$ is the Boussinesq density (that is, the mean density of dry air 1000 m above the sea level), $\nu$ is the kinematic viscosity of the air, $\kappa$ is the heat diffusivity of the air and $\kappa_v$ the mass diffusivity of the water vapor. $\mathcal{L}$ is the latent heat of evaporation of the water and $c_p$ is the specific heat of the air at the mean domain temperature $T_0$. All the physical constants in equations (\ref{boussinesq_NS}, a-d) are summarized in Table \ref{tab:physical_parameters}.
	The Boussinesq approximation allows us to take into account small perturbations of a parcel density of moist air due to local temperature and vapor density variations \citep{Tritton1988}. The buoyancy term, $\mathcal{B}$, in equation (\ref{momentum}) can be expressed as a function of the local values of $T$ and $\rho_v$ 
	\begin{equation}\label{buoyancy}
		\mathcal{B}=g\left[\frac{\Delta T}{T_0}-\frac{\Delta \rho_v}{\rho_0}\left(1-\frac{\mathcal{M}_a}{\mathcal{M}_w}\right)\right]
	\end{equation}
	where ${\mathcal{M}_a}$ and ${\mathcal{M}_w}$ are the molar masses of the air and water, respectively.
	We adopt periodic boundary conditions for the velocity and water vapor density fields in the three directions. The temperature field is non-periodic in the vertical direction and results from the superposition of a triply-periodic scalar field and a constant, negative, vertical temperature gradient. The temperature of the cloud region being higher than the clear-air one, the interfacial mixing layer is subject to an unstable stratification with a squared Froud number, $Fr^2_{int}$, approximately equal to $-7$. This leads to a local increase in the momentum and kinetic energy,  as a result of the Boussinesq body-force term in equation (\ref{momentum}).
	
	{The condensation term $C_d = C_d (x_i, t)$ in the energy and vapor density equation expresses the water vapor mass absorption (depletion) rate at the surface of all the spherical droplets contained in the cubic computational cell of volume  $( \Delta)^3$, 	\cite{Vaillancourt2001}. Since cloud droplets are advected by the turbulent flow, $C_d$ must be determined in the Lagrangian  reference frame used for the liquid water mixing ratio, which is described below in sub-section \textbf{II B}. 	However, in order to use $C_d$, in equations (\ref{energy}) and (\ref{vapor}),  it should be represented in the Eulerian frame of reference. The condensation rate field is determined as:}
	
	\begin{equation}
		C_d=\frac{1}{\Delta 
			x^3}\frac{dm_w}{dt}=\frac{4\pi\rho_w}{\Delta x^3}\sum_{j= 1}^{N_\Delta} R_j^2(\bm X_j(t))\frac{dR_j(\bm X_j(t))}{dt}
	\end{equation}
	
	\noindent where $\mathrm{R}_j(t)$ and $\bm X_j(t)$ are the radius and the coordinate of the $j-th$ drop contained within the grid cell, respectively, and  $N_\Delta$ represents the number of drops inside each grid cell. The interpolation of Eulerian field values at grid points to the positions occupied by the water droplets inside the cell is obtained via second-order Lagrange polynomials. An inverse procedure is used to calculate the condensation rate, which is determined at the first step at each droplet position and then relocated to the closest of the eight grid vertices. 
	The time derivative, $dR_j/dt$, expresses the droplet condensational shrinkage (growth) rate, as defined in Equation (\ref{condensational_growth}).
	
	\subsection{Lagrangian droplet dynamics and droplet populations}
	The Lagrangian motion of each k-th droplet in the physical system is modeled by a tracker of the type \citep{Vaillancourt2001,Ireland2012}: 
	\begin{subequations}\label{lagrangian_tracker}
		\begin{align}
			\displaystyle\frac{dX_{ki}}{dt}=&V_{ki} \label{drop_velocity}\\
			\displaystyle\frac{dV_{ki}}{dt}=&\frac{1}{\tau_{k}}\left[u_i\left(X_{ki},t\right)-V_{ki}\right]+g\delta_{i3}\left(1-\frac{\rho_0}{\rho_w}\right) \label{drop_acceleration}
		\end{align}
	\end{subequations}
	which features two vector equations for position, $X_{ki}$, and velocity, $V_{ki}$, of a droplet within the reference frame, where $i$ indicates the direction. The momentum equation is derived for low-Reynolds spherical droplets (e.g. \cite{Vaillancourt2001,Gotzfried2017}) and only accounts for the contribution of Stokes' drag and gravity, while the effects of Faxen and Basset's history force are negligible \citep{Shaw2003,Gotzfried2017,Golshan2021}. The inertia of a spherical droplet is proportional to its surface and is often expressed through a characteristic time scale, that is, the droplet response time ($\tau_d$) of the k-th droplet with radius $R_k$ \citep{Devenish2012} 
	\begin{equation}
		\tau_{dk}=\frac{2}{9}\frac{\rho_w}{\rho_0}\frac{R_k^2}{\nu}
	\end{equation}
	which is also the time constant of the solution to equation (\ref{drop_acceleration}) for a steady, homogeneous flow.
	
	{It should be noted that, in a similar way to what is done for the condensation rate field, Eulerian flow field quantities have to be determined at the droplet position to numerically proceed with Lagrangian equations. In this context, we adopted a simplified feedback on the droplet flow. The direct effect of the liquid droplet drag on the velocity field is neglected in the buoyancy term in the momentum equation. The feedback is therefore indirect and is confined to the coupling of the temperature field with the velocity field and the vapour mixing ratio through the condensation rate. The rationale behind this position depends  on the smallness Stokes' numbers of the drops and liquid mass and volume fractions $\sim 10^{-3}$ and $\sim 10^{-6}$, respectively. In fact, for radii in the range $ [1 - 30] \mu$m, the initial transient values of Stokes'numbers are in $ [0.02 - 0.7]$, while the end of transient values are in $ [0.002 - 0.066]$, which means Reynolds numbers of the drops much lower than 1.}
	
	Spherical cloud droplets are assumed to collide and coalesce with full collision and coalescence efficiency whenever their relative distance falls below the sum of the respective radii
	\begin{displaymath}
		\left[\sum_{i=1}^3\left(X_{li}-X_{ki}\right)^2\right]^{1/2}\leq R_l+R_k
	\end{displaymath}
	The single droplet resulting from coalescence conserves the total mass and momentum of the colliding drops.
	
	\begin{figure}[bht!]
		\centering
		\subfloat[a][monodisperse, cloud]{\includegraphics[width=0.48\linewidth]{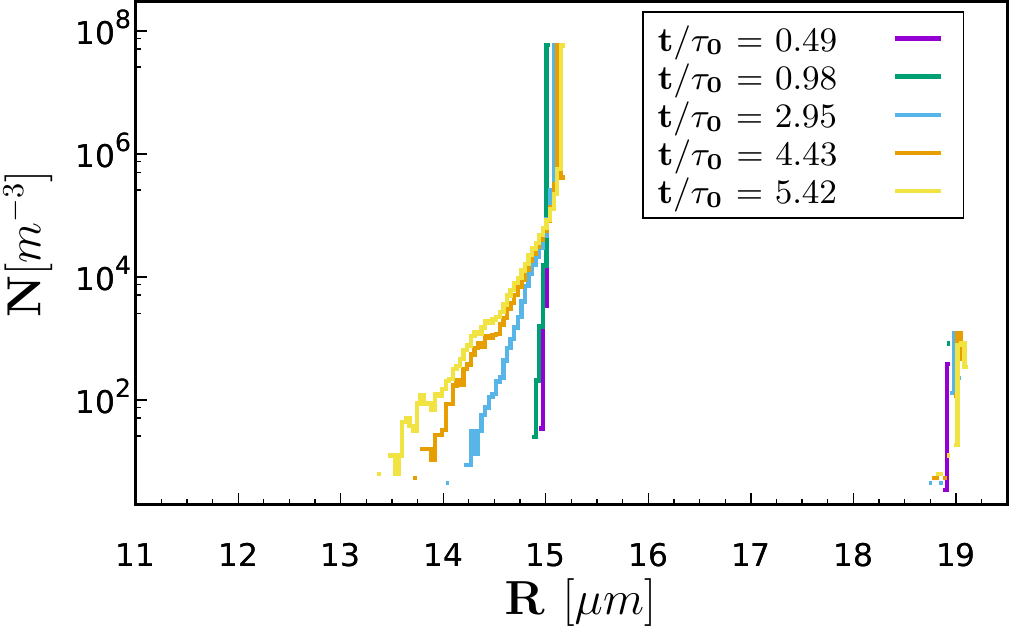}}\quad
		\subfloat[b][monodisperse, interface]{\includegraphics[width=0.48\linewidth]{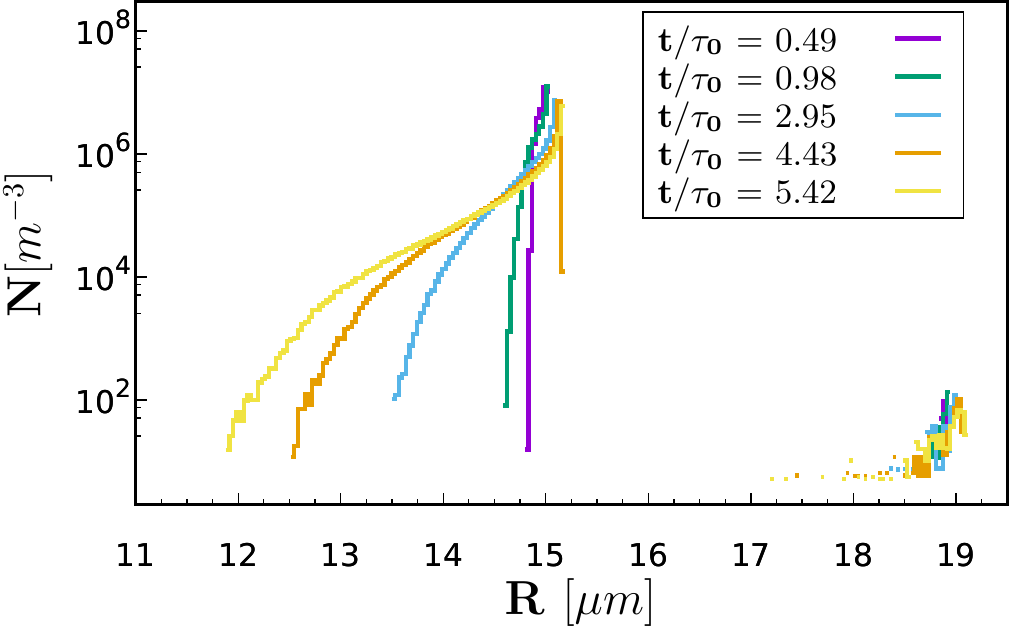}}
		\quad
		\subfloat[c][polydisperse, cloud]{\includegraphics[width=0.48\linewidth]{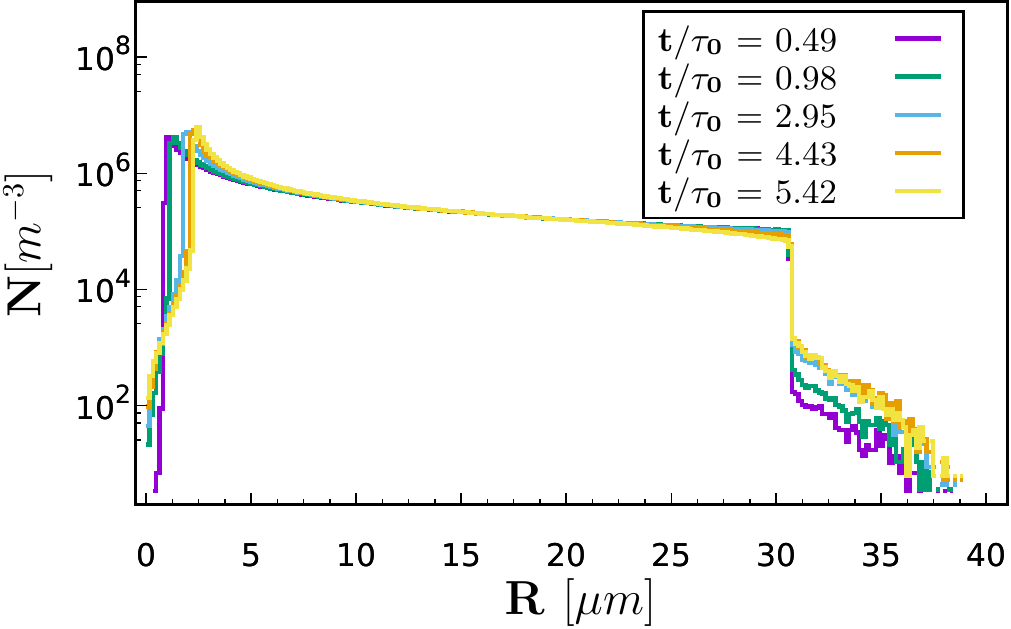}}\quad
		\subfloat[d][polydisperse, interface]{\includegraphics[width=0.48\linewidth]{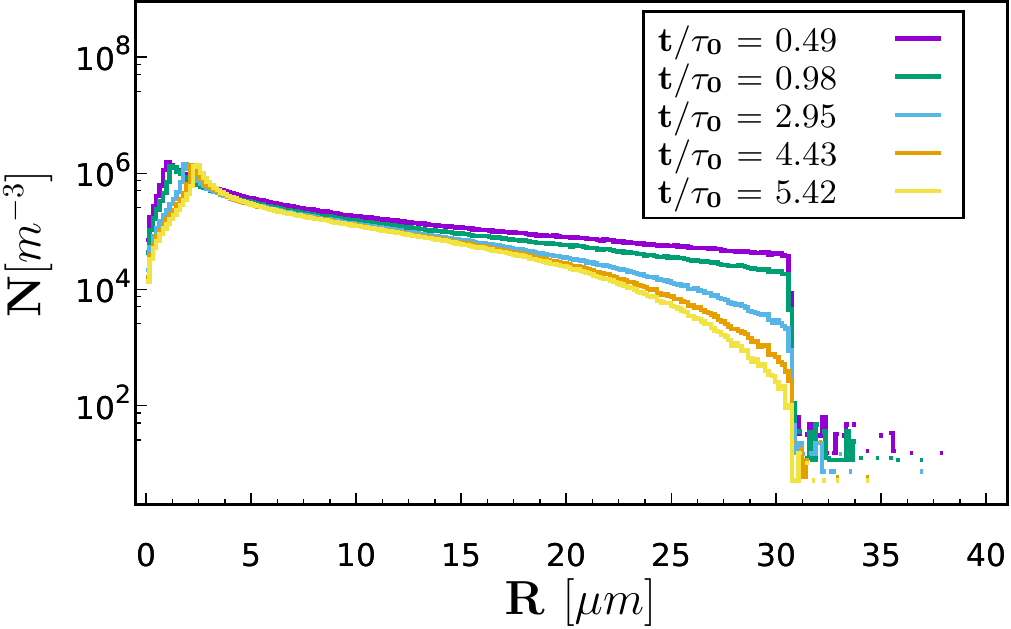}}
		
		\caption{Droplet size distributions as a function of the radius classes. Panels (a) and (b): size distribution for a monodisperse population (R = 15 $\mu$m) in the cloud and interface regions, respectively. Panels (c) and (d): size distribution for a polydisperse (equal mass in the droplet volume classes, R $\in$ [0.6 - 30] $\mu$m) population in the cloud and interface regions, respectively.}
		\label{fig:drop_distributions}
	\end{figure}
	
	Both monodisperse and polydisperse droplet size distributions are considered in the present simulation campaign (Fig. \ref{fig:drop_distributions}). At the beginning of the simulation, the droplets are randomly distributed in the cloud region of the computational domain, where the clear air region is initially void.  {The number of droplets, $N_{tot-mono} = 8 \cdot 10^6$, for the monodisperse population, is determined from the typical liquid water content $LWC_0 \sim 0.8 g/m^{3}$ encountered in warm cumulus clouds and  the choosen  initial monodisperse radius, $R_{0,mono}= 15 \mu$m,  
		\begin{equation}\label{LWC_definition}
			N_{tot-mono}=LWC_0 \frac{4}{3}\pi\rho_w R_{0,mono}^3.
		\end{equation}
		\noindent The faster dynamics of the droplet spectrum inside the highly intermittent mixing layer, with respect to that shown in the nearly Gaussian cloud turbulence, should be noted. 
		Panels a and b in Fig. \ref{fig:drop_distributions} highlight the intense acceleration of the broadening of the droplet spectrum in the interfacial layer (standard deviation time variation: $0.015(t/\tau_0) + 0.05$ in the cloud and $0.23 (t/\tau_0 ) + 0.003$ in the mixing, see figures 7,11 and table 3  in Golshan et al. (2021)\cite{Golshan2021}. Panels c and d in Fig. \ref{fig:drop_distributions}) show, for an initially flat polydisperse size distribution, a faster rate  of  modification toward the typical peaky shape in the interface than in the cloudy region. In fact, the temporal narrowing of the standard deviation goes like  $-0.19 (t/\tau_0 ) + 19.7 $ in the cloud and as $-0.74 (t/\tau_0) + 17.94$ in the mixing, see figures 8,12 and table 3 in Golshan et al. (2021) \cite{Golshan2021}.}
	
	Broad droplet size distributions have been observed in both in-situ measurements of forming shallow cumulus clouds \citep{Siebert2017} and in laboratory experiments \citep{Chandrakar2016}. These distributions usually show a peak for relatively small radii ($1\div10 \si{\micro\meter}$), which is accompanied by a monotonical decrease in concentration as the radius increases. However, the existence of a general and ubiquitous functional shape of the droplet size distribution in shallow cumulus clouds is still a matter of debate \citep{Chandrakar2020}. { Without any claim of generality, we introduce an initial polydisperse distribution in which the same mass is allocated to each class of radii. Each volume class gathers droplets that have roughly the same volume } \citep{Golshan2021}.
	
	\begin{figure}[bht!]
		\centering
		\includegraphics[width=0.6\linewidth]{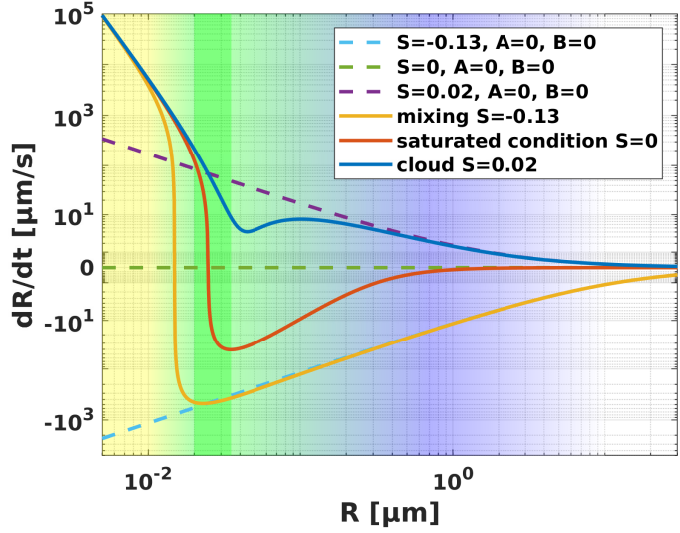}
		\caption{The plot shows the condensational growth rate (Equation \ref{condensational_growth}) for a drop of given radius $R$ for supersaturated ($S=0.02$, as in the cloud region of the present work), saturated ($S=0$) and subsaturated ($S=-0.3$) conditions. The contributions due to the effect of both surface tension (Kelvin) and curvature (Raoult) are negligible for droplets of radius above $1 \mu m $, such as the ones considered in the present work. The dashed lines only represent the effect of supersaturation, when the Kelvin and Raoult terms are set equal to zero. The vertical green bar highlights the range of radii below-above which the Raoult and  Kelvin terms dominate, respectively.}
		\label{fig:droplet_growth_model}
	\end{figure}
	A droplet is subject to ambient supersaturation, which is obtained through a polynomial interpolation with the neighboring cell values. The condensation-evaporation rate of the spherical j-th droplet can be estimated according to \cite{Squires1952,Cooper1989,Ghan2011}
	\begin{equation}\label{condensational_growth}
		\frac{dR_k}{dt}=\frac{K_s}{R_k}\left(S-\frac{A}{R_k}+\frac{Br_d^3}{R_k^3}\right)
	\end{equation}
	where $S$ is the supersaturation or saturation deficit (see section \ref{sec:FIELDS}), $A$ is the Kelvin coefficient, $B$ is the hygroscopicity parameter, $R_k$ is the k-th droplet radius, and $r_{d}$ is the accumulation radius. The second and the third terms on the right-hand side are known as the Kelvin and Raoult terms, respectively. The Kelvin term describes the effect of droplet curvature and surface tension, while the Raoult term indicates aerosol hygroscopicity. The diffusion coefficient $K_{s}$ is slightly sensitive to local equilibrium thermodynamics \citep{RogersYau1996,PruppacherKlett2010,Ghan2011}. It includes the self-limiting effects of latent heat release. This diffusion coefficient is  considered to be constant in the literature, for typical warm cloud conditions, where the characteristic heat flux due to latent heat from a small variation in the droplet temperature is of the same order as the heat flux due to thermal conduction for the same temperature difference. The temperature dependence of this constant is weak (the $K_{s}$ value in m$^2$ s$^{-1}$ ranges from $5.07 \cdot 10^{-11}$ at $T=270$ K, to $1.17 \cdot 10^{-10}$ at  $T=293$ K\citep{Kumar2014, Rogers1989}). In agreement with our volume averaged initial temperature of $281$ K, we used the value $8.6 \cdot 10^{-11}$ m$^2$ s$^{-1}$. The interpolation of Eulerian field values at grid points to the position occupied by the water droplets inside the cell is obtained via second-order Lagrange polynomials. An inverse procedure is then used for the calculate of the condensation rate, which is determined at the first step at each droplet position and then relocated to the closest of the eight grid vertices. A collision is hypothesized to occur when the distance between the centers is equal to or less than the sum of their radii. Such collisions are assumed to be completely inelastic.
	
	Supersaturation
	\begin{displaymath}
		S(\bm{x},t)=\frac{\rho_v(\bm{x},t)}{\rho_{vs}(T)}-1=RH-1
	\end{displaymath}
	is defined as the ratio between the water vapor and  the saturated vapor densities (i.e. the relative humidity) minus 1. The relative humidity $RH=\rho_v/\rho_{sv}$ and supersaturation (or saturation deficit) are functions of the saturated vapor density, $\rho_{vs}$, whose dependence on temperature is described by the Clausius-Clapeyron equation \citep{RogersYau1996}
	\begin{equation}\label{clausius_clapeyron}
		\frac{d\rho_{vs}}{\rho_{vs}}=\left(\frac{\mathcal{L}}{R_vT}-1\right)\frac{dT}{T}
	\end{equation}
	where $\mathcal{L}$ is the latent heat of evaporation (condensation) and $R_v$ is the gas constant of the water vapor. It is hypothesized that the droplet-gas interphase coupling is negligible \citep{Balachandar2010}, and the droplet motion therefore does not exert any relevant dynamical effect on the carrier field. Conversely, the turbulent field affects the motion of the water drops to a great extent. Phase transition at the droplet surface results in the exchange of water vapor and latent heat between the two phases, thus perturbing the buoyancy term in momentum equation (\ref{momentum}). 
	
	Since the coefficients $A$, $B$, and $r_d$ are hypothesized to be constant, the droplet growth rate mainly depends on the local value of $S$ and on the droplet radius $R$. The droplet growth (shrinkage) rate (\ref{condensational_growth}) is plotted in Figure \ref{fig:droplet_growth_model} for three constant values of supersaturation $S$, where the competing effects of the Kelvin and Raoult terms can be appreciated by observing the orange curve, which describes a saturated environment. In the present conditions, the Kelvin effect becomes important for $R<1\,\si{\micro\meter}$, and it is soon outweighed by the Raoult effect as the droplet (aerosol) radius falls below $\cong 23\,\si{\nano\meter}$. The Raoult term is the term that is prevalent below this threshold.
	
	\subsection{Numerical experiment. Computational domain, initial and boundary conditions and DNS algorithm}
	
	All the numerical experiments discussed in this paper have been performed over a 3D $512^2\times1024$ cartesian mesh grid \citep{Golshan2021}. This parallelepiped-shaped domain (see Figure \ref{fig:comp_domain}) is made up of two adjacent cubes of $512^3$ cells each. The volume of the domain is $L_{12}^3$, with $L_{12}$ being the length of each cube edge. Two initial zero-mean, homogeneous, isotropic air fluctuation fields are generated inside the two cubes. The turbulent spectra show the same functional shape and hence the same integral scale. {The cube in the lower half of the parallelepiped - which from now on is referred to as the cloud region - initially hosts a higher turbulent kinetic energy, $E$, and dissipation (\textit{decay}) rate, $\varepsilon$, than the upper cube, which models a clear air region, see Figures \ref{fig:comp_domain_scheme} and \ref{fig:kinetic}, Panels a and b. The initial integral scale is set equal in the two regions so as not to introduce a further control parameter - the integral length gradient across the layer - on the interface evolution and the related transport dynamics  \citep{Tordella_2006, Tordella2012}}.
	
	{The root mean square velocity in the more energetic region is $u_{rms}\cong0.11\si{\meter\per\second}$, which represents the large-scale energy in the cumulus spectral subrange of wavelengths  0.002 to 0.25 m.
		Since our system is time decaying, the initial dissipation rate was purposely set  high in order to reach  the commonly   values observed in cumulus clouds in the central part of the transient. However, the initial dissipation rate $\varepsilon\sim 500 \;  \si{\centi\meter^2\per\second^3}$ is  of the same order as those measured by \cite{MacPherson1977} in cumulus clouds  in the proximity of the top (cloud \# 1 measurement, 100 m below the cloud top, height of the top 4800 m) although in the presence of a much higher kinetic energy of the air fluctuation ($rms \sim 2$ m/s). Lower values ($10 \sim 20 \; \si{\centi\meter^2\per\second^3}$) have been reported by \cite{Siebert2006,Siebert2009,Lehmann2009,Siebert2010}, and they are obtained  during the transient decay, see Fig.s \ref{fig:kinetic}, Panels c and d. The estimated Kolmogorov scale is $\eta_1\sim\left(\nu^3/\varepsilon_1\right)^{1/4}\cong 0.5\si{\milli\meter}$. The edge of a cube grid cell in the physical domain is $\Delta x=1\si{\milli\meter}$, and the highest resolved wavenumber is $k_{max}=\pi/\Delta x=\pi\cdot10^3\si{\meter^{-1}}$ \citep{Pope2000}. Since we have $k_{max}\eta_1\cong 1.6$, the resolution is acceptable for the problem at hand \citep{Ishihara2009,Ireland2012}. An appropriate time step for advancement is computed from the initial $u_{rms}$ values of the more energetic cloudy region $\Delta t=4.7\cdot10^{-4}\si{\second}$. }
	
	\subsection{Initial and boundary conditions for the flow velocity, temperature and vapor fields}\label{sec:FIELDS}
	
	Two homogeneous isotropic solenoidal turbulent fields, with zero-mean velocity and different kinetic energies, that mix at a common interface, are studied in this numerical experiment. A smoothing function, $p(x_3)$, is applied  to modulate the velocity and scalar vapor fields along $x_3$ \citep{Tordella_2006, Gallana_2022}
	\begin{displaymath}
		u_i(x_j)=u_{i1}(x_j)p(x_3)+u_{i2}(x_j)\sqrt{1-p^2(x_3)}
	\end{displaymath}
	\begin{displaymath}
		\rho_v=\rho_{v1}p(x_3)+\rho_{v2}\sqrt{1-p^2(x_3)}
	\end{displaymath}
	\begin{displaymath}
		p(x_3)=1+\tanh{\left[a\frac{x_3}{L_3}\right]}\tanh{\left[a\left(\frac{x_3}{L_3}-\frac{1}{2}\right)\right]}\tanh{\left[a\left(\frac{x_3}{L_3}-1\right)\right]}
	\end{displaymath}
	where $\rho_{v1}=\rho_{vs}(T_1)RH_1$ and $\rho_{v2}=\rho_{vs}(T_2)RH_2$ were chosen to obtain the desired level of supersaturation in both regions (see Table \ref{tab:physical_parameters}). 
	Direction $ x_3$ is the inhomogeneous direction and $L_3$ is the width of the computational domain in the $x_3$ direction. Constant $a$ determines the initial mixing layer thickness $\Delta$, which is  conventionally defined as the distance between the points with normalized energy values of 0.25 and 0.75, whenever the low energy side is mapped to zero and the high energy side to one. When
	$ a = 12 \pi$, the initial  $\Delta / L_3 $ ratio is about 0.026, a value that was chosen so
	that the initial thickness would be large enough to be resolved but small enough to have large regions of homogeneous turbulence during the simulations.
	
	The initial distributions of the velocity, temperature and water vapor density fields in the vertical direction are plotted in Figure \ref{fig:comp_domain} a.
	
	The same initial values of $T$ and $\rho_v$ are defined for all the cells of a horizontal plane and are thus functions of their vertical position with respect to the interface. As in \cite{Iovieno2014,Golshan2021}, the vapor field is periodic and continuous in the three directions, whereas the temperature field %is composed of a continuous term and a linear, discontinuous term
	
	\begin{equation}\label{temperature_full}
		T\left(x_3,0\right)=T'\left(x_3,0\right)+T_0+G\frac{x_3}{L_3}
	\end{equation}
	is composed of the sum of a vertical, triple-periodic fluctuating temperature $T'(x_3,t)$, a static component $Gx_3$, and a global average temperature $T_0$. The periodic term $T'$ in equation (\ref{temperature_full})  is defined with a hyperbolic tangent
	\begin{equation}\label{temperature_continuous}
		T'\left(x_3,0\right)=\frac{T_2-T_1}{2}\cdot \left[ \tanh\left(a (\frac{x_3}{L_3} -\frac{1}{2})\right)-\frac{2x_3}{L_3}+1\right]
	\end{equation}
	However, the code is required to solve the periodic field $T'$. Equation (\ref{energy}) then becomes 
	\begin{displaymath}
		\displaystyle \frac{\partial T'}{\partial t}+u_1\frac{\partial T'}{\partial x_1}+u_2\frac{\partial T'}{\partial x_2}+u_3\frac{\partial \left(T'+Gx_3\right)}{\partial x_3}= \kappa\nabla^2T'+\frac{\mathcal{L}C_d}{\rho_0c_p}
	\end{displaymath}
	The cloud-clear air interface is located in the center $(x_3-x_c)/L_3\cong 0$, with $x_c=L_{12}$. We define  the distance between the points whose normalized temperature $(T-T_{min})/(T_{max}-T_{min})$ is 0.75 and 0.25, respectively, as the width of mixing layer region $\Delta$ \citep{Veeravalli1989, Tordella_2006, Tordella2011}. %The vertical temperature gradient across the mixing layer is $\left.\nabla T\right\vert_{mix}\cong-5\,\si{\kelvin\per\meter}$. 
	The squared Brunt-V\"{a}is\"{a}l\"{a} frequency, $\mathcal{N}^2=g \frac{\delta T}{T_0} \frac{1}{\Delta}  \cong-0.69\,\si{\hertz^2}$, is negative, and thus describes an unstable environment. The internal Froude number associated with this stratification  is initially %\citep{Tritton1988}
	\begin{displaymath}
		\textrm{Fr}_{int}^2=\frac{u_{rms,av}^2}{\mathcal{N}^2\Delta^2}\cong-7
	\end{displaymath}
	The saturated vapor density, the relative humidity and the supersaturation are computed with the values of $T$ expressed by equation (\ref{temperature_full}). %Hence, the supersaturation field is periodic but discontinuous in the vertical direction.
	
	% TABLES
	% TABLES
	% TABLES
	\begin{table*}
		\caption{The key physical parameters used in the numerical experiments.}
		\footnotesize
		\centering
		\begin{tabular}{p{0.55\textwidth} m{0.1\textwidth} m{0.12\textwidth} m{0.12\textwidth}}
			\hline
			\textbf{Quantity} & \textbf{Symbol} & \textbf{Value} & \textbf{Unit}\\
			\hline
			Latent heat of evaporation & $\mathcal{L}$ & $2.48\cdot10^6$ & $\si{\joule\per\kilo\gram}$ \\
			Heat capacity of the air at a constant pressure & $c_p$ & 1005 & $\si{\joule\per\kilo\gram\per\kelvin}$ \\
			Gravitational acceleration & $g$ & 9.81 & $\si{\meter\per\second^2}$ \\
			Molar mass of the water & $\mathcal{M}_w$ & 18 & $\si{\kilo\gram\per\kilo\mole}$ \\
			Gas constant of the water vapor & $R_v$ & 461.5 & $\si{\joule\per\kilo\gram\per\kelvin}$ \\
			Molar mass of the dry air & $\mathcal{M}_a$ & 29 & $\si{\kilo\gram\per\kilo\mole}$ \\
			Gas constant of the air & $R_a$ & 286.7 &  $\si{\joule\per\kilo\gram\per\kelvin}$ \\
			Diffusivity of the water vapor mass & $\kappa_v$ & $2.52\cdot10^{-5}$ & $\si{\meter^2\per\second}$ \\
			Thermal conductivity of the dry air & $K$ & $2.5\cdot10^{-2}$ & $\si{\watt\per\meter\per\kelvin}$ \\
			Liquid water density & $\rho_w$ & 1000 & $\si{\kilo\gram\per\meter^3}$ \\
			Dry air density at an altitude of 1000 m & $\rho_0$ & $1.11$ & $\si{\kilo\gram\per\meter^3}$ \\
			Dry air kinematic viscosity & $\nu$ & $1.5\cdot10^{-5}$ & $\si{\meter^2\per\second}$ \\
			Average temperature of the whole domain & $T_0$ & $281.16$ & $\si{\kelvin}$ \\
			Average temperature of the cloud region & $T_1$ & $282.16$ & $\si{\kelvin}$ \\
			Average temperature of the clear air region & $T_2$ & $280.16$ & $\si{\kelvin}$ \\
			Background temperature gradient (unstable) & $G$ & -2/1.024 & $\si{\kelvin\per\meter}$ \\
			Brunt-V\"{a}is\"{a}la amplification factor & $\mathcal{N}^2$ & -0.69 & $\si{\second^{-2}}$ \\
			Droplet growth coefficient & $K_s$ & $8.6\cdot10^{-11}$ & $\si{\meter^2\per\second}$ \\
			Accumulation mode (radius) & $r_d$ & 0.01 & $\si{\micro\meter}$\\
			Kelvin coefficient & $A$ & $1.15\cdot10^{-9}$ & $\si{\meter}$ \\
			Raoult solubility parameter for inorganic, hygroscopic substances such as ammonium sulfate, lithium chloride etc... & $B$ & 0.7 & - \\
			Initial relative humidity in the cloud region & $RH_1$ & 1.02 & - \\
			Initial relative humidity in the clear air region & $RH_2$ & 0.7 & - \\
			Initial liquid water content & $LWC_0$ & $7.9\cdot10^{-4}$ & $\si{\kilo\gram\per\meter^3}$ \\
			\hline
		\end{tabular}
		\label{tab:physical_parameters}
	\end{table*}
	\begin{table*}
		\caption{The key simulation parameters and initial conditions, which are the same for all the runs.}
		\footnotesize
		\centering
		\begin{tabular}{p{0.46\textwidth} m{0.22\textwidth} m{0.15\textwidth} m{0.07\textwidth}}
			\hline
			\textbf{Quantity} & \textbf{Symbol} & \textbf{Value} & \textbf{Unit}\\
			\hline
			Domain size & $L_{1,2}^2\cdot L_3$ & $0.512^2\cdot1.024$ & $\si{\meter^3}$ \\
			Domain discretization & $n_{1,2}^2\cdot n_3$ & $512^2\cdot1024$ & - \\
			Grid step & $\Delta x$ & $10^{-3}$ & $\si{\meter}$ \\
			Initial rms velocity (cloud) & $u_{rms,1}$ & 0.11 & $\si{\meter\per\second}$ \\
			Initial integral scale & $\ell_0$ & $2.65\cdot10^{-2}$ & $\si{\meter}$ \\
			Initial dissipation rate (cloud) & $\varepsilon_1$ & 0.05 & $\si{\meter^2\per\second^3}$ \\
			Initial energy ratio (cloud-clear air) & $E_1/E_2=u_{rms,1}^2/u_{rms,2}^2$ & 6.7 & - \\ 
			Initial Kolmogorov time (cloud) & $\tau_{\eta0}=\left(\nu/\varepsilon_1\right)^{1/2}$ & $1.74\cdot10^{-2}$ & $\si{\second}$ \\
			Initial Kolmogorov length scale (cloud) & $\eta_0=\left(\nu^3/\varepsilon_1\right)^{1/4}$ & $5.1\cdot10^{-4}$ & $\si{\meter}$ \\
			Initial eddy turnover time & $\tau_0=2\ell/(u_{rms,1}+u_{rms,2})$ & 0.35 & s \\
			Initial Reynolds number (cloud) & $\textrm{Re}_\ell=u_{rms}\ell/\nu$ & 196 & - \\
			Droplet Response timescale ($ 1\mu$m,\; $30\mu$m) & $\tau_d = 2 \rho_v R / 9 \rho_0 \nu$ & $4.4 10^{-4}, \; 1.3 10^{-2} $ & s  \\
			Initial droplet Stokes numbers (R $\in 1-30 \; \mu$m)&
			$St = \tau_d / \tau_{\eta_0}$& 0.025 - 0.7 & \\
			Final droplet Stokes number (R $\in 1-30 \; \mu$m)&
			$St = \tau_d / \tau_{\eta_f}$& 0.002 - 0.066 & \\
			Initial Taylor microscale Reynolds number & $\textrm{Re}_\lambda=u_{rms}\lambda/\nu$ & 52 & - \\ 
			Integration time step & $\Delta t=1/20\cdot\Delta x/u_{rms}$& $4.64\cdot10^{-4}$ & \si{\second} \\
			Initial number of droplets (monodisperse distribution) & $N_{tot-mono}$ & $8\cdot10^6$ & -\\ 
			Initial number of droplets (polydisperse distribution) & $N_{tot-poly}$ & $10^7$ & - \\
			Initial droplet radius (monodisperse distribution) & $r_{0,mono}$ & 15 & $\si{\micro\meter}$ \\
			Initial droplet radius (polydisperse distribution) & $r_{0,poly}$ & $0.6\div30$ & $\si{\micro\meter}$ \\
			
			\hline
		\end{tabular}
		\label{tab:simulation_parameters}
	\end{table*}
	
	\subsection{DNS algorithm}\label{SUBSECTION_algo}
	The DNS algorithm is based on the dealiased pseudospectral Navier-Stokes solver described in \cite{Iovieno_2001}. Code versions and releases are available on the official  \href{https://areeweb.polito.it/ricerca/philofluid/software/95-turbulent-flows.html}{Philofluid Research Group website}. This software has been used in several works conducted by the group \citep{Gallana_2022, Golshan2021, Iovieno2014, Tordella2011, Tordella_2008, Tordella_2006} to investigate turbulence self-diffusion in shearless mixings, with passive or active scalars, and water drop populations. Spectral discretization is achieved by means of the Fourier-Galerkin method with pseudo-spectral treatment of the advection terms in the momentum (\ref{momentum}) equation, and scalar transport ones (\ref{vapor} and \ref{energy}. Time integration is performed, according to Ireland (2012)\cite{Ireland2012}, with a second-order explicit Runge-Kutta method  \citep{Brucker2007}. The diffusive terms for the momentum, internal energy (\ref{energy}) and vapor density fields (\ref{vapor}) are computed by means of exponential integration. Droplet velocities \ref{drop_velocity} and accelerations \ref{drop_acceleration} are integrated with a second-order explicit methodand a second-order implicit trapezoidal method, respectively. The implicit structure of the integration scheme used for the equations \ref{drop_acceleration} ensures numerical stability for arbitrary values of $\Delta t$. 
	
	The code stores the velocity, temperature and vapor fields in three-dimensional arrays and distributes them along one direction in both physical and Fourier spaces. The three-dimensional discrete Fourier transform is performed with the FFTW library. A slab-like parallelization is implemented with Message Passage Interface (MPI) standard libraries.
	
	\section{Results. Velocity and supersaturation fluctuations, and turbulence broadening of the droplet size distribution.}\label{SECTION_RESULTS}
	\begin{figure}[bht!]
		\centering
		\subfloat[a][]{\includegraphics[width=0.48\linewidth]{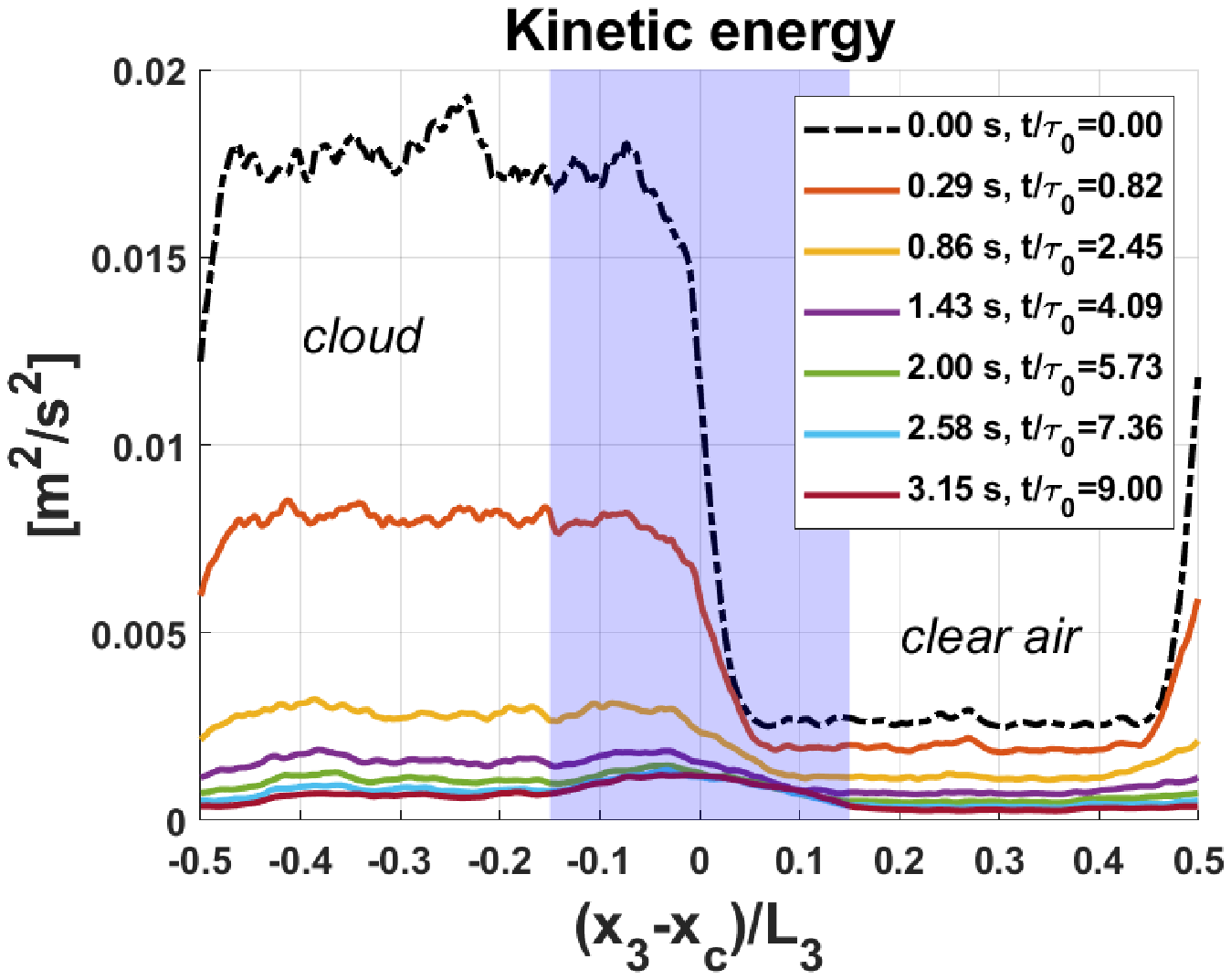}\label{fig:kinetic_a}}
		\subfloat[b][]{\includegraphics[width=0.48\linewidth]{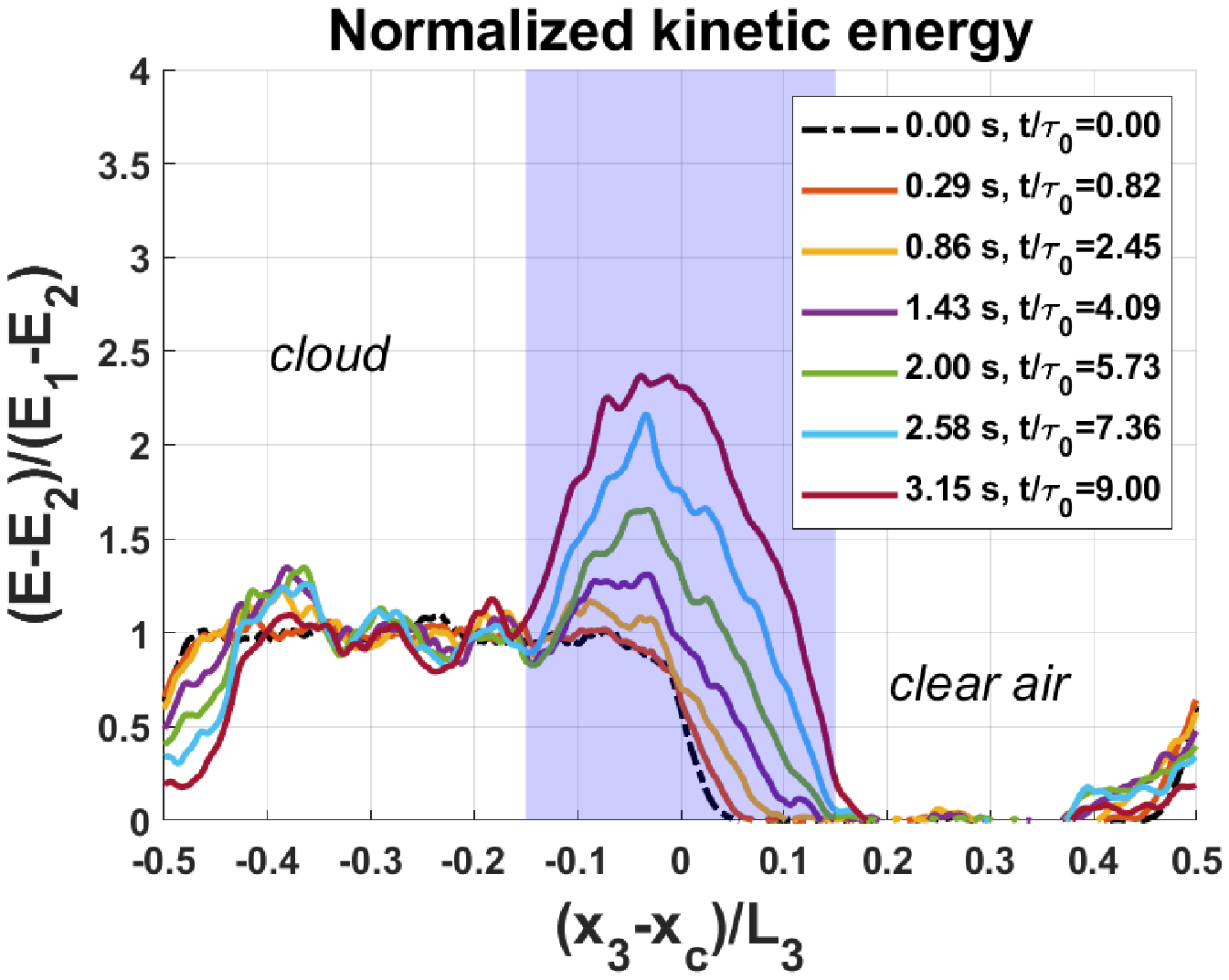}\label{fig:kinetic_b}}\\
		\subfloat[c][]{\includegraphics[width=0.48\linewidth]{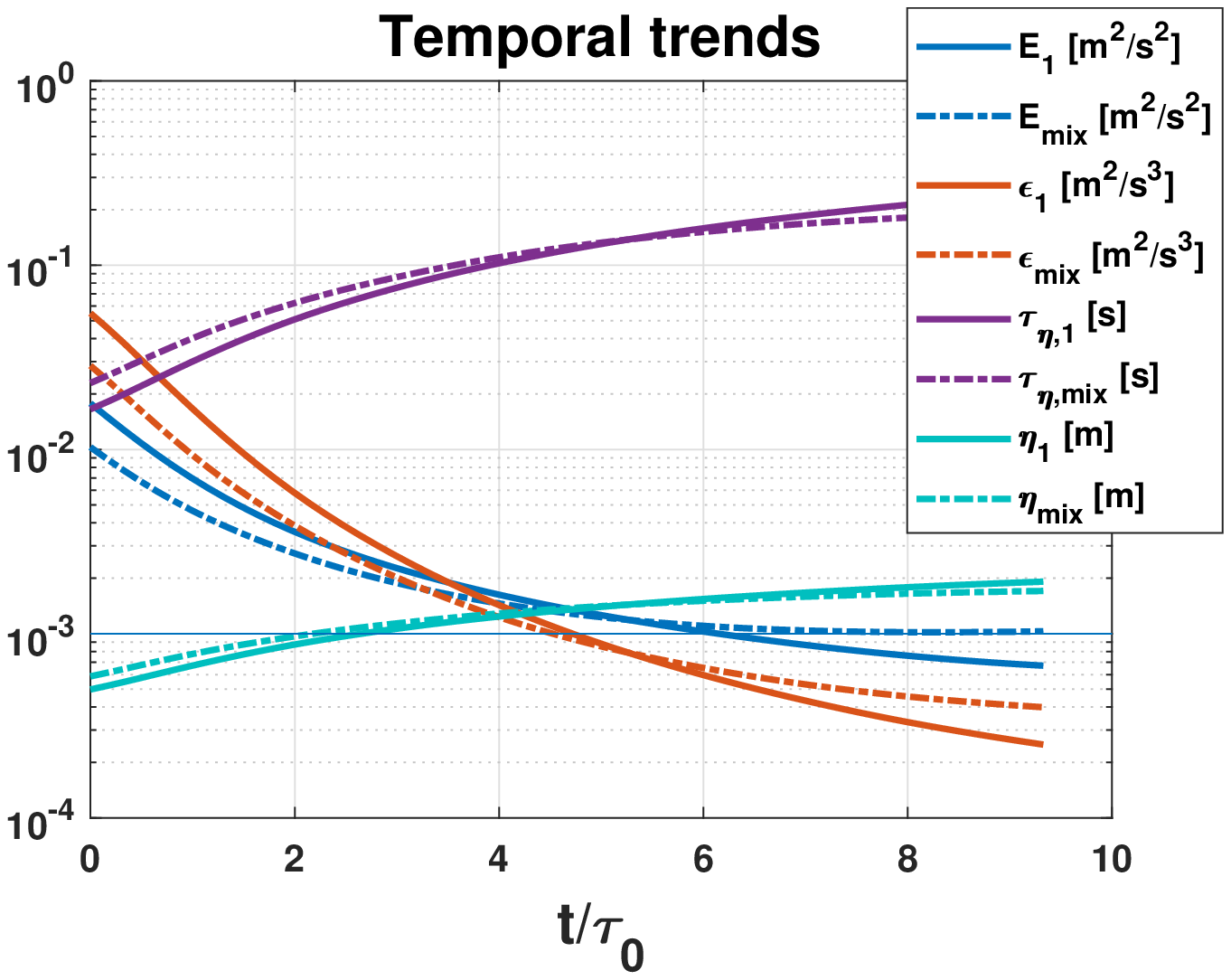}\label{fig:kinetic_c}}
		\subfloat[d][]{\includegraphics[width=0.48\linewidth]{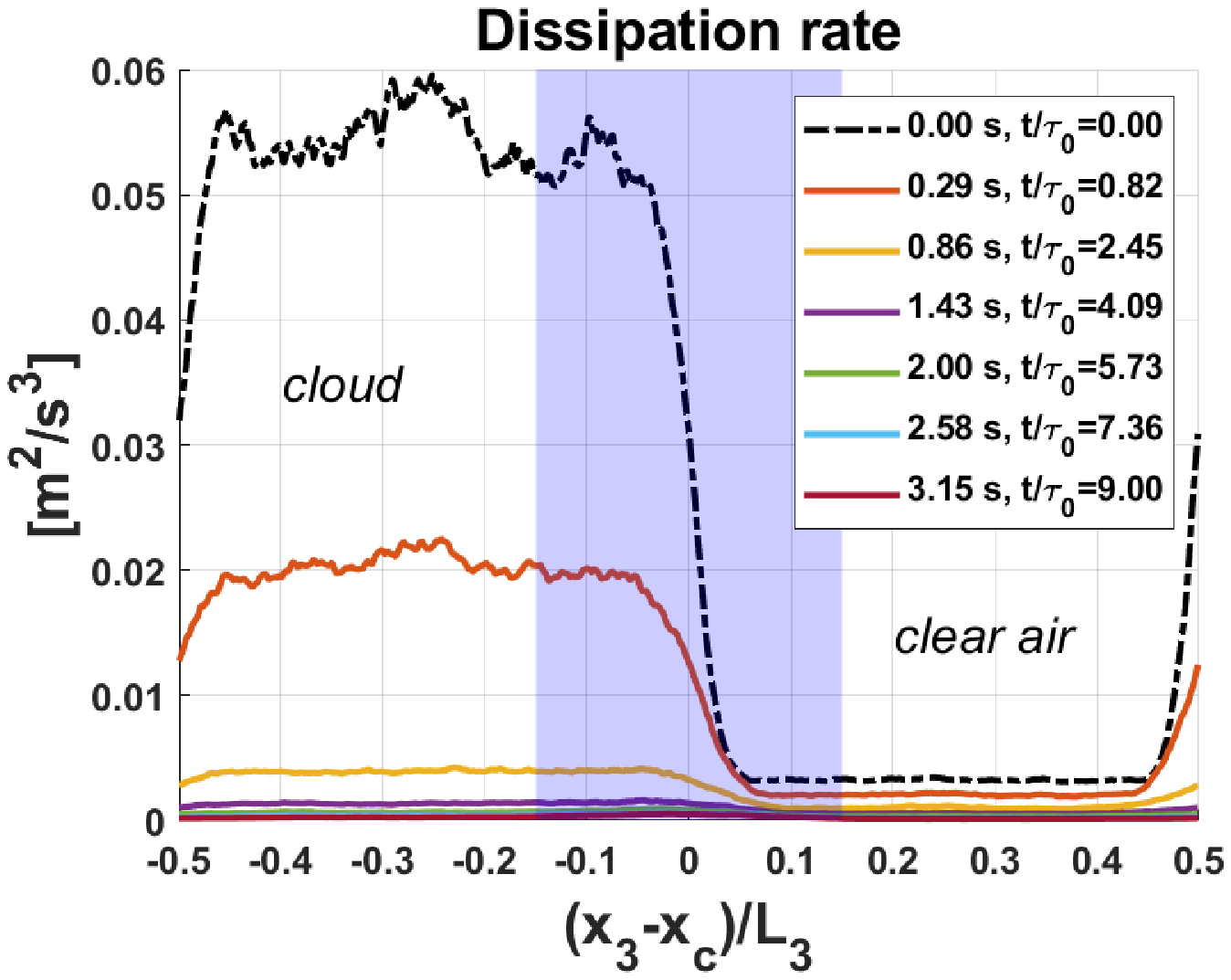}\label{fig:kinetic_d}}
		\caption{\textbf{Trends of the kinetic energy and turbulent dissipation rate}. \protect\subref{fig:kinetic_a} Evolution of the turbulent kinetic energy $E$ across the domain for an initial unstable background temperature gradient. The interface is located at the center of the figure (that is, for $x_3\cong x_c$). Mildly unstable stratification ($\mathrm{Fr}_{int}^2\cong-7$). \protect\subref{fig:kinetic_b} Normalized values of $E(t)$ with respect to the mean kinetic energy in the cloud $E_1(t)$ and clear air $E_2(t)$ regions. The initial energy ratio across the interface is $E_1/E_2=6.7$. The plotted values of $E$ are the planar averages of each horizontal plane. \protect\subref{fig:kinetic_c} Transient evolution of the kinetic energy $E$, the dissipation rate $\varepsilon$, the Kolmogorov time scale $\tau_\eta$ and the Kolmogorov length scale $\eta$ in the cloud and mixing regions (subscripts 1 and 2, respectively). The thin horizontal line indicates the grid width, $\Delta x$, in meters. \protect\subref{fig:kinetic_d} Evolution of the dissipation rate $\varepsilon$ across the domain.}
		\label{fig:kinetic}
	\end{figure}
	
	\begin{figure}[bht!]
		\centering
		%\large\par\medskip
		\subfloat[b][]{\includegraphics[width=0.48\linewidth]{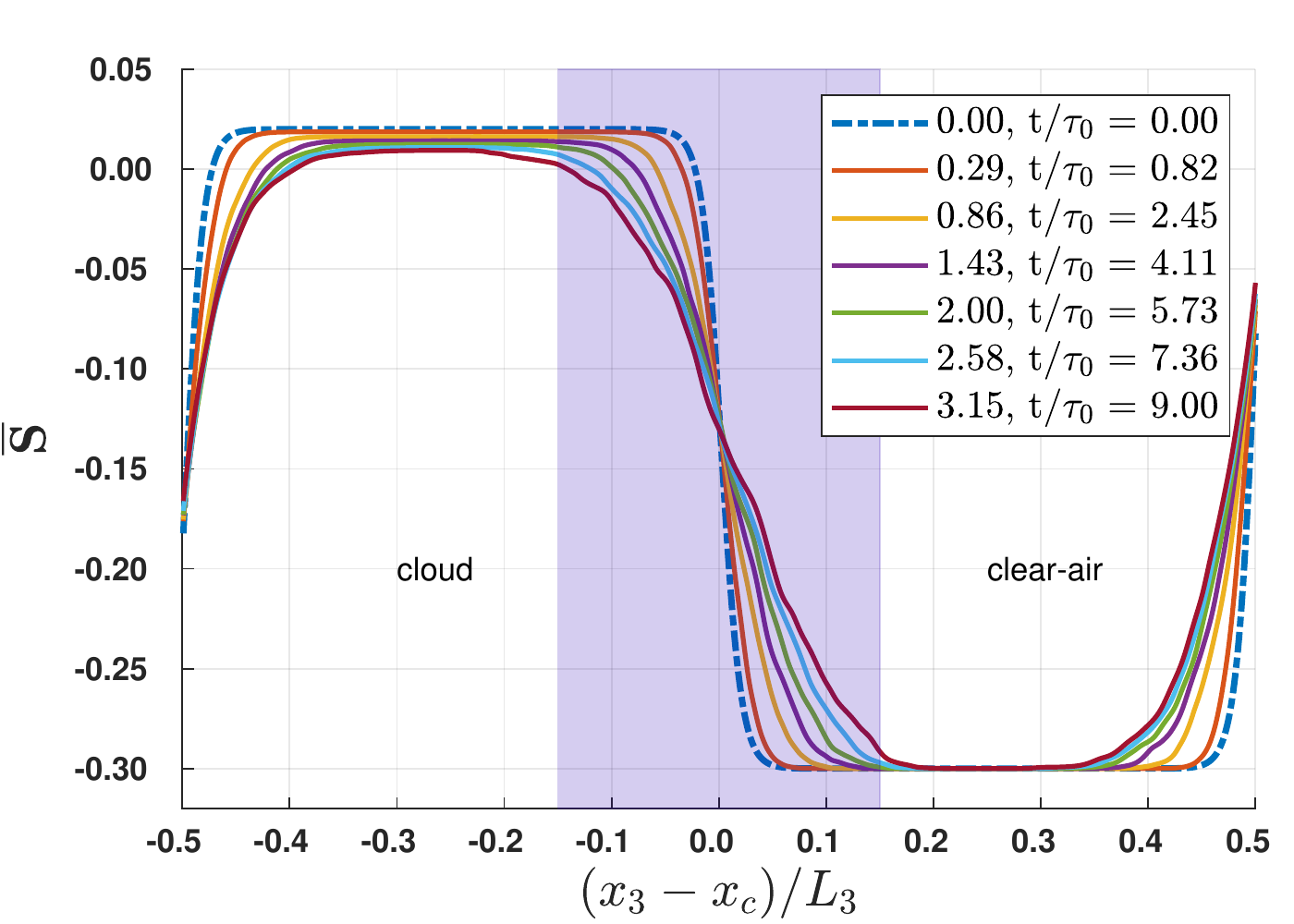}\label{fig:supersaturation_a}}
		\subfloat[b][]{\includegraphics[width=0.48\linewidth]{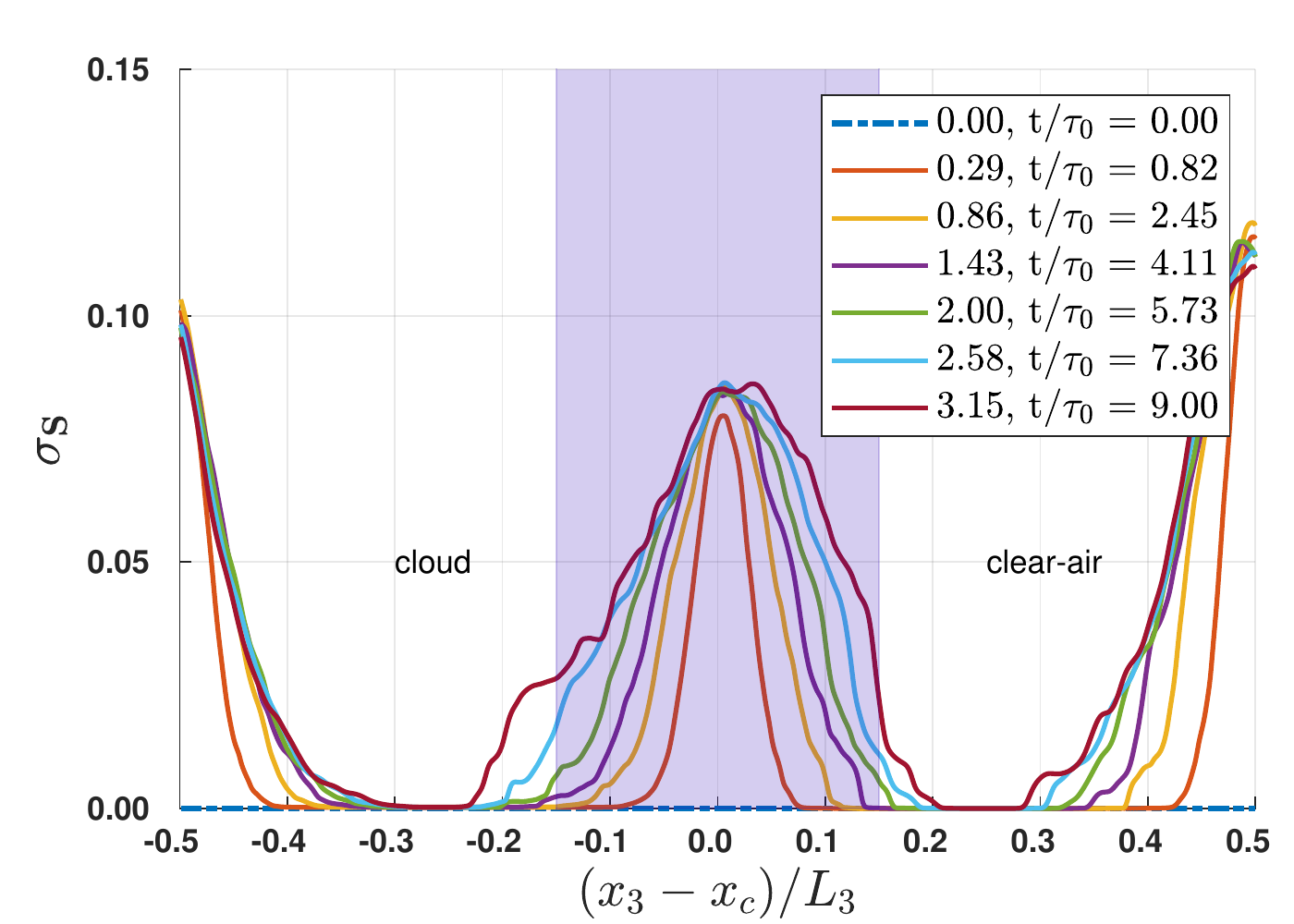}\label{fig:supersaturation_b}}\\
		\subfloat[c][]{\includegraphics[width=0.48\linewidth]{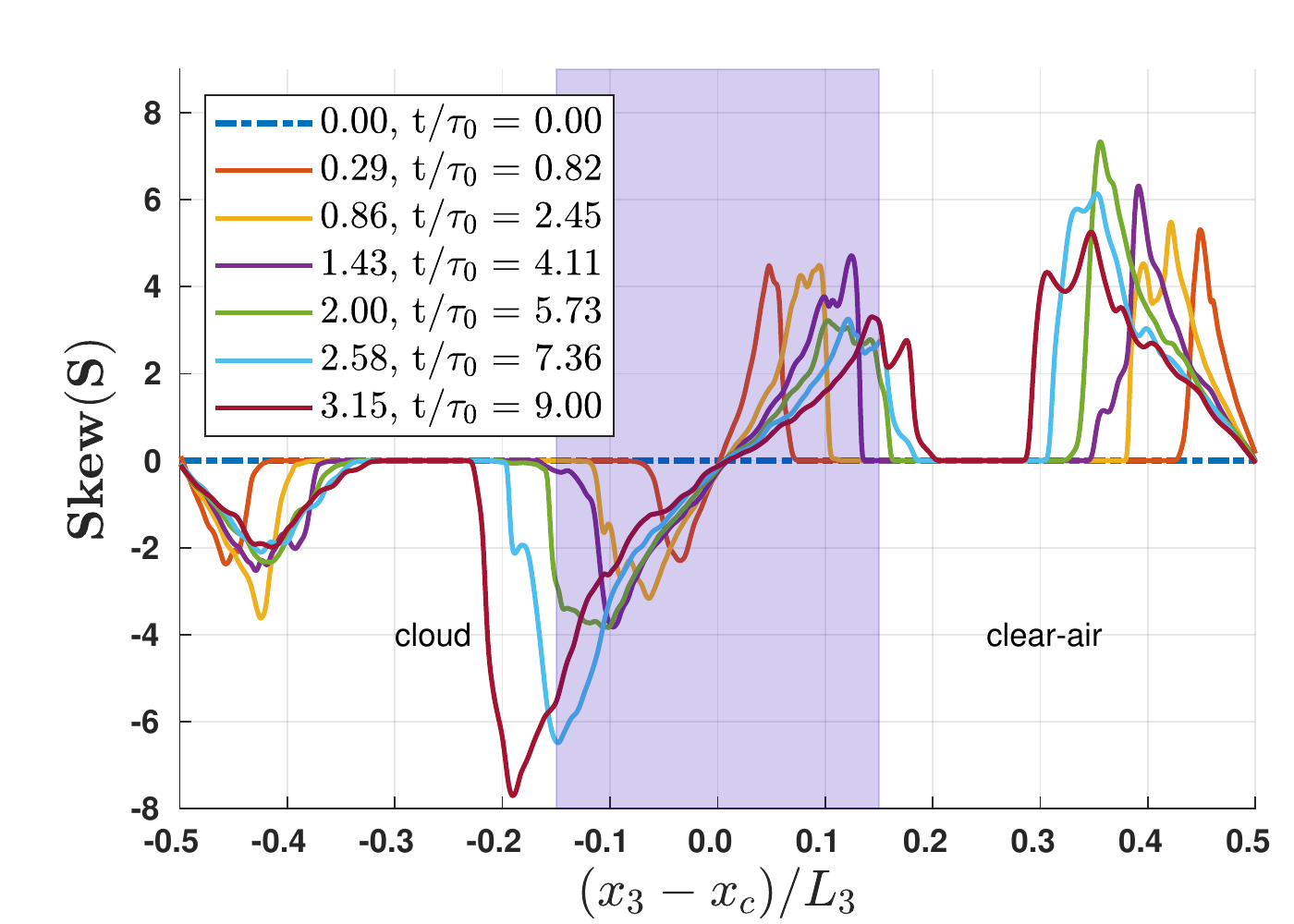}\label{fig:supersaturation_c}}
		\subfloat[d][]{\includegraphics[width=0.48\linewidth]{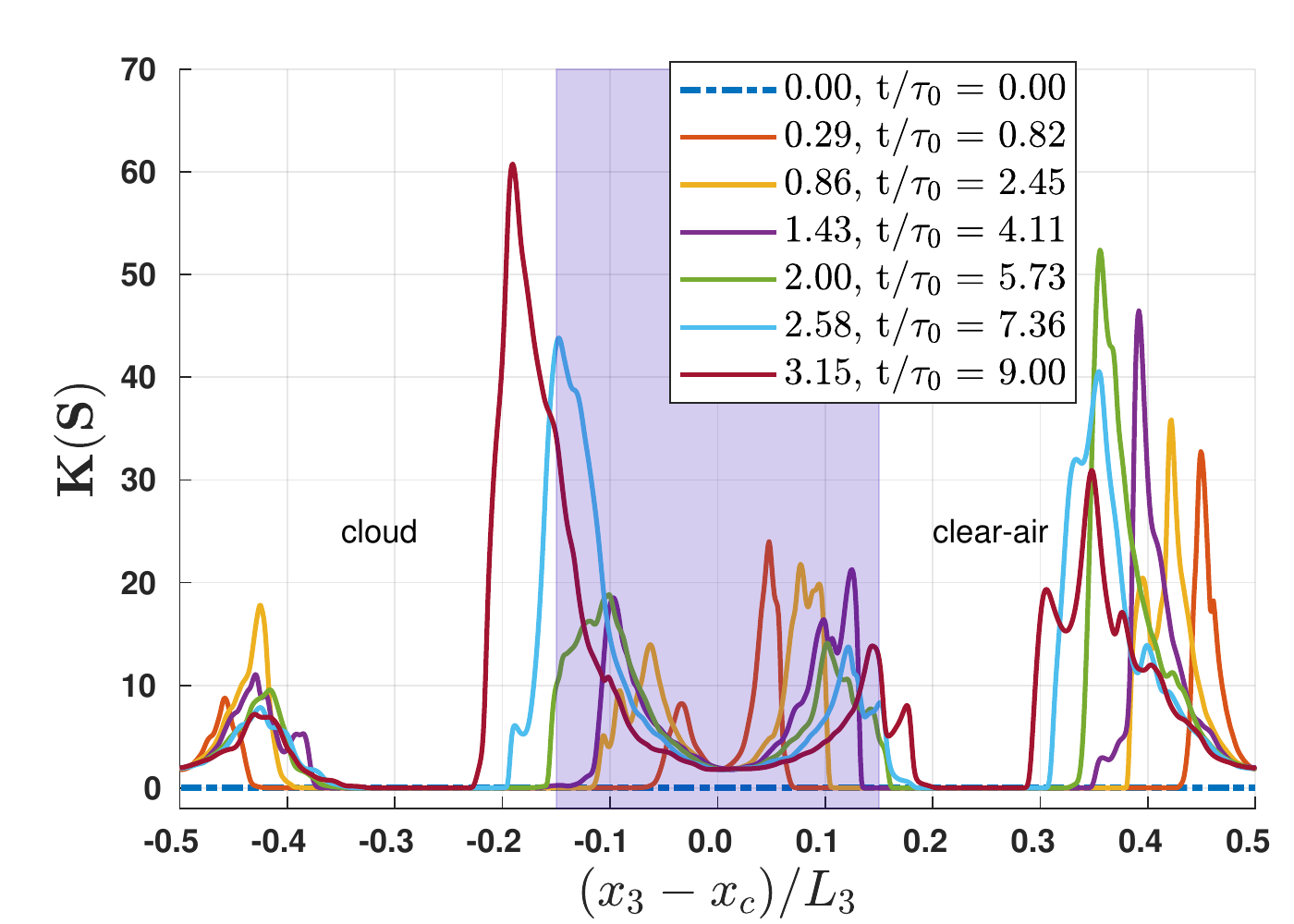}\label{fig:supersaturation_d}}
		\caption{{\textbf{Monodisperse droplet population. Planar averages and statistical moments of supersaturation across the cloudy - under-saturated ambient air interface layer.} \protect\subref{fig:supersaturation_a} Supersaturation (or saturation deficit) across the layer. \protect\subref{fig:supersaturation_b} Standard deviation.  \protect\subref{fig:supersaturation_c} Skewness. \protect\subref{fig:supersaturation_d} Kurtosis. The initial distributions are plotted with black dash-dotted lines. The approximate extension of the interface mixing layer is indicated as the blue-shaded area between the cloudy and clear air regions.}}
		\label{fig:supersaturation}
	\end{figure}
	
	\begin{figure}[bht!]
		\centering
		%\large\par\medskip
		\subfloat[b][]{\includegraphics[width=0.48\linewidth]{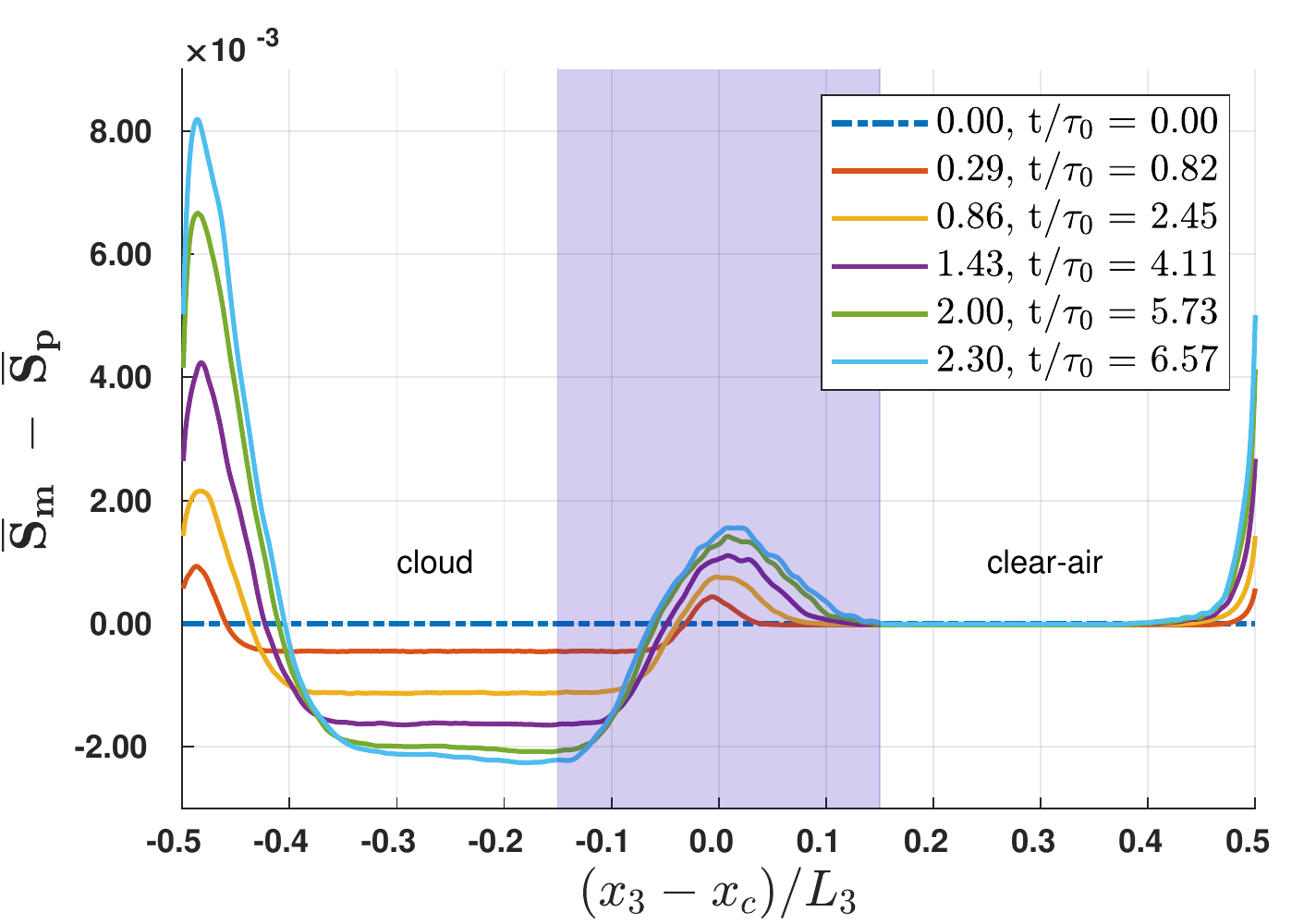}\label{fig:diff_ss_mean}}
		\subfloat[b][]{\includegraphics[width=0.48\linewidth]{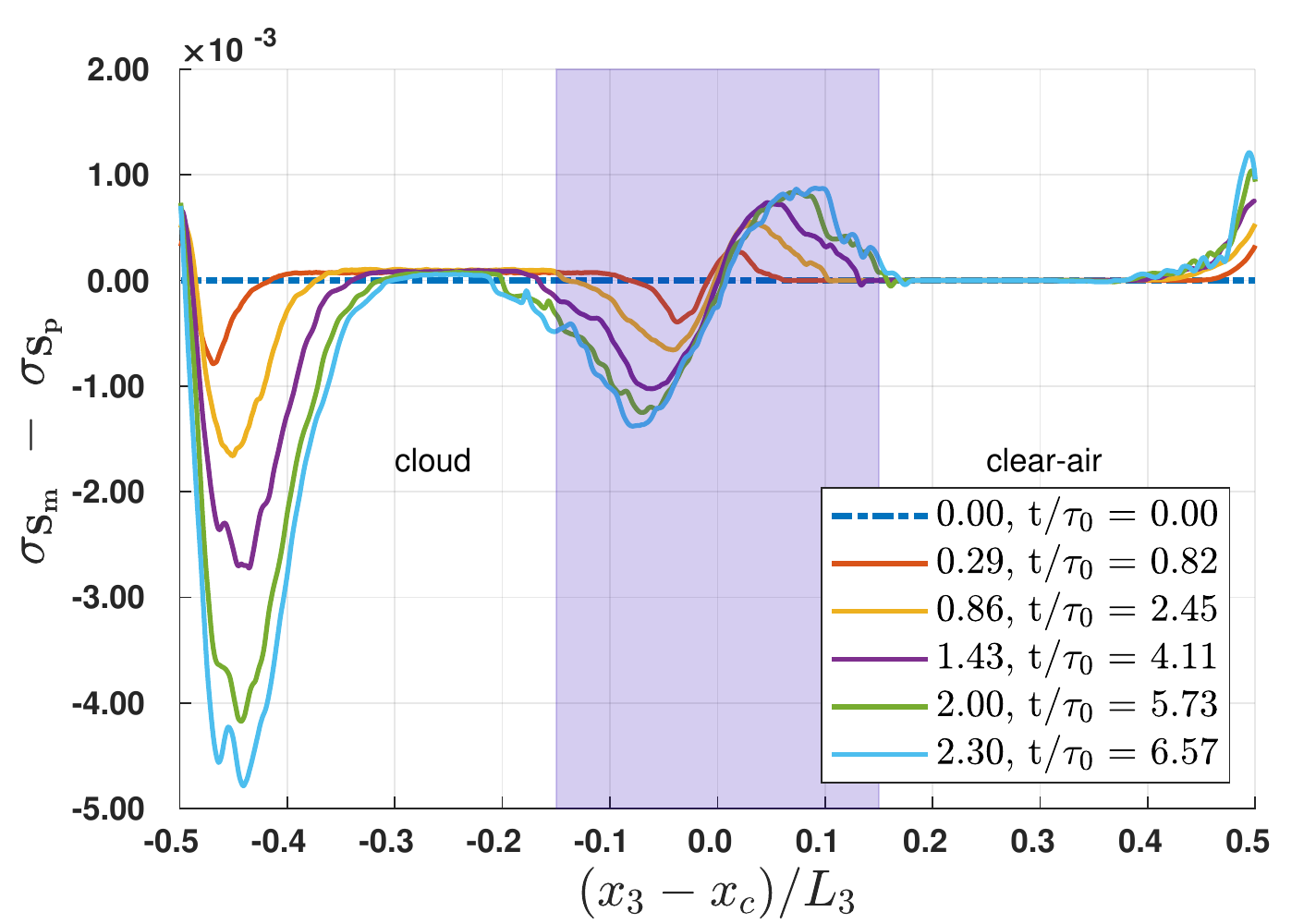}\label{fig:diff_ss_std}}\\
		\subfloat[c][]{\includegraphics[width=0.48\linewidth]{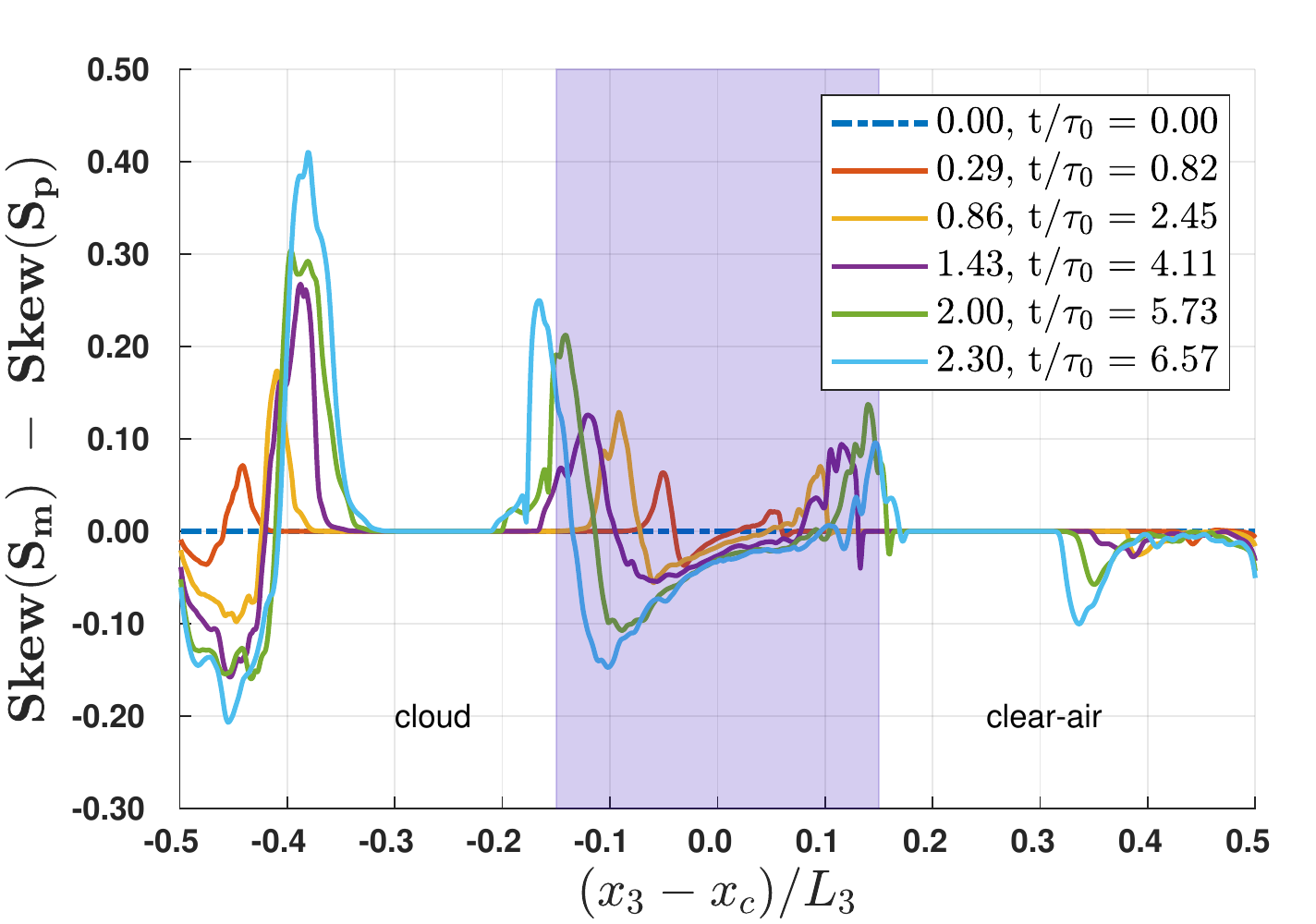}\label{fig:diff_ss_ske}}
		\subfloat[d][]{\includegraphics[width=0.48\linewidth]{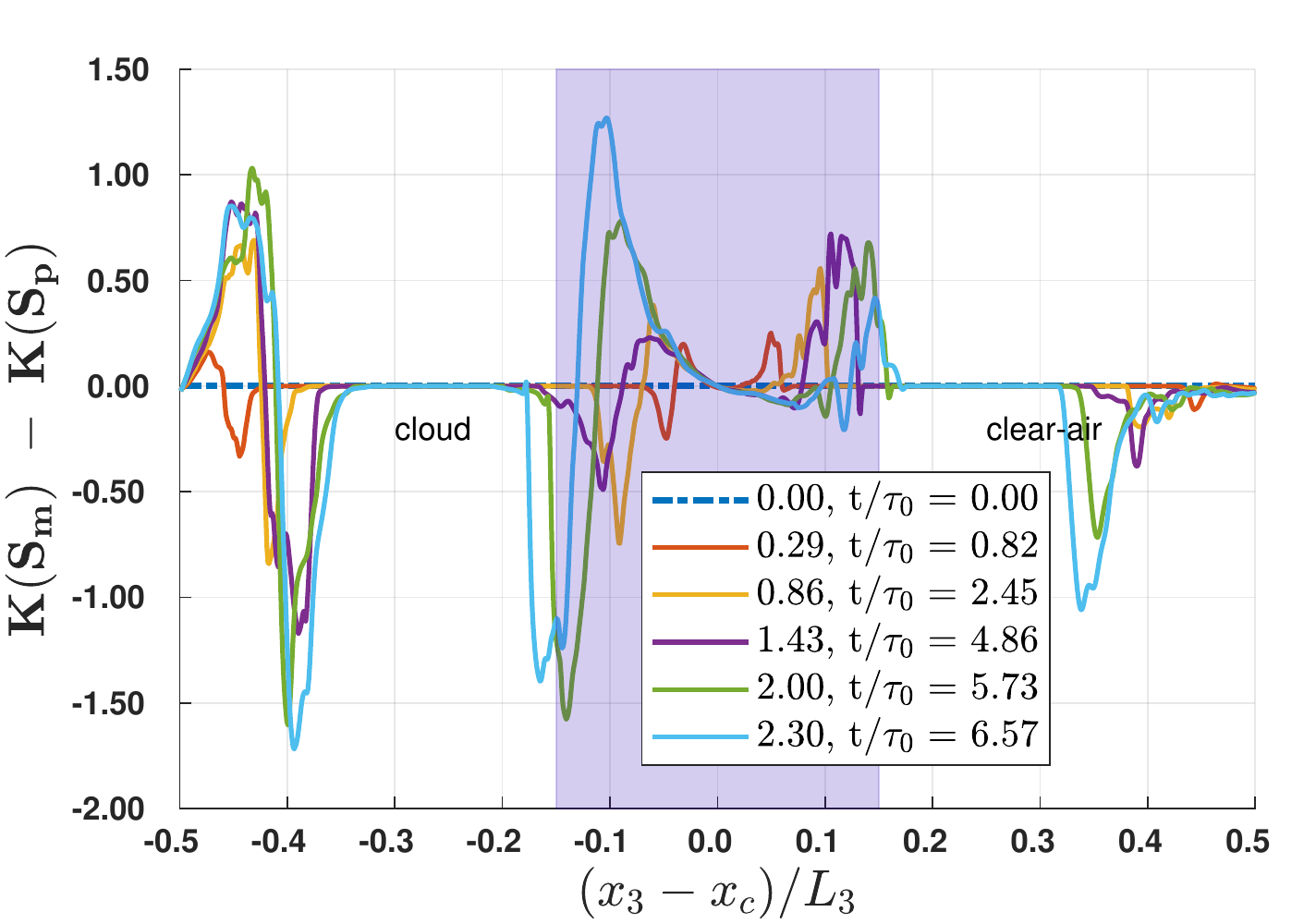}\label{fig:diff_ss_kur}}
		\caption{\textbf{Difference in the supersaturation statistics between monodisperse ($\mathbf{S_m}$) and polydisperse ($\mathbf{S_p}$) droplet populations across the cloudy - under-saturated ambient air interface layer.} \protect\subref{fig:diff_ss_mean} Difference in the mean supersaturation (or saturation deficit) across the layer. \protect\subref{fig:diff_ss_std} Difference in the standard deviation. \protect\subref{fig:diff_ss_ske} Difference in the skewness. \protect\subref{fig:diff_ss_kur} Difference in the kurtosis. 
			The approximate extension of the interface mixing layer is indicated as the blue shaded area between the cloudy and clear air regions.}
		\label{fig:diff-stats-supersaturation}
	\end{figure}
	
	Droplet and flow statistics are taken from horizontal $x_1-x_2$ planes at a constant $x_3$, and plotted with respect to the normalized height $(x_3-x_c)/L_3$, with $x_c$ being the position of the cloud-clear air interface, and $L_3=2L_{1,2}$ being twice the length of the edge of the cube. To observe the interface cloud - clear air dynamics, it is necessary to focus on the evolution of the statistics along the non-homogeneous (vertical) direction of the domain. The mean, standard deviation and higher order-moments are computed over the cells in the same horizontal plane and associated with the corresponding vertical coordinate $x_3$. The covariance for each horizontal plane 
	
	\begin{equation}\label{cov}
		cov_{X,Y}(x_3, t) = \frac{1}{n_1 n_2}  \sum_{i,j=1}^{n_1,n_2} \left(X(x_1, x_2; x_3, t) -\overline{X}\right)\left(Y(x_1, x_2; x_3, t) - \overline{Y}\right)
	\end{equation}
	
	\noindent where the over-line indicates the average of a given physical quantity in the $x_1,x_2$ planes and 
	\begin{equation}\label{planar_average}
		\overline{X}(x_3,t) = \frac{1}{n_1 n_2}\sum_{i=1, j=1}^{n_1, n_2} X(x_1,x_2;x_3,t)
	\end{equation}
	
	\noindent The Pearson product-moment correlation coefficient of two planar averaged quantities, $\overline{X}(t)$ and $\overline{Y}(t)$, only depends on the time and when used to correlate variations across the interface layer $\Delta(t)$, it can be written as
	\begin{equation}\label{Pearson}
		\rho_{{X,Y}_{\Delta}(t)} =  \sum_{k=1}^{n_{\Delta}} \frac{(\overline{X}(x_3 , t) -\overline{X}_{\Delta}(t))}{\sigma_{\overline{X}}} \;\; \frac{(\overline{Y}(x_3 , t) -\overline{Y}_{\Delta}(t))}{\sigma_{\overline{Y}}}
	\end{equation}
	
	\noindent where subscript ${\Delta}$ stands for the quantity averaged inside the interface and $n_{\Delta}$ is the number of planes inside the interface.
	
	The kinetic energy inside the homogeneous cloudy and clear air regions, decays over time with a power-law exponent (see Figure \ref{fig:kinetic}) of the $E/E_0\sim (t/\tau_0)^{-n}$ type, where $n$ ranges from 1.6 to 2.15 \cite{Tordella_2006, Djenidi_2015}. The initial values of the root mean square velocity of the flow, of the longitudinal integral length scale and of the eddy turnover time are reported in Table \ref{tab:simulation_parameters}. The eddy turnover time $\tau_0=2\ell_0/(u_{rms,c0}+u_{rms,a0})$, is computed from the initial integral length scale and root mean square velocity of the flow, averaged over the domain, and has an initial value of $0.35\,\si{\second}$. 
	
	The decay of both kinetic energy $E$ and  dissipation rate $\varepsilon$ can be observed in Figure \ref{fig:kinetic}). During the decay, the integral scale grows homogeneously over the entire domain.  %approximately two initial large eddy turnover times, $\tau_0$,
	
	The system relaxes to a quasi steady-state condition as values of $\varepsilon$ of the order of 10 \si{\centi\meter^2\per\second^3} are reached inside the mixing layer. Values of this order of magnitude have already been measured in shallow cumulus clouds \citep{Siebert2006, Devenish2012}.
	
	The average value of $E$ quickly decreases  during the transient. However, the effects of the unstable stratification are highlighted by the normalized kinetic energy, $(E-E_2)/E_1 - E_2)$, which in fact shows a hump that amplifies in time (see Figure \ref{fig:kinetic}b, and Figures 11 and 12 in Gallana et al. \citep{Gallana_2022}). 
	The warmer air close to the interface is convected upward and gains vertical velocity, thus increasing the kinetic energy locally. This injection of kinetic energy at the small scales of the turbulence affects the mixing process by enhancing the vertical advection of the dispersed water droplets, water vapor and internal energy up to the subsaturated region.
	High values of higher moments of the spatial  longitudinal derivatives of the velocity indicate the high anisotropy and intermittency of the small-scale of the carrier flow in the mixing region. Small-scale intermittency in the mixing region is associated with accelerated droplet population dynamics and an increased collision-coalescence rate, see Figure 2 in section II and also Table 3 and Figures 11 and 12 in Golshan et al. 2021 \citep{Golshan2021}. The time required by the two populations to reach the same width for the evaporation and condensation processes is estimated by equating the time variations of the standard deviations of the monodisperse and polydisperse size distributions. The estimate is about $100 \tau_0$ in the cloudy region, which is homogeneous and isotropic. The estimate is about $18,5 \tau_0$ in the interface region, i.e. more than 5 times faster. A remarkable acceleration of the broadening of the droplet size distribution, due to turbulent fluctuations, is therefore observed in the shear-free mixing layer that separates the cloud from the sub-saturated environmental air.
	
	Planar averages and higher order statistical moments of supersaturation across the cloudy - undersaturated ambient air interface layer are shown in Fig. \ref{fig:supersaturation}. The high intermittency of the  distributions should be notes. Very high values of both  skewness and kurtosis are reached on the two sides of the mixing layer. Moreover, comparatively higher absolute values can be ovserved at the border with the cloudy region (S down to -8, K up to 60), where the vapor flux is spatially increasing, with respect to that observed at the border with sub-saturated air (S up to 4, K up to 20), where the vapor flux is spatially decreasing\cite{Gallana_2022}.
	
	%\subsection{Droplet size and concentration across the interface}
	\begin{figure}[bht!]
		\centering
		\large\textbf{Mean droplet radius and concentration across the cloud and interfacial mixing regions}\par\medskip
		{\subfloat[a][]{\includegraphics[width=0.49\linewidth]{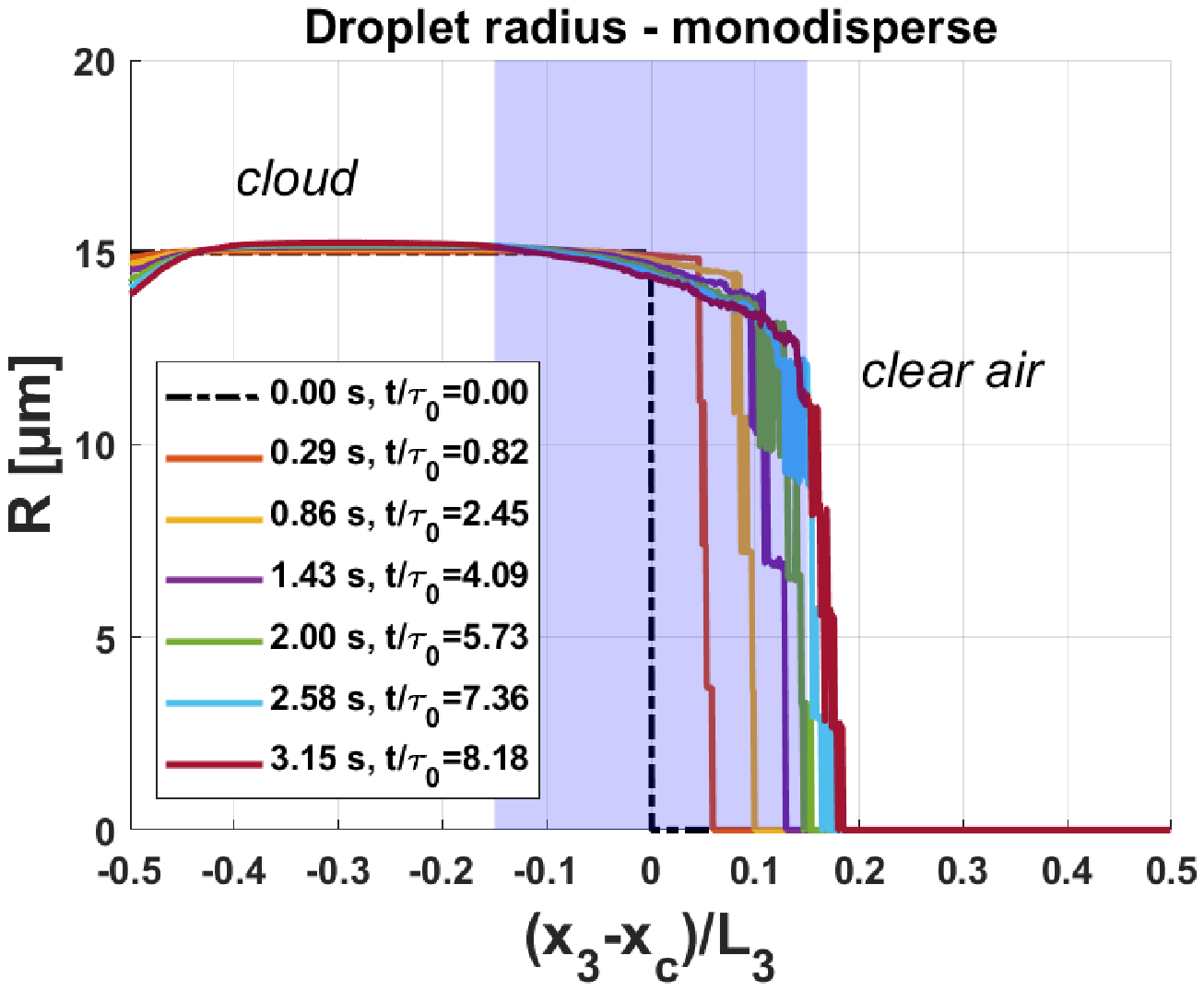}\label{fig:droplet_stats_a}}}
		{\subfloat[b][]{\includegraphics[width=0.49\linewidth]{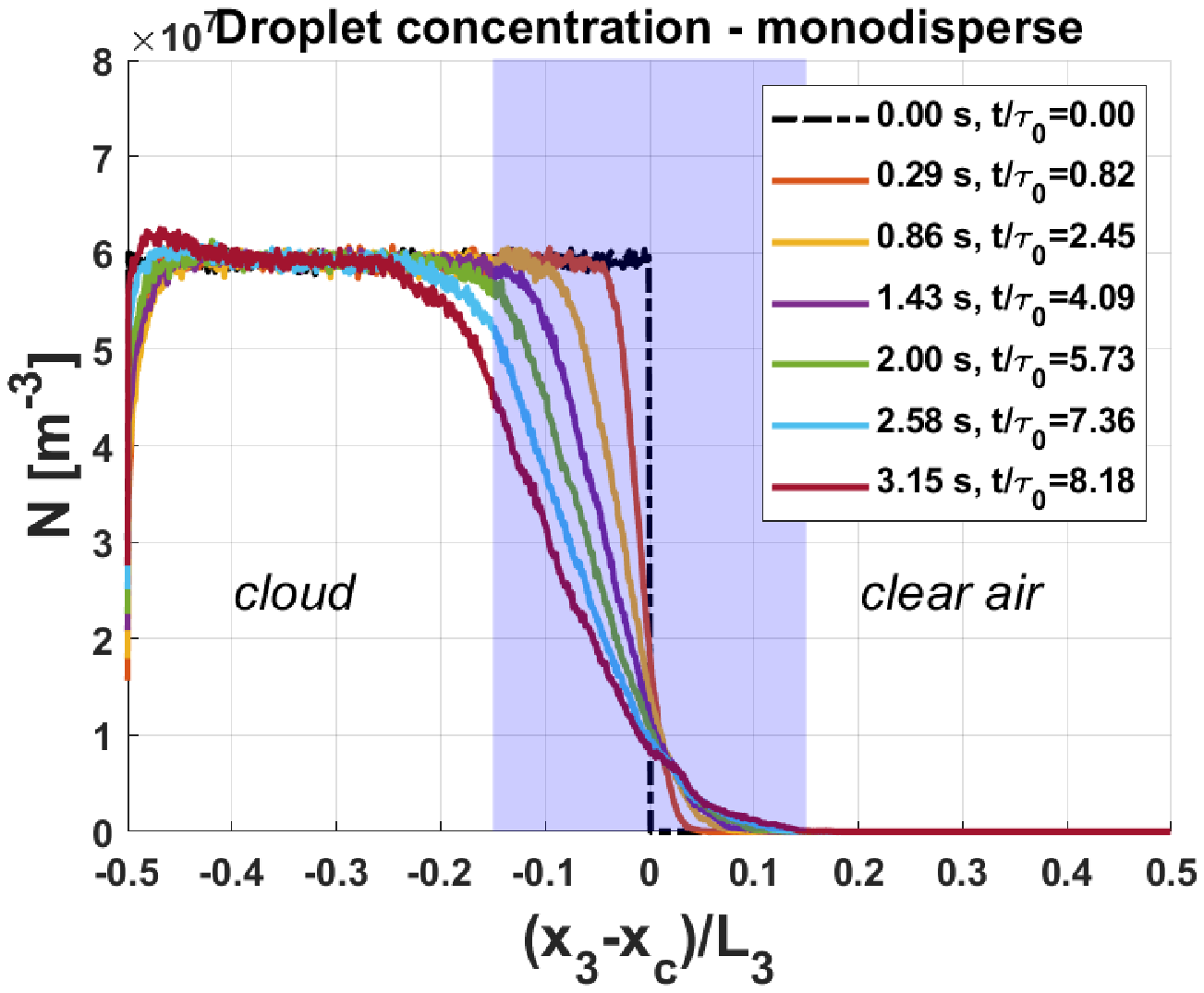}\label{fig:droplet_stats_b}}}
		{\subfloat[c][]{\includegraphics[width=0.49\linewidth]{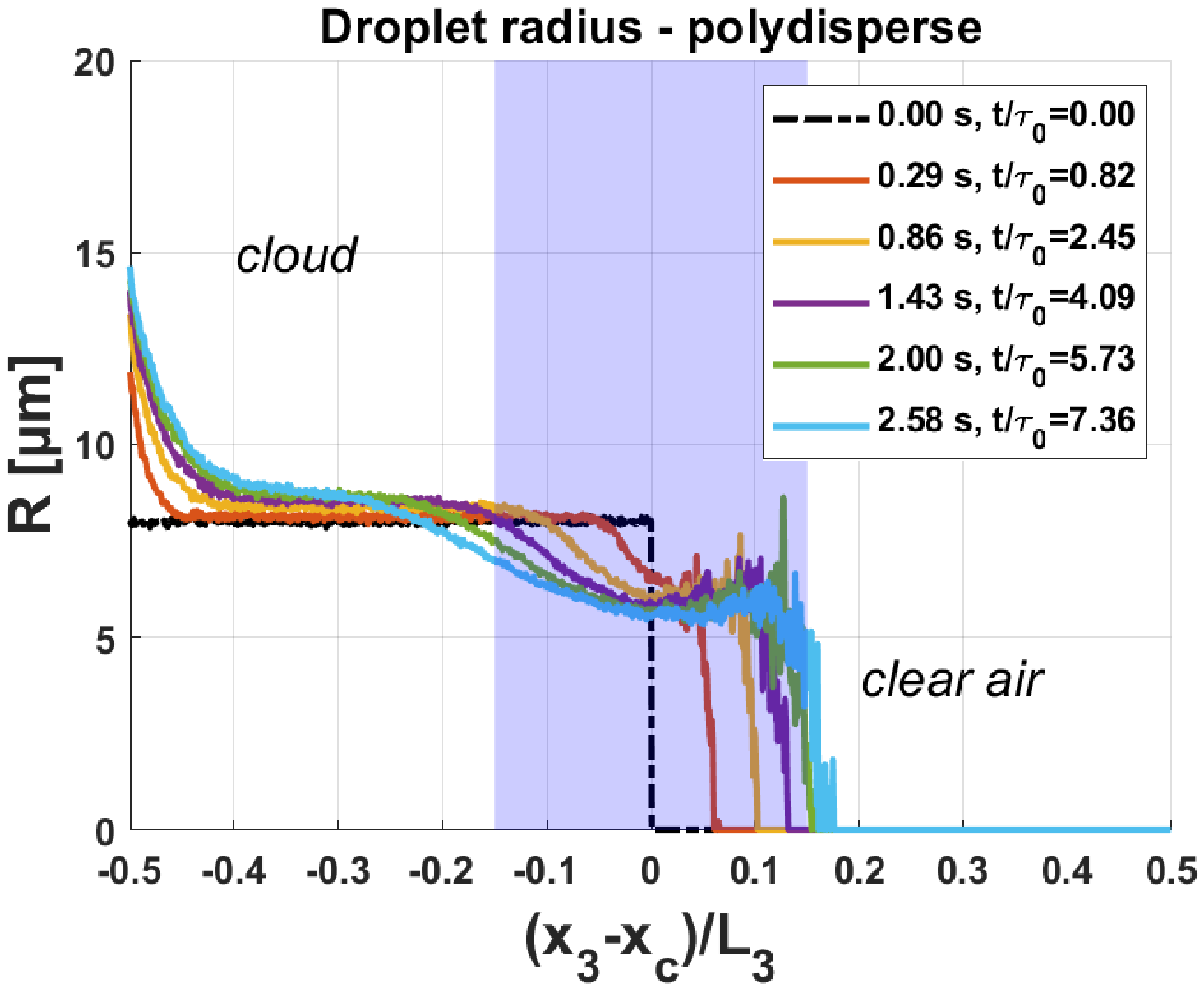}\label{fig:droplet_stats_c}}}
		{\subfloat[d][]{\includegraphics[width=0.49\linewidth]{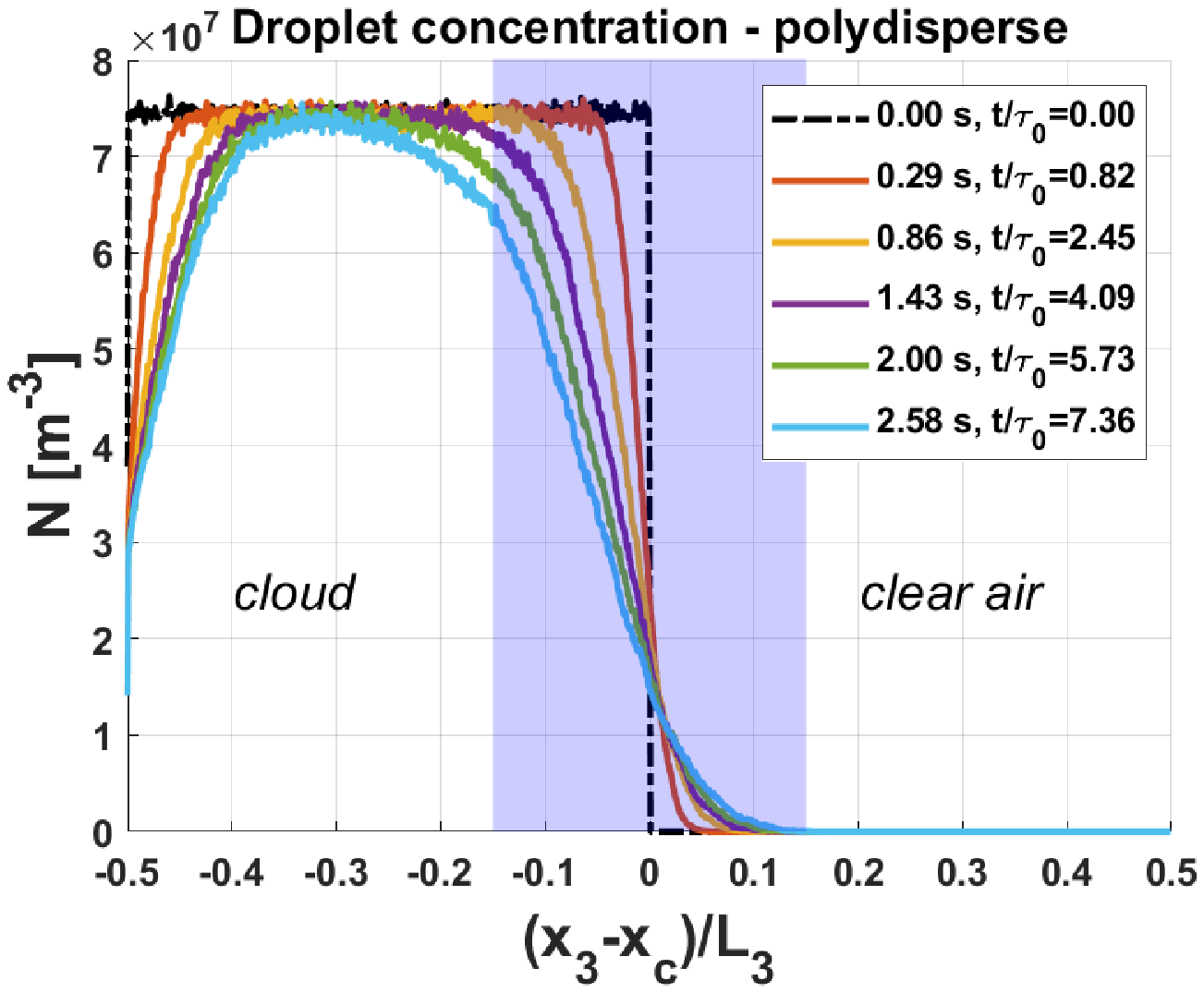}\label{fig:droplet_stats_d}}}
		\caption{Average mean droplet radius \protect\subref{fig:droplet_stats_a} \protect\subref{fig:droplet_stats_c}, and concentration \protect\subref{fig:droplet_stats_b} \protect\subref{fig:droplet_stats_d}. Each value represents the planar average computed for a horizontal plane (see Figure \ref{fig:comp_domain}). Shearless mixing takes place in the shaded area. The dash-dotted black line shows the initial conditions.}
		\label{fig:droplet_stats}
	\end{figure}
	
	The droplet statistics have been computed over the horizontal planes {to complement the data in Fig. \ref{fig:drop_distributions}}. From now on, we denote the droplet numerical concentration with the symbol $N$. 
	The results of the monodisperese and polydisperse distributions across the mixing layer are shown in Figure \ref{fig:droplet_stats}, where both the droplet radius and the concentration are plotted along $x_3$. At the beginning of the transient, the droplets populate the lower part of the domain and are randomly distributed within the cloud. The core environment of the cloud is supersaturated (see Figure \ref{fig:supersaturation_a}), and this permits a uniform condensation growth of the droplets to take place within the cloud. As the central mixing proceeds, a few drops are advected in the upper subsaturated clear-air region. Here, smaller drops will rapidly evaporate and eventually be eliminated by the algorithm. Dissipation rate $\varepsilon$ decreases during the transient and heavier droplets are likely to settle, as the small-scale Froude number scales sublinearly with the dissipation rate $\mathrm{Fr}_\eta\sim\varepsilon^{3/4}$ \citep{Devenish2012}.
	
	The mean radius plot for the monodisperse case (see Figure \ref{fig:droplet_stats_a}) is almost flat in the cloud core region. The extension of this constant-radius plateau becomes more and more reduced as the decaying shearless mixing proceeds. The blue-shaded area represents an approximate extension of the mixing region at the end of the simulated transients. The concentration plots (Figure \ref{fig:droplet_stats_b}) displays analogous trends.
	{In the polydisperse case, Figures \ref{fig:droplet_stats_c} and \ref{fig:droplet_stats_d}, the flat region of nearly constant radii is narrower and presents a peak close to the very top of the mixing layer. This is because collisions are much more frequent in this case. Moreover, given the concomitant presence of very different droplets, the volume ratio between the largest to the smallest droplet is of the order of $1.25 \cdot 10^5$, thus the number of collisions is large. Out of a total of $10^7$ droplets, we in fact observe about $5 \cdot 10^4$ collisions over about $8$ physical time scales. Information on the collision kernel inside the cloudy and mixing layer regions can be found in Figures  13 and 14 of Golshan et al. 2021\cite{Golshan2021}}. 
	
	\subsection{Supersaturation evolution equation, and the microphysical time scales.}
	\begin{figure}[bht!]
		\large\textbf{Microphysical time scales and mean supersaturation in the cloud and interfacial mixing regions}\par\medskip
		{\subfloat[a][]{\includegraphics[width=0.49\linewidth]{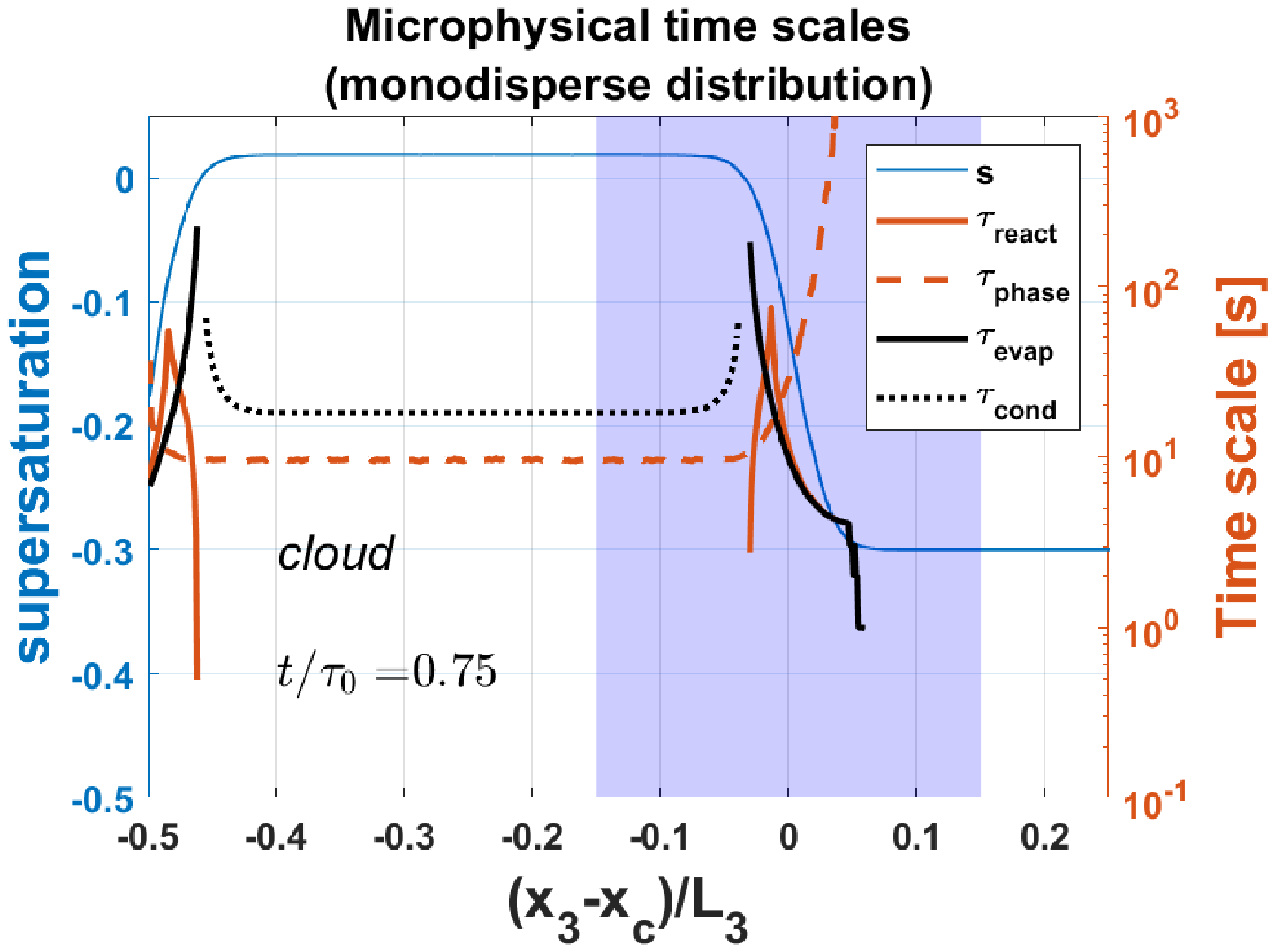}\label{fig:temporal_scales_a}}}
		{\subfloat[b][]{\includegraphics[width=0.49\linewidth]{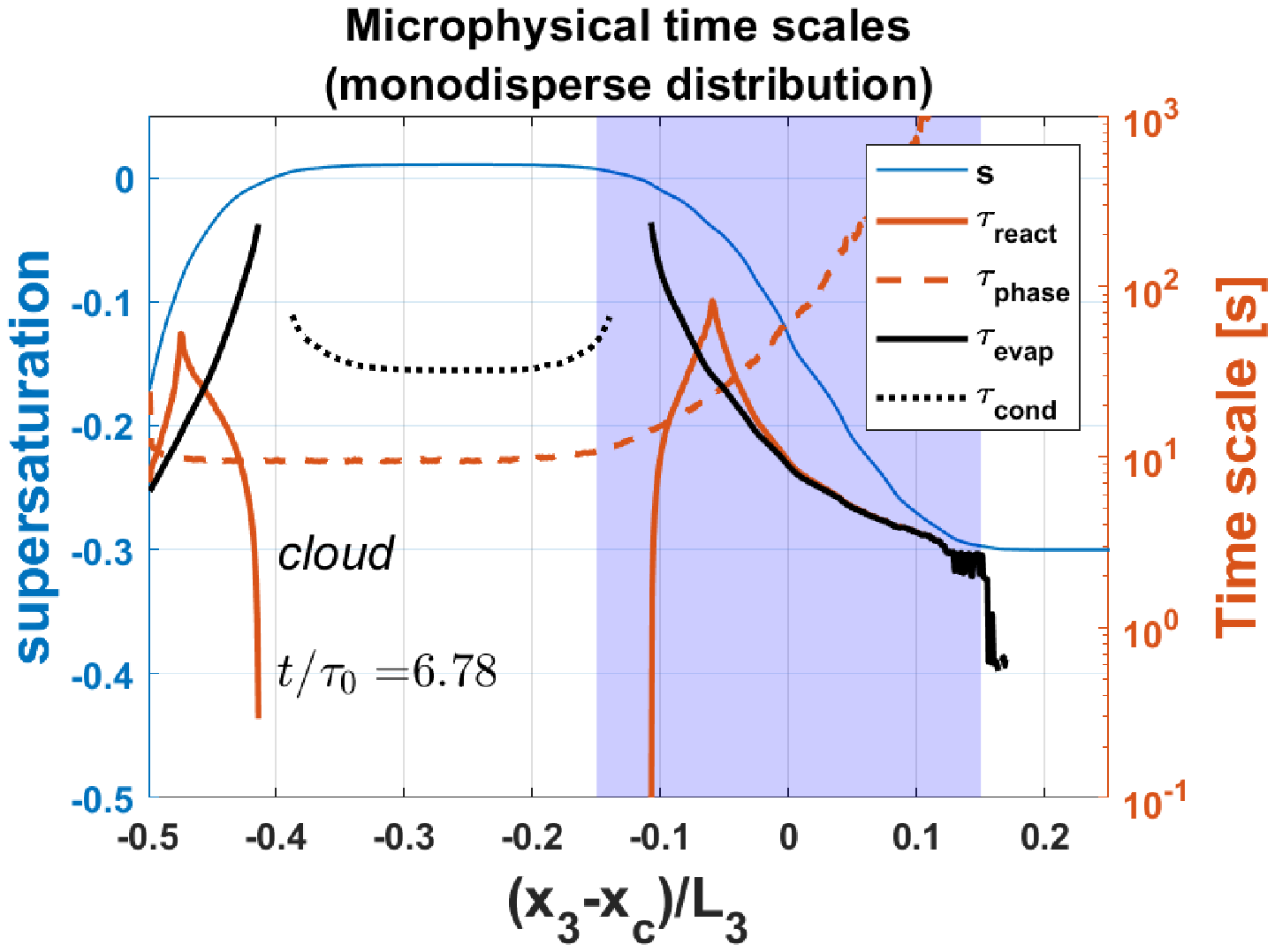}\label{fig:temporal_scales_b}}}\\
		{\subfloat[c][]{\includegraphics[width=0.49\linewidth]{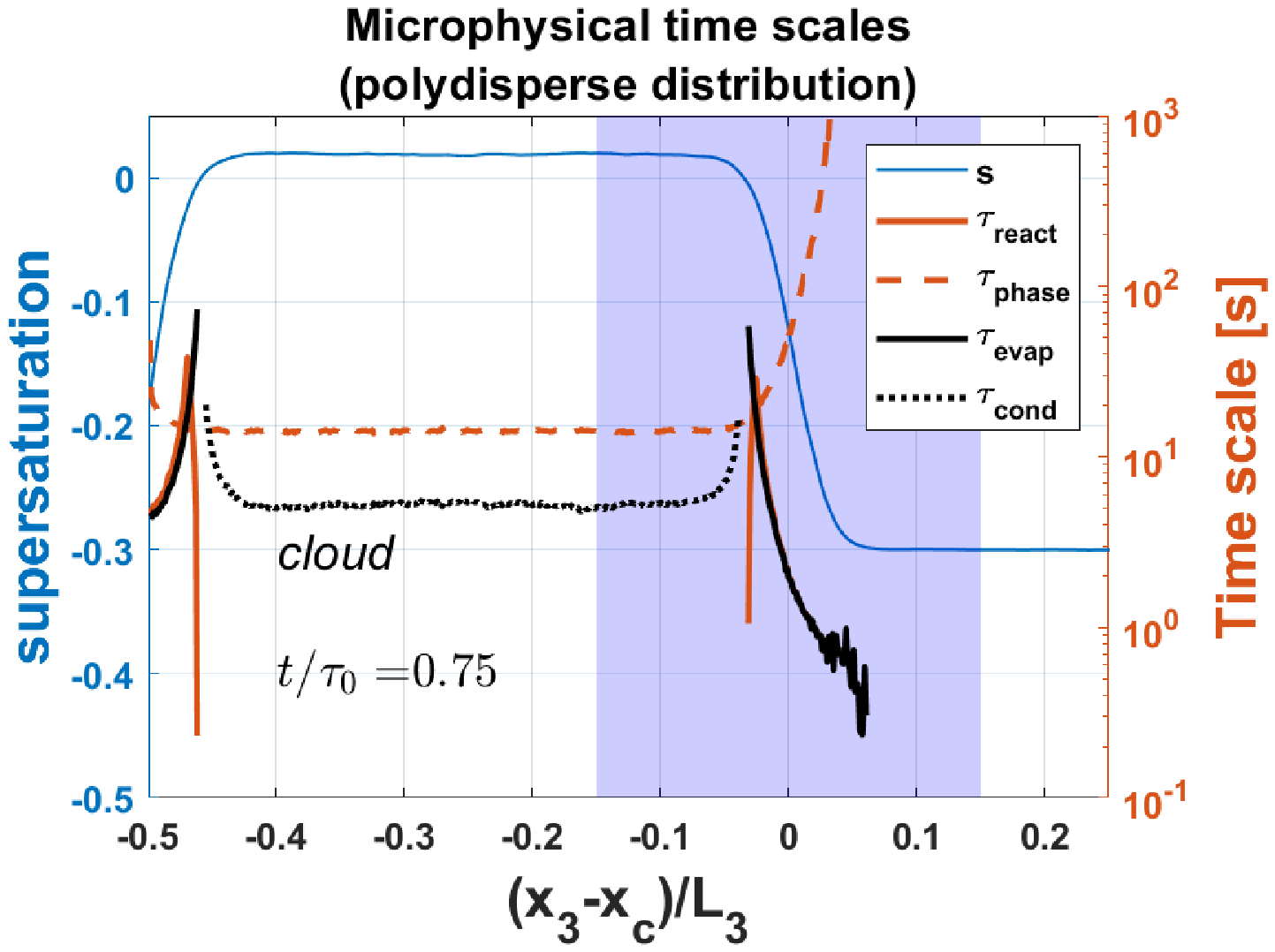}\label{fig:temporal_scales_c}}}
		{\subfloat[d][]{\includegraphics[width=0.49\linewidth]{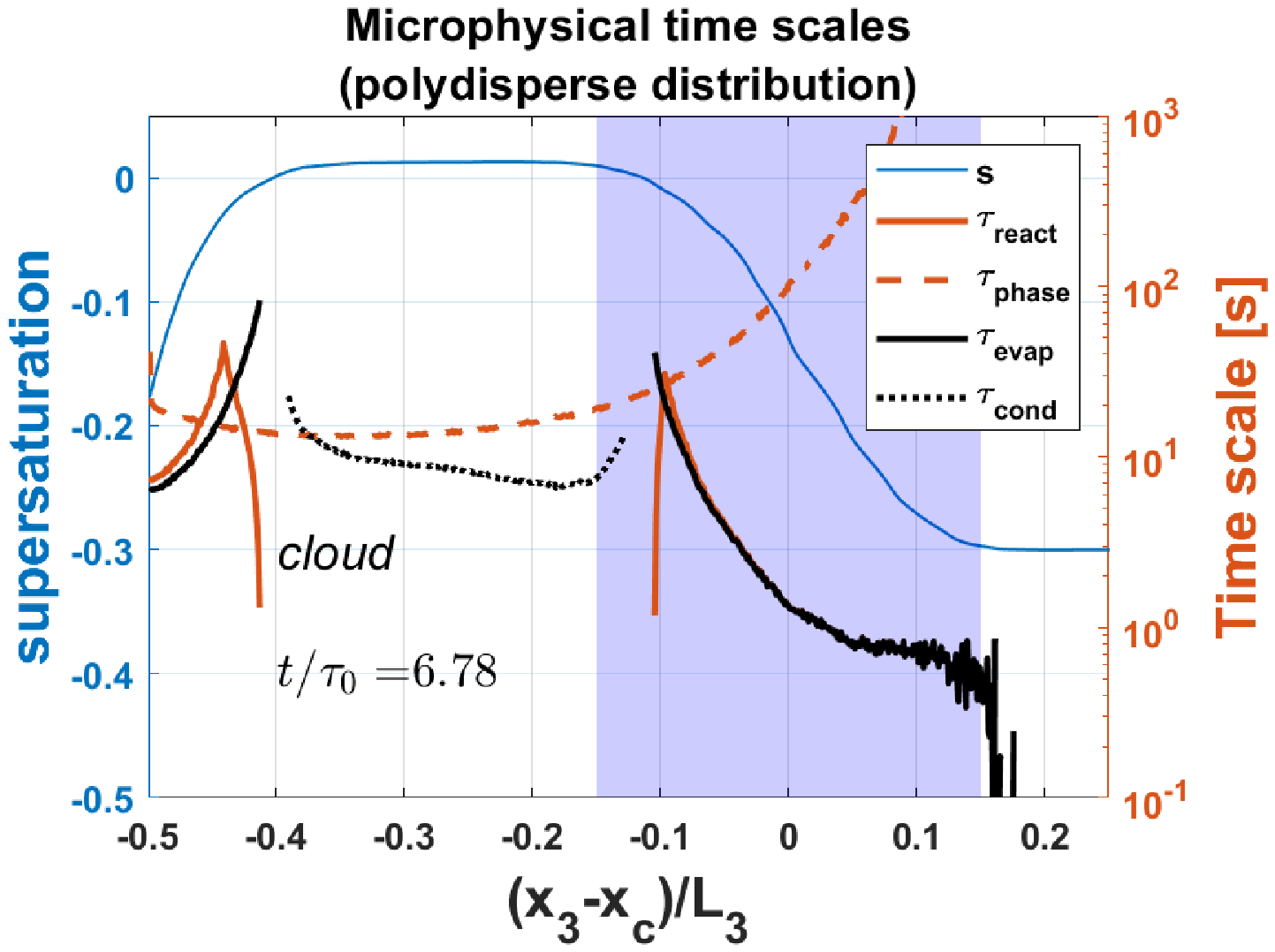}\label{fig:temporal_scales_d}}}\\
		\caption{Vertical distribution of the evaporation $\tau_{evap}$, phase relaxation $\tau_{phase}$ and reaction  $\tau_{react}$ time scales computed inside each grid cell and then averaged on horinzontal planes. The data are displayed for the monodisperse \protect\subref{fig:temporal_scales_a} \protect\subref{fig:droplet_stats_b}, and the polydisperse cases \protect\subref{fig:temporal_scales_c} \protect\subref{fig:droplet_stats_d} for two different time steps at the beginning and the end of the transient. The planar average of supersaturation $\overline{S}$ (Figure \ref{fig:supersaturation_a}) is also plotted for comparison purposes.}
		\label{fig:temporal_scales}
	\end{figure}
	
	The supersaturation evolution equation has often been used to model a water vapor budget on a developing cloud \citep{Devenish2012}. This equation is based on a production-condensation model, where the time derivative of supersaturation is determined by balancing a production term, $\mathcal{P}$, and a condensation term, $\mathcal{C}$ \citep{Squires1952,Twomey1959}
	\begin{equation}\label{ssat_eq}
		\frac{dS}{dt} = \mathcal{P} + \mathcal{C}
	\end{equation}
	The condensation term accounts for the depletion of water vapor and the release of latent heat during condensation at the surface of a spherical droplet, and it is a function of the local level of supersaturation
	
	\begin{equation}\label{condensation}
		\mathcal{C} = -\frac{S}{\tau_{phase}} = 4 \pi \kappa_v N \overline{R} S. \end{equation} 
	
	The source term, $\mathcal{P}$, has often been modeled as a linear function of the vertical mean velocity of the updraft\citep{Cooper1989,Siebert2017}, or identified as the net flux of supersaturated water vapor through the parcel boundaries \citep{Prabhakaran2020}. In the present analysis, updraft is absent, thus $\mathcal{P}=0$. Equation (\ref{ssat_eq}) does not account for the advection and diffusion of water vapor and internal energy in the environment surrounding the droplet ($\textrm{Re}_{drop}\ll1$), and considers supersaturation $S$ as a rather global, bulk property of an adiabatic cloud parcel \citep{RogersYau1996}. In their study on cloud cores, \cite{Sardina2015} generalized the supersaturation evolution equation (\ref{ssat_eq}) to a transport model by assuming a linear dependency between the diffusive term of the supersaturation and the Laplacian of the temperature and vapor density fields.
	They showed that, under steady-state conditions and within the limit of a real-cloud Reynolds' number, the diffusive term of the supersaturation variance becomes negligible.
	
	In a homogeneous, nearly isotropic cloudy layer that is statistically in equilibrium, a zero-mean vertical velocity field would imply a null net vertical transport of cooling vapor parcels. It should be noted that whenever an updraft can be neglected, equation (\ref{ssat_eq}) can be solved by separating the variables \citep{Khvorostyanov2014}
	\begin{equation}\label{ssat_eq_isobaric}
		\frac{dS}{dt}\cong -4\pi \kappa_vN\overline{R}S=-\frac{S}{\tau_{phase}}
	\end{equation}
	Therefore, an initially subsaturated (supersaturated), droplet-laden environment experiences an increase (decrease) in the vapor concentration, which results in $S$ relaxing exponentially to 0. The time constant of this solution is the phase relaxation time 
	\begin{equation}\label{TIME_phase_relaxation}
		\tau_{phase}=\left(4\pi \kappa_vN\overline{R}\right)^{-1}
	\end{equation}
	The definition of $\tau_{phase}$ depends on the assumption of the droplet population having a constant integral radius, $N\overline{R}$, and it is able to describe the temporal variation of the supersaturation and the liquid water content \citep{Lehmann2009,Lu2018} in a homogeneous context. The phase relaxation time was chosen from the microphysical time scales used in several DNS studies that focused on entrainment-mixing processes \citep{Kumar2013,Kumar2014,Chandrakar2016,Gotzfried2017,Kumar2018}, and was used to define the Damk\"{o}hler number
	\begin{displaymath}
		\textrm{Da}=\frac{\tau_{turb}}{\tau_{microphysics}}
	\end{displaymath}
	Depending on whether the choice of $\tau_{turb}$ falls into large or small-eddy time scales, a vast range of Damk\"{o}hler numbers can be defined for the same microphysical time scale in a turbulent flow \citep{Kumar2013}. Large and small values of $\textrm{Da}$ are associated with a fast and slow microphysical response of the droplet population to entrainment and turbulent mixing, respectively \citep{Andrejczuk2009}. Large $\textrm{Da}$ are also associated with inhomogeneous mixing, whereas small $\textrm{Da}$ often indicate homogeneous mixing \citep{Latham1977,Baker1980}.
	
	% reaction time with mean radius and supersaturaiton
	\begin{figure}[bht!]
		{\subfloat[a][monodisperse, $t/\tau_0$ = 0.75]{\includegraphics[width=0.49\linewidth]{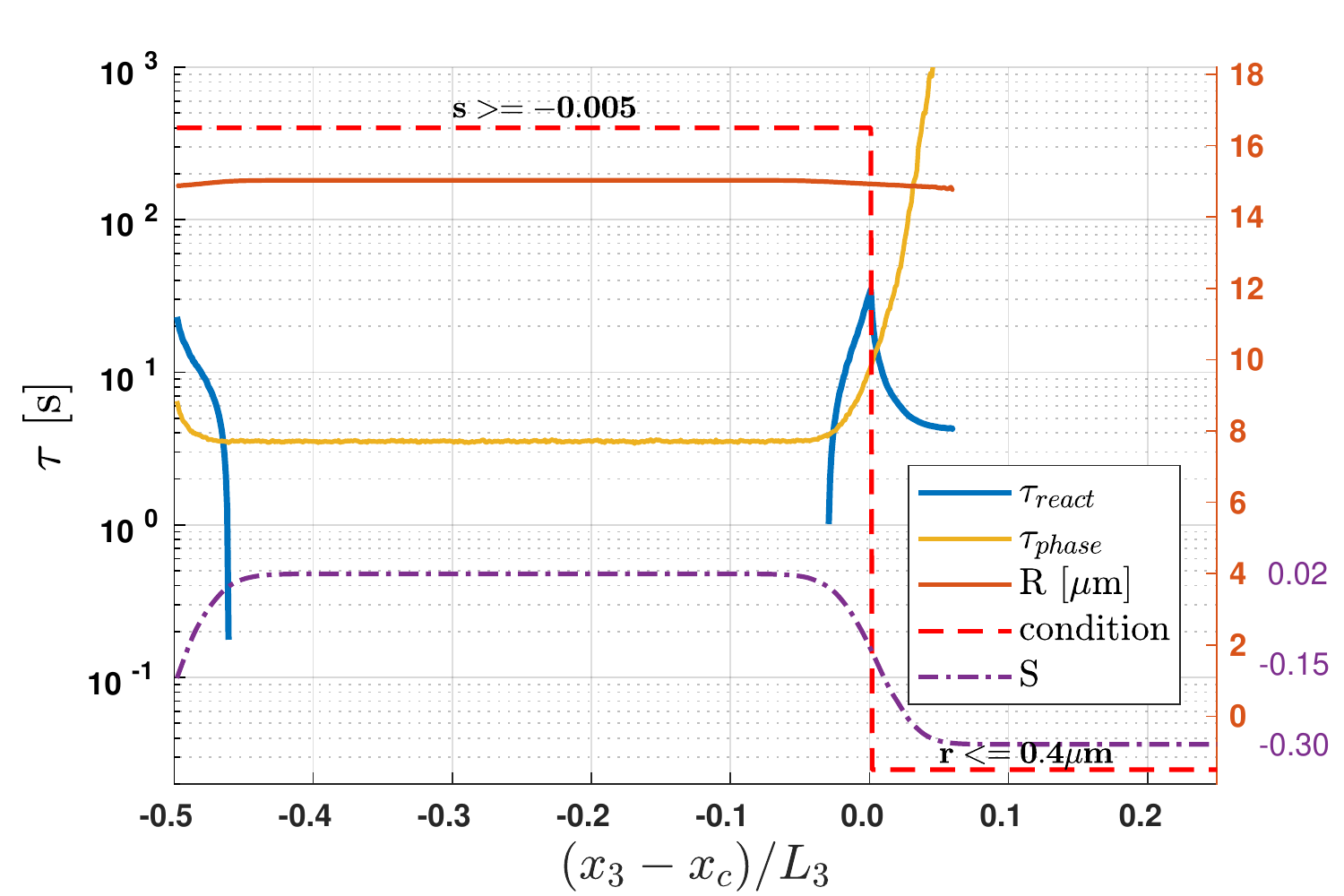}\label{fig:react_time_a}}}
		{\subfloat[b][monodisperse, $t/\tau_0$ = 6.61]{\includegraphics[width=0.49\linewidth]{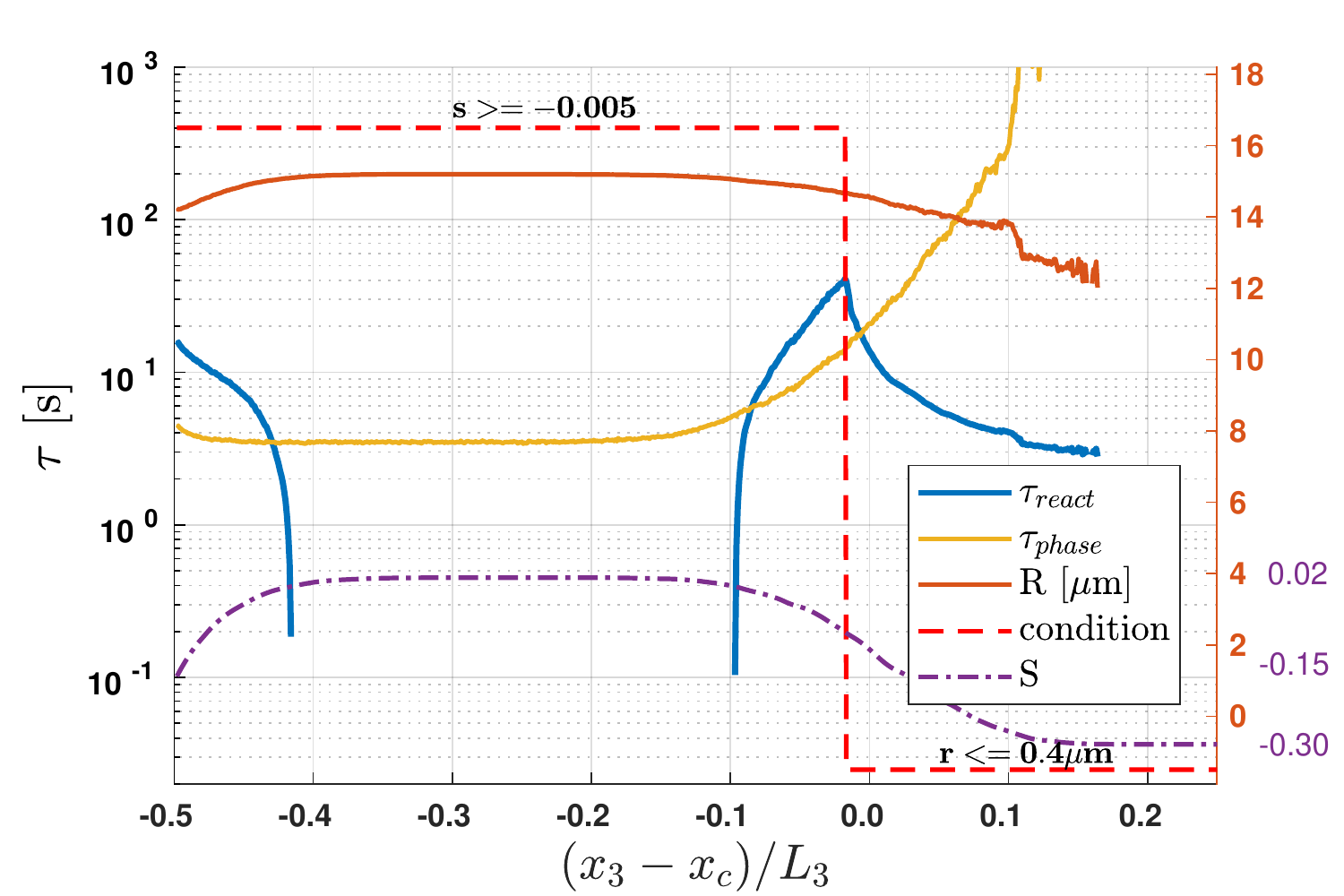}\label{fig:react_time_b}}}\\
		{\subfloat[c][polydisperse, $t/\tau_0$ = 0.75]{\includegraphics[width=0.49\linewidth]{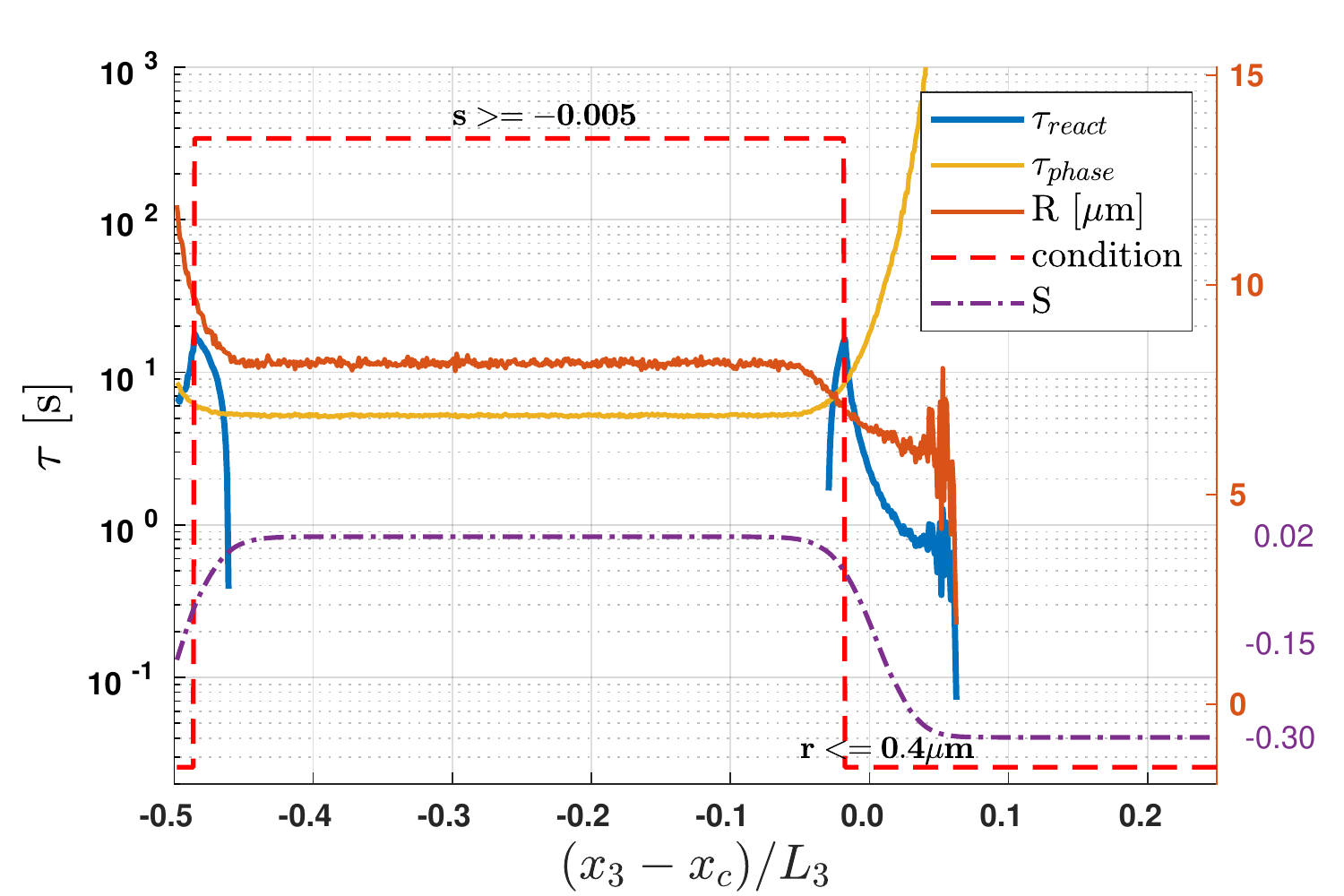}\label{fig:react_time_c}}}
		{\subfloat[d][polydisperse, $t/\tau_0$ = 6.61]{\includegraphics[width=0.49\linewidth]{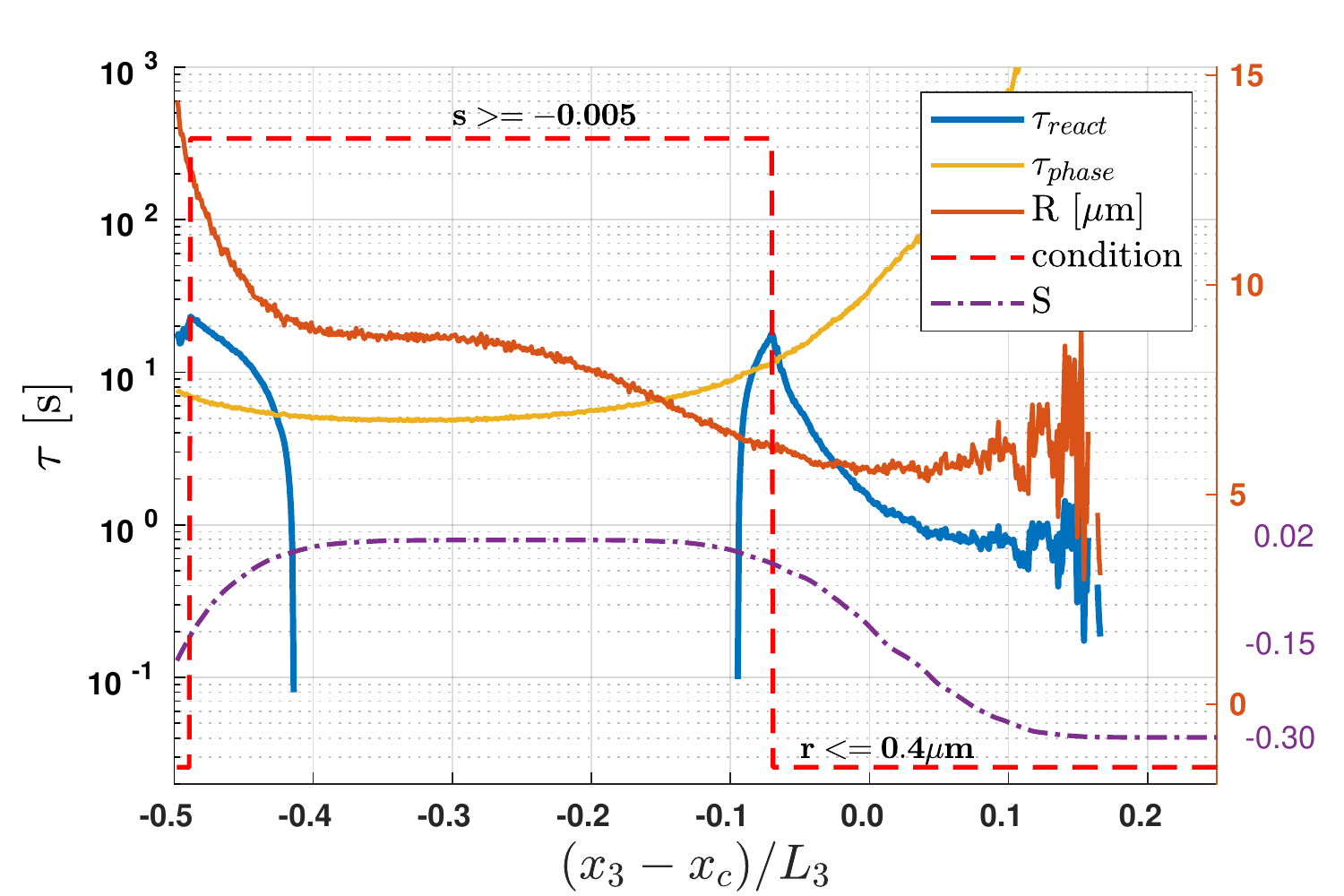}\label{fig:react_time_d}}}\\
		\caption{For comparison with Fig. \ref{fig:temporal_scales}, the reaction time and relaxation phase time statistics, computed using planar averaged quantities, are here shown for both polydipserse and monodisperse populations and for two different time instances. The red dashed line represents the condition that is first reached at each vertical location when numerically solving the coupled system of Equations 8 and 17. Supersaturation S and the planar averaged R are included for reading convenience purposes.    }
		\label{fig:react_time_detailed}
	\end{figure}
	
	{However, it should be noted that, in a highly anisotropic, in-homogeneous situation, such as inside the mixing layer that separates the cloud from the subsaturated environemental air, the momentum, internal and kinetic energy fluxes and the water vapor are not zero. The fluxes are positive and rising, forming a peak value that is almost centered in the middle of the layer. Beyond this point, the fluxes decrease and become zero inside  the isotropic homogeneous  subsaturated ambient air, \cite{Veeravalli1989}, \cite{Tordella_2008}, \cite{Gallana_2022}.
		In such a situation, a mismatch between the supersaturation time derivative and the condensation term can be expected.} 
	
	On the other hand, if the focus is on the evolution of the droplet size and the number concentration, the evaporation time scale offers a good practical description of the process, and should be taken as the relevant microphysical time scale, $\tau_{microphysics}$. By neglecting the Kelvin and Raoult terms in equation (\ref{condensational_growth}), and integrating for a constant $S_0<0$, one can obtain in each computational cell, an estimate of the time required for a single droplet, with an initial radius of $R_0,$ to evaporate completely in a uniform subsaturated environment
	\begin{equation}\label{TIME_evaporation}
		\tau_{evap}=-\frac{R_0^2}{2K_sS_0}
	\end{equation}
	
	Both $\tau_{phase}$ and $\tau_{evap}$ rely on the assumption of constant supersaturation and integral radius. However, since both quantities vary concurrently inside a mixing layer, it is better to define a reaction time $\tau_{react}$ \citep{Lehmann2009} that considers variations of both $S$ and $N\overline{R}$. The reaction time is defined as the shortest time that has elapsed since either the droplet has evaporated completely or the parcel has become saturated, and it is obtained by numerically solving the coupled system of differential equations, that is, Equations (\ref{condensational_growth}) and (\ref{ssat_eq_isobaric}), for initial non-zero values of positive $R_0$ and negative $S_0$.
	It should be noted that $\tau_{evap}$ and $\tau_{react}$ are only defined for the subsaturated regions, whereas $\tau_{phase}$ is defined for non-zero values of the integral radius, and can also be used in supersaturated regions. In order to describe a characteristic time of the condensation process in supersaturated regions of the domain, we introduce a condensation time $\tau_{cond}$, which we arbitrarly define as the time it takes a droplet to double its radius for a constant local supersaturation $S$:
	\begin{equation}\label{condensation_time}
		\tau_{cond}=\frac{3}{2}\frac{R^2_0}{K_sS_0}.
	\end{equation}
	
	{The horizontal planar average values of all these microphysical time scales are plotted for two different time instants and the initial droplet size distribution type in the cloud and mixing regions. See Figure \ref{fig:temporal_scales}, where the computation is performed in each computational grid cell and then averaged over the horizontal planes. For comparison, see also Figure \ref{fig:react_time_detailed}, where the computation is performed by directly using the averaged quantities, $\overline{R}$ and $\overline{S}$. It should be noted that the differently computed quantities are very close, except for the case of the monodisperse population at $t/\tau= 6.78$, where the location of the maximum reaction time changes from  $(x_3-x_c)/L_3 = - 0.06$, Fig. \ref{fig:temporal_scales}, to $(x_3-x_c)/L_3 = - 0.025$, Fig. \ref{fig:react_time_detailed}.
		\\
		The condensation and evaporation times diverge toward values of the order of $10^3$ seconds at the saturation location, $S=0$,  where they are not defined, see Figure \ref{fig:supersaturation_a} to observe the displacement in time of the spatial points where $S=0$. 
		The phase relaxation time, $\tau_{phase}$, elongates across the mixing layer as the mean radius and the droplet concentration (numerical density) decrease. In time, the $\tau_{phase}$ growth rate smoothes out as the layer widens. The fact that $\tau_{phase}$ grows indefinetely in the diluted interfacial region is not surprising and was also observed during the in-situ measurements of shallow cumulus clouds by Siebert and Shaw (2017)\cite{Siebert2017}.}
	
	{It is interesting to observe that, in the monodisperse case, the reaction time is converging to the saturation time (ratio $\rho_v/\rho_{vs}=0.995$), where the skewness of $S$ is negative, while it is converging to the evaporation time when $S$ is positively skewed.}
	
\begin{figure}[bht!]
		{\subfloat[a][]{\includegraphics[width=0.49\linewidth]{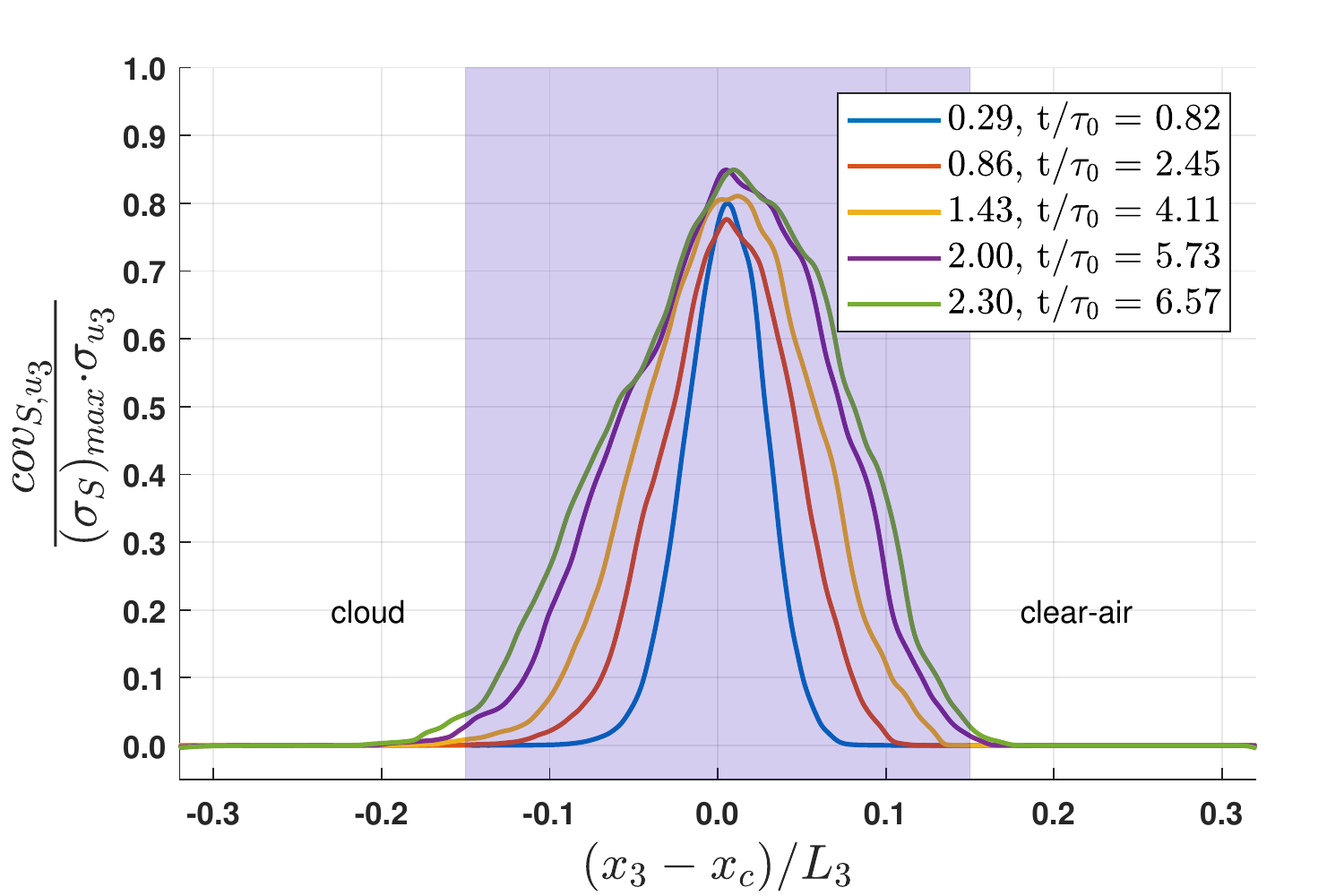}\label{fig:flux_time}}}
		{\subfloat[b][]{\includegraphics[width=0.49\linewidth]{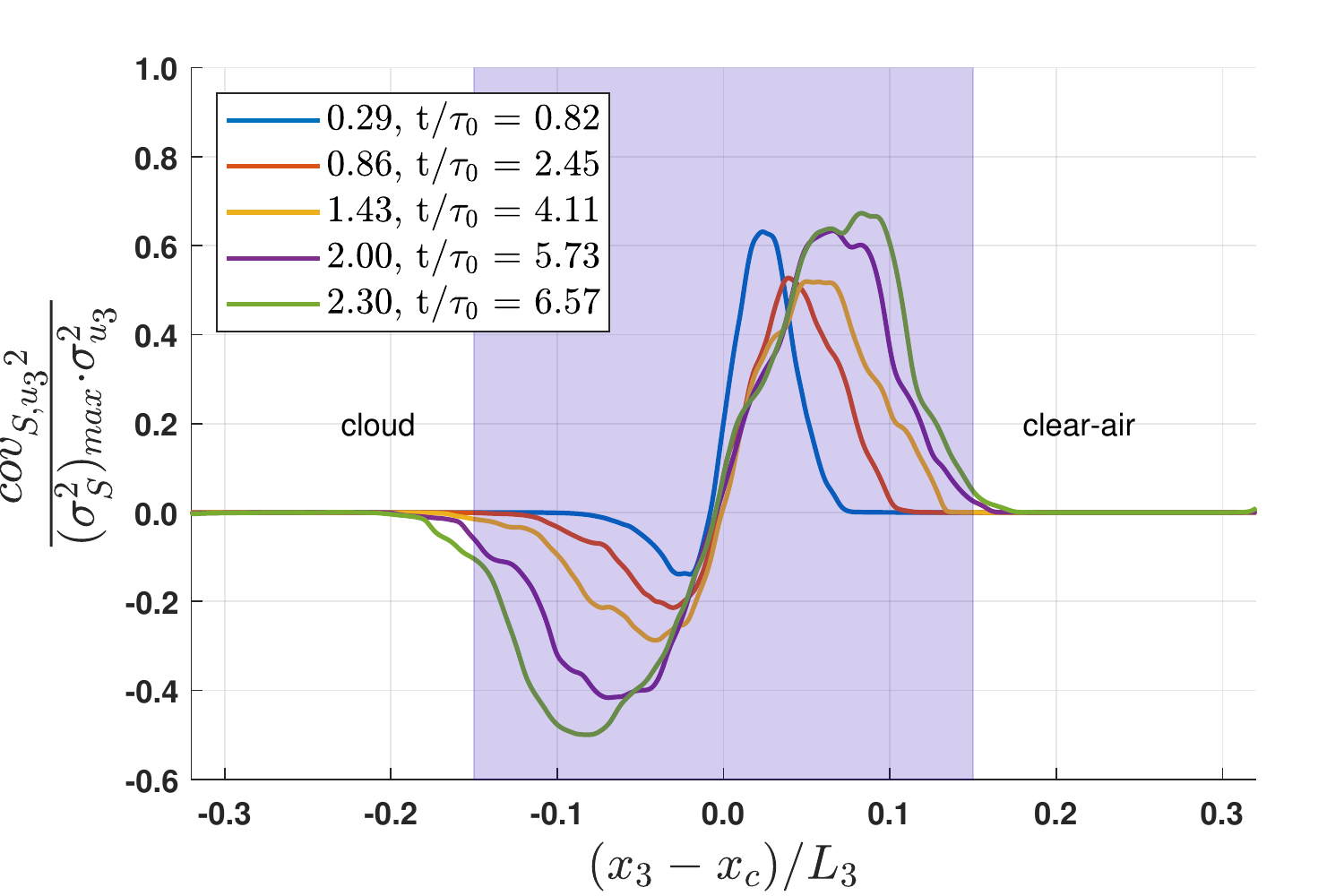}\label{fig:flux_flow_time}}}\\
		{\subfloat[c][]{\includegraphics[width=0.49\linewidth]{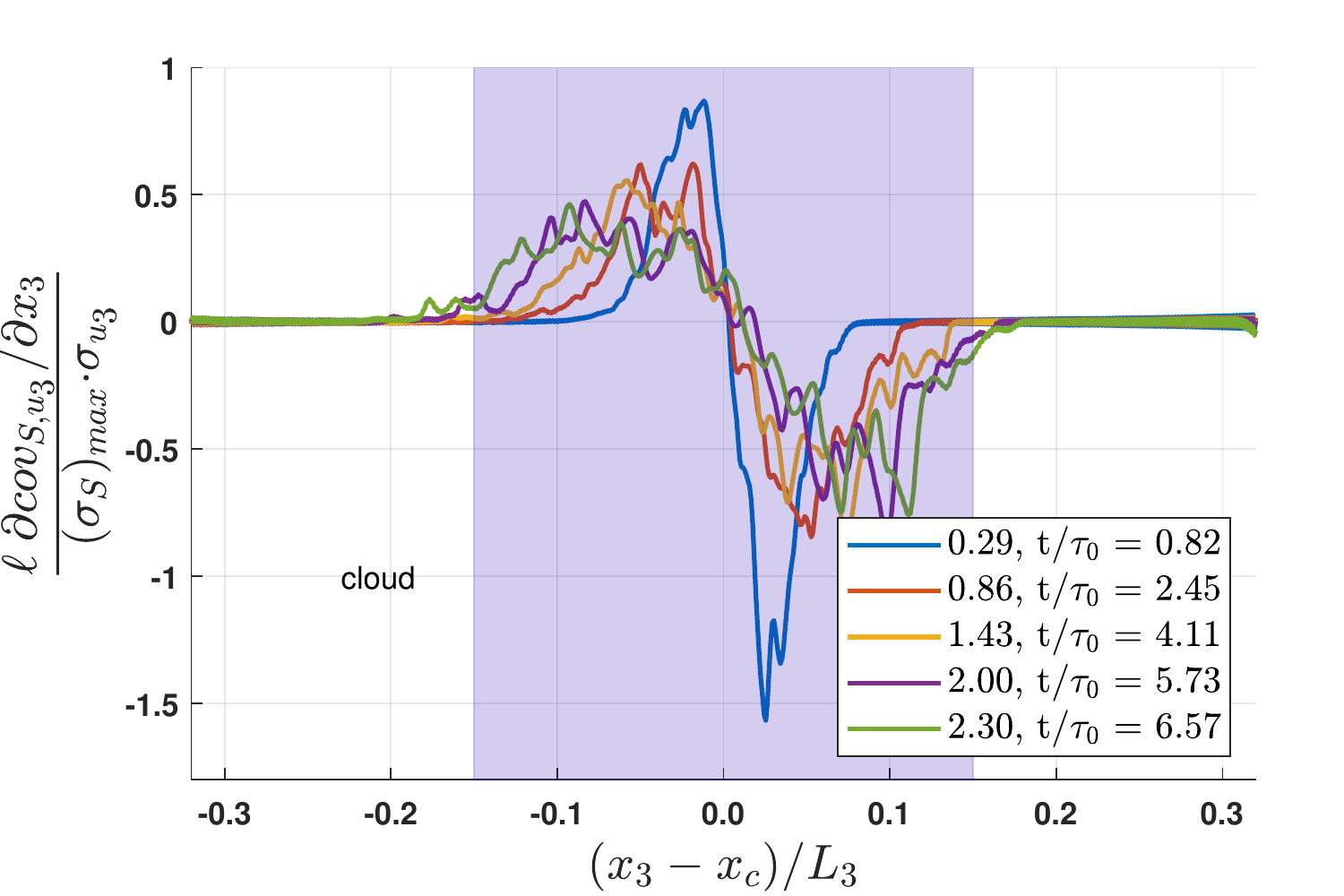}\label{fig:flux_der_time}}}\\
		\caption{Supersaturation flux statistics for the monodisperse drop population. \protect\subref{fig:flux_time} Normalized covariance (flux) of the supersaturation with vertical velocity component. \protect\subref{fig:flux_flow_time} Normalized covariance of the supersaturation and square of the vertical velocity component. \protect\subref{fig:flux_der_time} Normalized derivative of the covariance (flux) of the supersaturation with vertical velocity component. The difference of values between the monodisperse and polydisperse population distributions is negligible. The same comparative situation shown in Figures 5 and 6 holds true.}
		\label{fig:fluxes}
	\end{figure}

	{The droplet condensation time is considerably higher everywhere than the phase relaxation time in the monodisperse case, and increases in time. The condensation time is instead shorter than the phase relaxation time in the cloudy region in the polydisperse case, but it becomes of the same order as $\tau_{phase}$ where the mixing process starts. A rise of $\tau_{cond}$ is observed at the end of the transient in the bottom region of the domain where the sedimentation due to gravity becomes substantial. 
		\\
		However, the most interesting observation is that there is a location inside the mixing layer where the phase relaxation, the reaction time and the evaporation time cluster together. This location precedes the location where the turbulent fluxes maximize. By comparing the distributions in Figure \ref{fig:temporal_scales} with the distribution of the turbulent supersaturation flux, see Figure \ref{fig:fluxes}, it is possible to see that the  clustering of the microphysical times takes place at almost the same location, where the flux rate is close to a maximum. The microphysical times separate before and after this location.  In particular, the  the reaction time is much shorter than the evaporation and phase relaxation times before this location. The reaction time then collapses to the evaporation time, which is much shorter that the phase relaxation time. In the polydisperse case, the clustering of the microphysical times also includes the condensation time, as expected, since condensation often occurs rapidly in the spectral range of the drops with a small radius for these populations.
		The evaporation time on the right hand side of the panels in Figure \ref{fig:temporal_scales} oscillates to great extent, particularly for the two panels showing the last part of the transient, due to the higher collision rate there. This result should be contrasted with Figure \ref{fig:drop_distributions}. }
	
	\subsection{Turbulent transport effects on the supersaturation balance}
	
	The observed acceleration of the population dynamics in the same cloud-clear air interface region as those studied here, as well as the rapid differentiation in the size of the droplets, due to the different weights that evaporation, condensation and collision have in these highly intermittent mixing region\cite{Golshan2021}, can, at least in part, explain the rapid increase in the size of the droplets that is observed in some cumulus cloud formations, in particular in maritime ones, and which is considered capable of locally inducing rain-fall, Mason and Chien (1962)\cite{mason1962cloud}, Li et al. (2020)\cite{li2020}. These finding have been  observed  despite the fact that beyond the temporal decay of the turbulence present in the whole system, the interface also hosts the spatial decay of the kinetic energy. It should be considered that the large scales of turbulence vary very little in this flow system, because the computational domain is fixed and because the ratio of the large scales and the ratio of the kinetic energies between the cloudy and ambient air regions vary slowly in time \citep{Tordella_2006, Tordella2011}. All things considered, these observations lead to the conclusion that the observed accelerated dynamics is associated with the particular small-scale anisotropy and intermittency of the interfacial layer.
	
	\begin{figure}[bht!]
		\centering
		\includegraphics[width=0.7\linewidth]{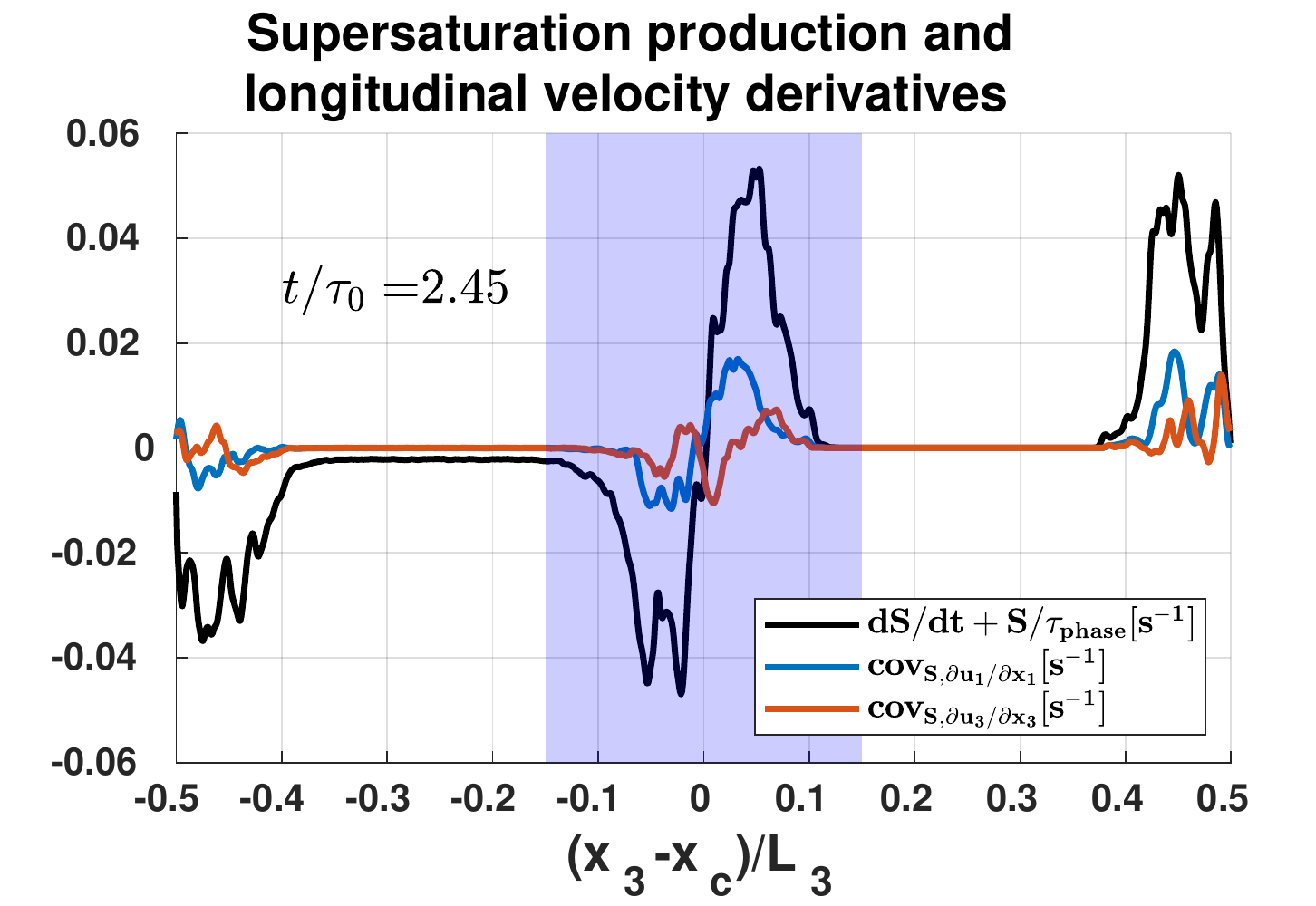}
		\caption{{Distribution of the planar horizontal averages along the vertical direction of the difference between the time derivatives of the supersaturation and condensation terms, $\overline{dS/dt} - \overline{C} = \overline{dS/dt} + \overline{S/\tau_p}$ and of the covariances $cov_{S,\partial u_1/\partial x_1}$, $cov_{S,\partial u_3/\partial x_3}$, see equation (\ref{cov}). These quantities vary considerably inside the mixing layer, and the two kinds of curves are both almost antisymmetric with respect to the center of mixing layer $x_c$. The data were retrieved from a monodisperse simulation at $t/\tau_0=2.45$.
		}}
		\label{fig:prod_long_vel_der}
	\end{figure}
	
	\begin{figure}[bht!]
		\centering
		\large \textbf{Supersaturation and longitudinal velocity derivatives in the mixing layer}\par\medskip
		\includegraphics[width=0.9\linewidth]{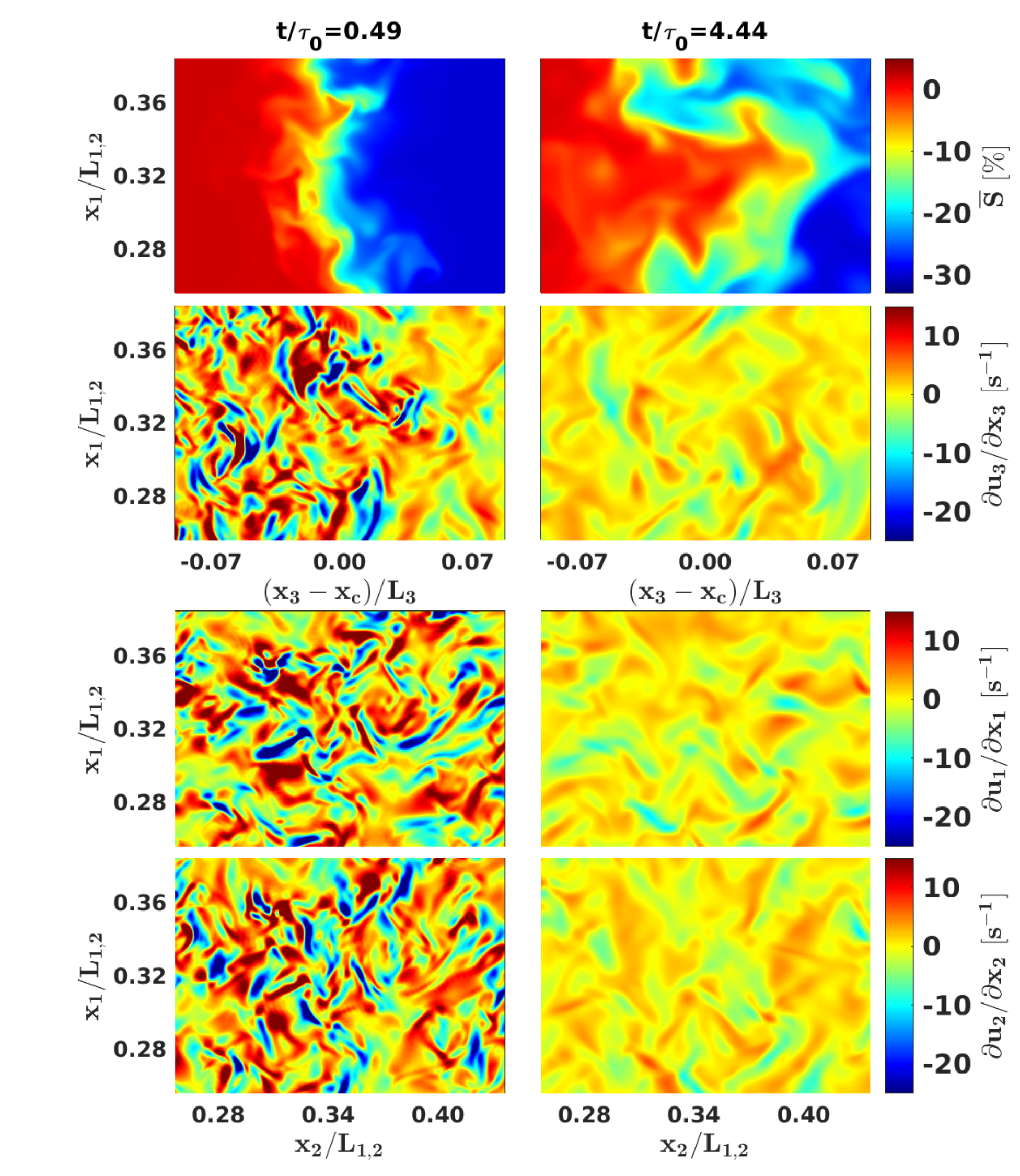}
		\caption{Visualizations of the supersaturation (top row) and longitudinal velocity derivatives in the mixing region (second, third and fourth rows). The plots display only a portion of the domain, as shown by the normalized coordinates at two time instances $t/\tau_0=0.49$ (a) and $t/\tau_0=4.44$ (b). $x_c$ denotes the initial position of the mixing layer. The variance of the longitudinal velocity derivatives is of the order of $10\,\si{\second^{-1}}$ at the beginning of the transient, but rapidly decreases. The values of the inverse Kolmogorov times in the mixing region are plotted in Figure \ref{fig:kinetic_c} (indigo dash-dotted line) and are of the same order of magnitude as the derivatives shown in this figure.}
		\label{fig:snapshots}
	\end{figure}
	
	{In conditions of zero updraft, under almost statistical equilibrium conditions (steady state, homogeneity and isotropy), the planar averages of the difference between the time derivative of the supersaturation and the condensation terms must be null}
	\begin{equation*}
		\overline{dS/dt} - \overline{\mathcal{C}} = \overline{dS/dt} + \overline{S/\tau_p}  \cong 0. 
	\end{equation*}
	{However, in more general turbulence situations, as in the present system, which is unsteady (a temporal decay follows an initial transient kinetic energy growth due to an unstable stratification), highly in-homogenous, and thus anistropic, $\overline{dS/dt}$ may not necessarily balance  $\overline{S/\tau_p}$.
		\noindent  This inbalance can be referred to as a turbulence supersaturation fluctuation production,  $\overline{\mathcal{P}_t}$. As can be seen in Fig. \ref{fig:prod_long_vel_der}, $\overline{dS/dt} + \overline{S/\tau_p} $ %= \overline{\mathcal{P}_t}$ 
		and the covariances between the supersaturation and the longitudinal velocity derivatives along the vertical direction, $\mathrm{cov}_{S,\partial u_3/\partial x_3}$, as well as along the horizontal direction $\mathrm{cov}_{S,\partial u_1/\partial x_1}$, are plotted across the entire $(x_3 - x_c)/ L_3 \in [-0.5, 05]$ domain.  All the quantities become larger and almost antisymmetric in the mixing region, and they appear qualitatively skew-symmetric with respect to the central plane $x_3 \cong x_c$. It is thus evident that, under spatially averaged (planar averages) conditions, the condensation term alone in the supersaturation evolution equation (\ref{ssat_eq}) is not able to account for the value of the time derivative of the supersaturation that takes place inside the mixing region. 
		\\
		Provided that the Kolmogorov time, $\tau_\eta$, scales with dissipation rate $\varepsilon$, the former is found to be much smaller ($10^{-2}\div10^{-1}\si{\second}$) than the evaporation and phase relaxation time scales reported in  Fig. \ref{fig:temporal_scales}. Large values of $\tau_{phase}$ at the interface result in low small-scale Damk\"{o}hler numbers, and should therefore enhance turbulent supersaturation fluctuations \citep{Siebert2017}. It is therefore reasonable to assume that supersaturation fluctuations, due to turbulence, are prevalent with respect to those generated by phase transition at the droplet surface.} There are two reasons for this hypothesis. First, the statistical moments of the vapor density in an analogous unstable  mixing layer with identical initial and boundary conditions and an identical set of control parameters for the carrier flow, but with a subsaturated cloud region where droplets are absent, are close in shape and value to those of mixings that contain droplets coming from a  supersaturated cloudy region (see Figure 8 in \cite{Gallana_2022})
	\\
	Second, as can be seen in Figures \ref{fig:supersaturation} and \ref{fig:diff-stats-supersaturation}, the effects of the supersaturation statistics associated with the different size distributions of the drop populations are negligible, and the differences are in fact well below 1\%. It should be noted that the opposite is not true, that is, that the dynamics of the populations is very sensitive to the shape of the droplet size distribution.
	\\
	\indent We therefore hypothetize that the amplitude of the local supersaturation 
	is modulated by small-scale turbulent fluctuations and that such turbulent fluctuations may contribute to the overall local supersaturation balance.
	In order to assess this hypothesis, we looked for the proportionality relation between: i) the difference in the planar averages of the supersaturation time derivative and the condensation term, and ii) the covariance, eq. (\ref{cov}), 
	of the supersaturation and the intermittency of the small-scale, as represented by the fluctuations of the longitudinal  derivatives of the velocities.
	In other words, we put 
	\begin{equation}\label{ssat_model_eq}
		\overline{\frac{dS}{dt}}+\overline{\frac{S}{\tau_{phase}}} = \overline{\mathcal{P}_t} \sim  \mathrm{cov}_{S,\partial u_i/\partial x_i} 
		%= \it{const} \;\;\mathrm{cov}_{S,\partial u_1/\partial x_1}
	\end{equation}
	This is conceptually equivalent to modeling supersaturation production as the product of the supersaturation fluctuations and the characteristic frequency of small-scale turbulent structures, $\sim\tau_\eta^{-1}$, where the characteristic frequency of small-scale turbulent motions can be represented by the longitudinal velocity spatial derivatives. 
	
	The generation of small-scale anisotropy in turbulent shearless mixing has recently been investigated numerically. Data from direct numerical simulations for Taylor microscale Reynolds numbers between 45 and 150 \citep{Tordella2011, Tordella2012, Iovieno2014, Golshan2021, Gallana_2022} show that there is a significant departure of the longitudinal velocity derivative moments from the values found in homogeneous and isotropic turbulence and that the variation of skewness has the opposite sign for the components across the mixing layer and parallel to it. 
	The anisotropy induced by the presence of a kinetic energy gradient has a very different pattern from the one generated by homogeneous shear. The transversal derivative moments in the mixing are in fact found to be very small, which highlights that smallness of the transversal moments is not a sufficient condition for isotropy. This intermittency is characterized by a large departure of the longitudinal derivative moments (as shown in Fig. \ref{fig:snapshots} together with the supersaturation for two time instances), which are different in direction across and parallel to the layer from the typical values of the isotropic condition, even in such a flow, where there is no energy production(due to the lack of mean flow gradients). The structure of the anisotropy is such that the skewness departure from isotropy reduces the compression on the fluid filaments parallel to the mixing layer and enhances that of the filaments orthogonal to it.
	
\begin{figure}[bht!]
		\centering
		\large\textbf{Supersaturation production and small-scale velocity statistics}\par\medskip
		\subfloat[a][]{\includegraphics[width=0.49\linewidth]{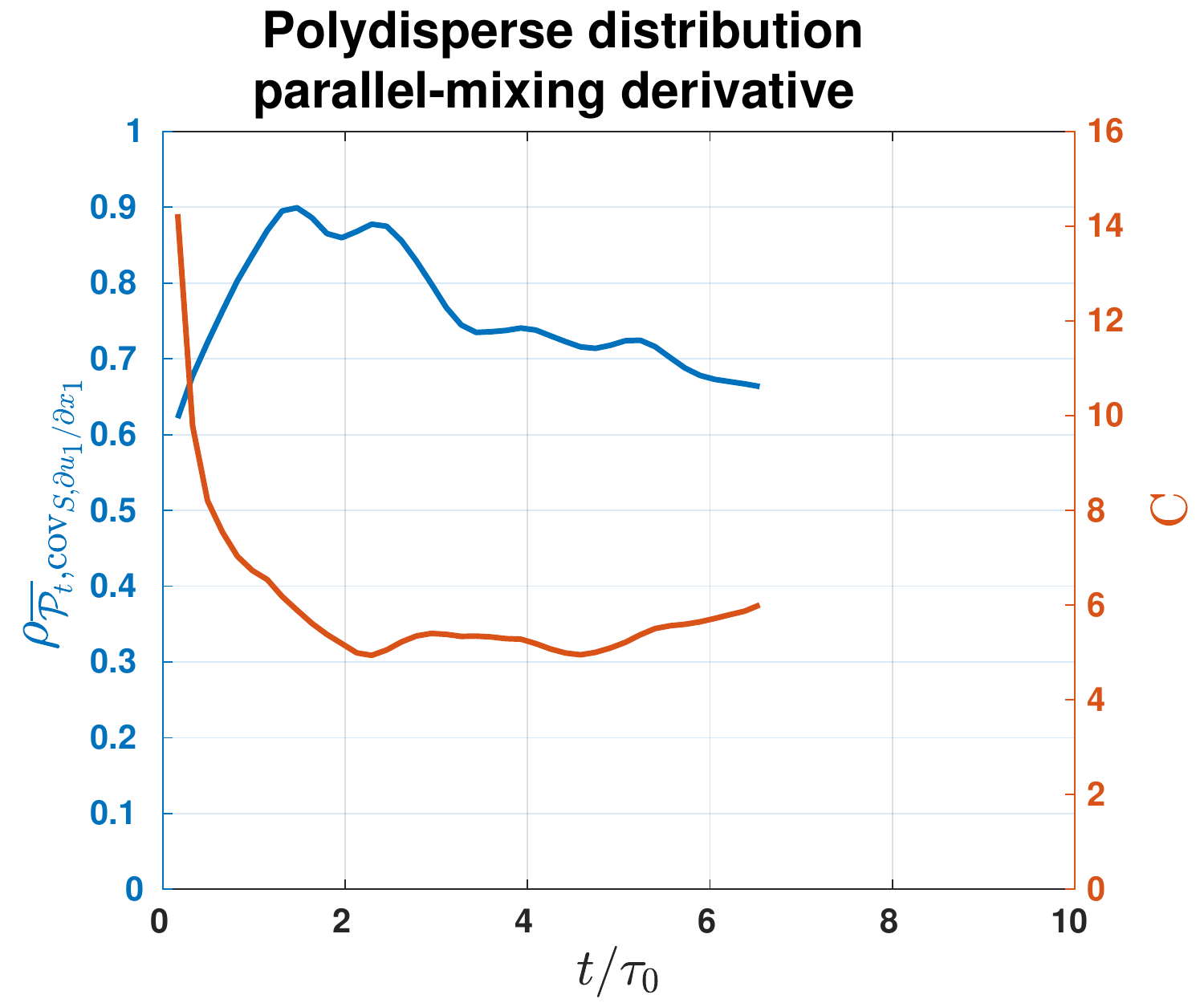}}
		\subfloat[b][]{\includegraphics[width=0.49\linewidth]{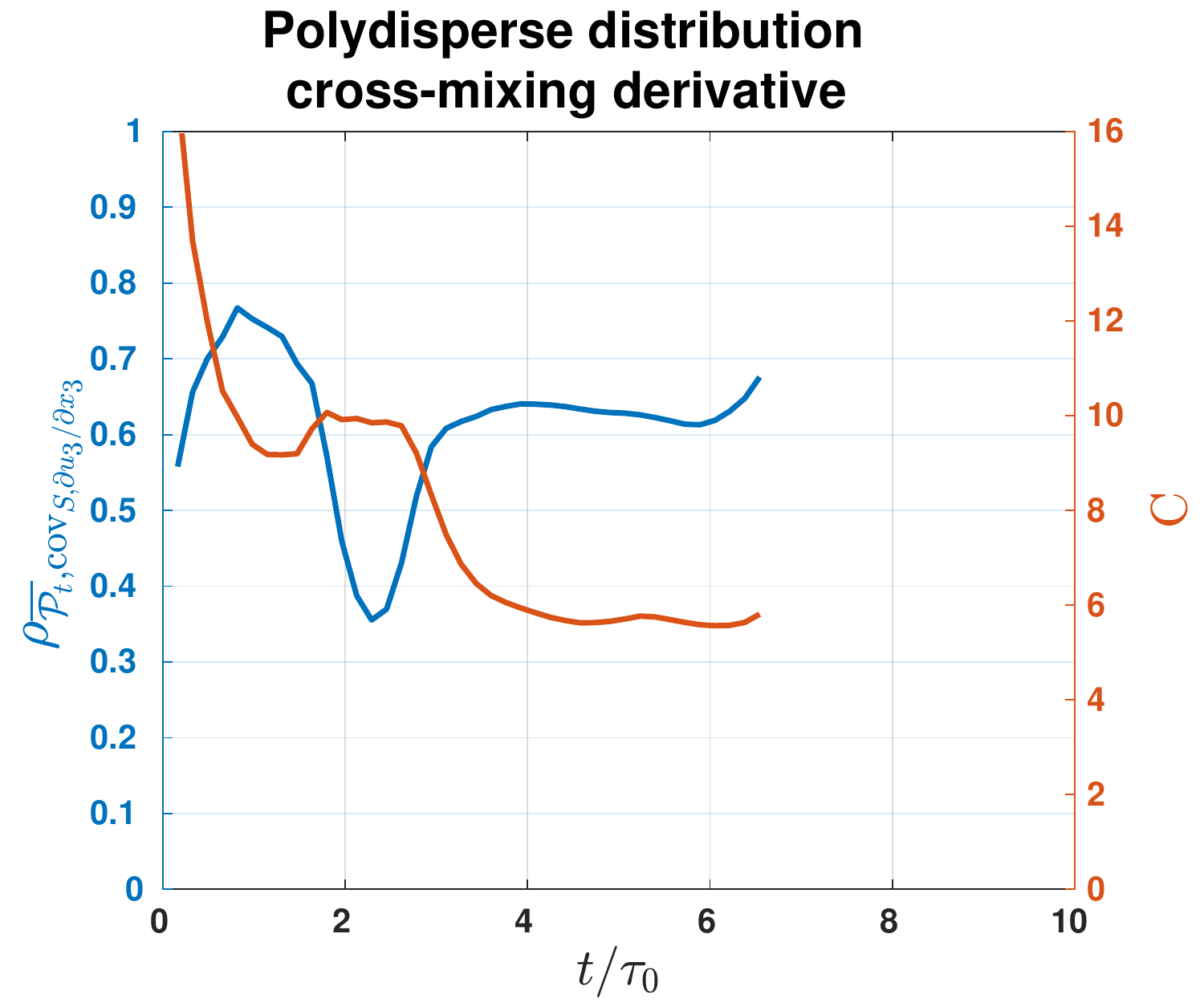}}\\
		\subfloat[c][]{\includegraphics[width=0.49\linewidth]{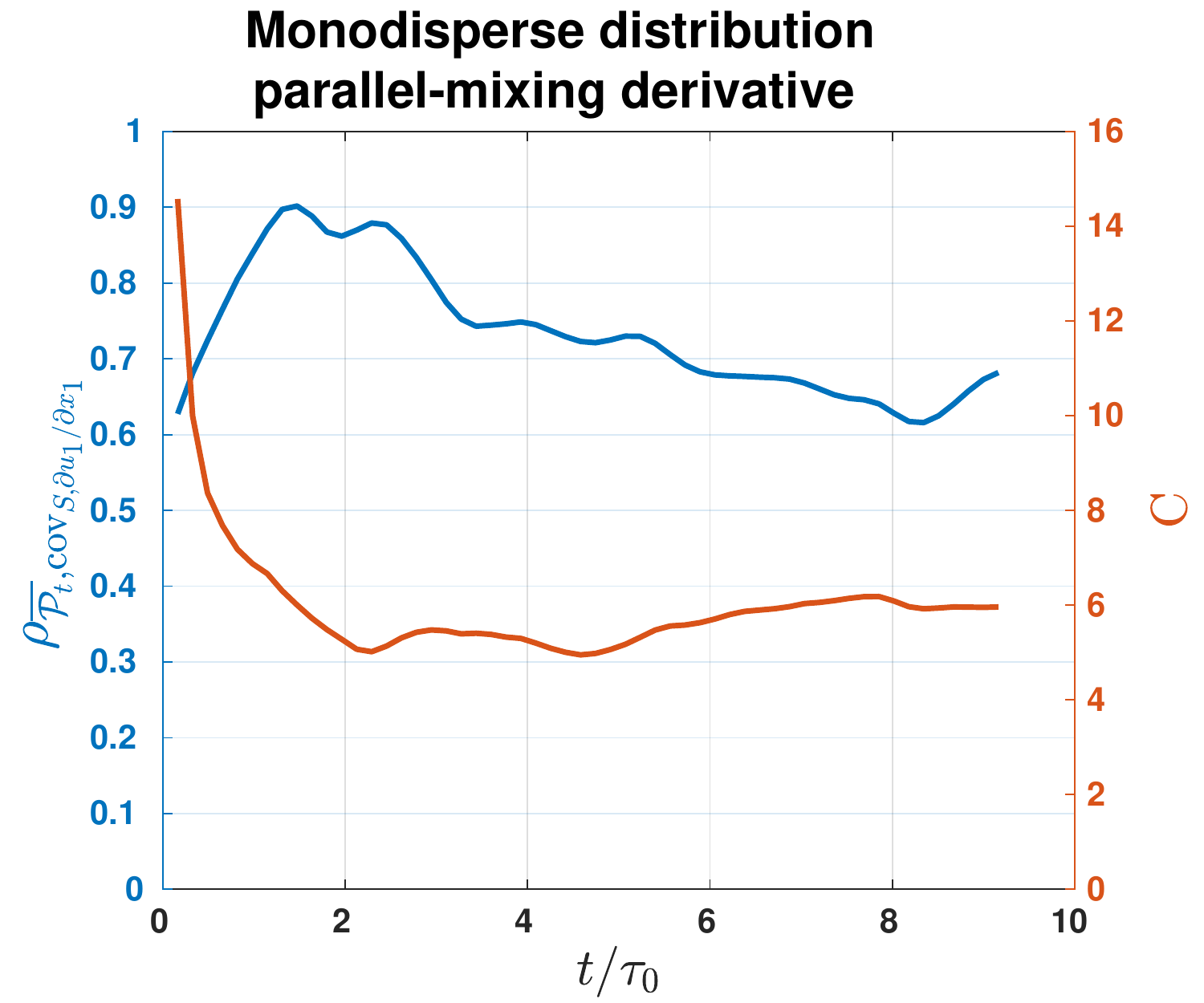}}
		\subfloat[d][]{\includegraphics[width=0.49\linewidth]{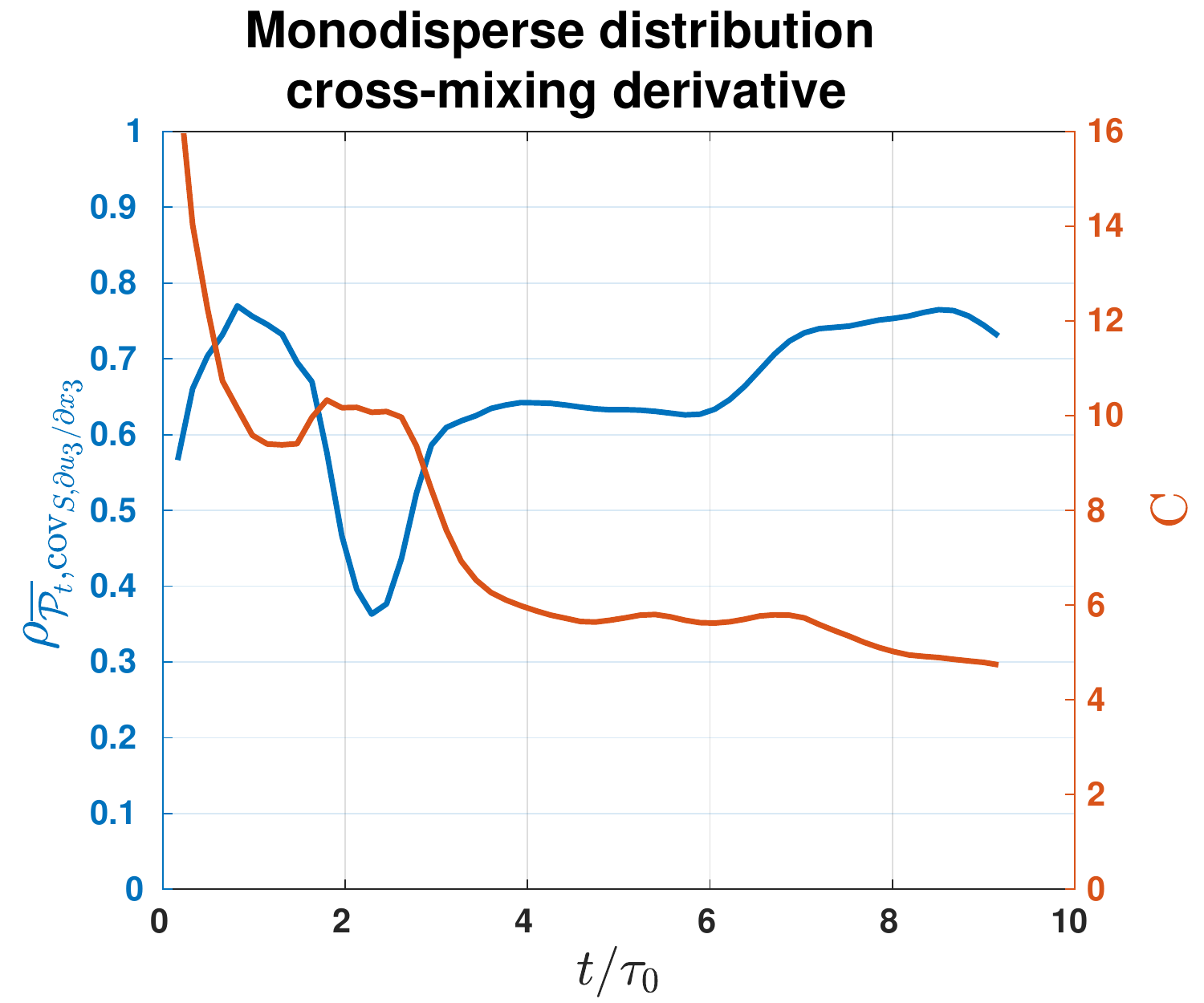}}
		\caption{\textbf{Time evolution of the estimated production-planar covariance correlation coefficient and the proportionality constant}. The Pearson correlation coefficient between the turbulence production term $\mathcal{P}_t$, see equation(\ref{ssat_model_eq}),  and the supersaturation-velocity longitudinal derivative covariance (blue curves) plotted for the horizontal (left) and vertical (right) longitudinal derivatives during the transient. The correlation coefficient peaks around the first initial eddy turnover time and slowly decreases  in magnitude to an asymptote $\sim 0.7$ as the transient progresses.}
		\label{fig:P_sdudx_corr}
	\end{figure}

	The Pearson correlation coefficient, $\rho_{\overline{\mathcal{P}_t},\mathrm{cov}_{S, \partial u_i/\partial x_i}}$, inside the layer of thickness ${\Delta}(t)$, see eq. \ref{Pearson}, was computed along the transient for i=1,2,3. The results are shown in Figure \ref{fig:P_sdudx_corr}, where the data points have been collected for a time increment, that is approximately one half of the initial eddy turnover time.
	
	When the  correlation coefficient, that is, the linear correlation between two sets of data, is above 0.7, the correlation is defined as strong. However, we do not expect the numerical simulation to describe the first  initial eddy turnover time of the transient accurately. The  correlation coefficient  decreases slightly along the transient, beyond the first eddy turn over time, as the transient proceeds from values as high as 0.9, when the longitudinal velocity derivative is horizontal, and as high as 0.8, when the longitudinal velocity derivative is vertical, to values close to 0.7. This is true for both monodisperse and polydisperse drop populations.
	
	The relatively large absolute values of the correlation coefficients confirm that a quasi-linear relation should hold between the source term $\overline{\mathcal{P}_t}$ and $\mathrm{cov}_{S \partial u_i/\partial x_i}$.
	An alternative way of estimating the proportionality constant, $C$, relevant to the dimensional quantities, along the transient is to integrate across the mixing layer of $\overline{\mathcal{P}_t}$ and  $\mathrm{cov}_{S \partial u_i/\partial x_i}, i=1,2$:
	\begin{displaymath}
		\int_{\Delta}\left\vert\frac{d\overline{S}}{dt}+\frac{\overline{S}}{\tau_{phase}}\right\vert dx_3\cong C\int_{\Delta}\left\vert\mathrm{cov}_{S \partial u_i/\partial x_i}\right\vert dx_3
	\end{displaymath}
	The estimated values of the non-dimensional constant $C$ are reported in Figure \ref{fig:P_sdudx_corr} (orange curve). Once again, the shape of the initial droplet size distribution does not seem to affect either the evolution of the correlation coefficient or the non-dimensional constant $C$ during the transient. In all these cases, the estimated value is $5$, an asymptotic value, that is rapidly reached after the first initial eddy turnover. We can observe a different pre-asymptotic trend for the horizontal and vertical longitudinal derivative  correlation coefficients, which is due to the intrinsic small-scale anisotropy of the mixing layer between the cloudy region and the clear-air, see the discussion above.
	
	\section{Conclusions}\label{SECTION_CONCLUSIONS}
	We have considered the relationship between supersaturation fluctuations and turbulent small-scale dynamics in the context of an inhomogeneous, anisotropic, shearless,  turbulent air mixing layer, which is often used to model the carrier flow at the interface between warm clouds and unsaturated environmental air. Two initial droplet population types, that is, a $15\si{\micro\meter}$-monodisperse one and constant-mass-per-volume-class polydisperse one, were tested. 
	
	The various time scales pertaining to the microphysics of a droplet population were compared inside the top of the cloudy region,  the layer where the turbulent transport process toward the environmental subsaturated air takes place. The evaporation, reaction, and phase relaxation time scales match for a value close to $20\sim 30$ s inside the layer just before the location where the supersaturation flux reaches its maximum rate of variation. In the case of a polydisperse population, this match includes the condensation time. The time scales before  this spatial location are different, with differences of the order of one minute. Beyond this location, the evaporation and reaction times overlap, while the relaxation phase and condensation time scales asymptotically diverge, since the environment becomes more and more undersatured. 
	
	Under the  hypothesis of the supersaturation fluctuation depending to a great extent on the small-scale intermittency of the carrier flow that hosts the vapor and liquid water phases, we have analyzed the supersaturation balance equation with the aim of evincing their reciprocal correlation. In order to assess this hypothesis, we compared the estimated planar averages of the time derivative of the supersaturation and the condensation terms with the planar covariance of the supersaturation and the longitudinal velocity derivatives. The statistics of the velocity derivatives are in fact particularly relevant for small-scale dynamics. For the specific shearless turbulent structure considered here, the longitudinal velocity derivatives are more significant for small-scale intermittency than the transversal ones, which are null. Moreover, the longitudinal velocity derivative can be considered as a characteristic measure of the small-scale frequency, $\tau_\eta^{-1}$. We have found a high value of the Pearson  correlation coefficient, $\rho_{\overline{\mathcal{P}_t},\mathrm{cov}_{S \partial u_i/\partial x_i}}$ $\sim 0.7$ for the droplet populations, both inside the interfacial layer and along the entire simulation transient, which leads to the conclusion that, in the absence of an updraft,  the mismatch of the time derivative of the supersaturation and the condensation terms is linearly related to the covariance of the suparsaturation and the longitudinal velocity derivatives of the carrier flow.
	
	\section*{Acknowledgements}
	This project has received funding from the Marie-Sklodowska Curie Actions (MSCA ITN ETN COMPLETE) under the European Union’s Horizon 2020 research and innovation program. Grant agreement 675675, \href{http://www.complete-h2020network.eu}{http://www.complete-h2020network.eu}. 
	
	We acknowledge \href{http://www.hpc.cineca.it/}{Cineca - Supercomputer Applications and Innovations (SCAI)} for providing the computational resources and technical support. All the runs were performed on the \href{http://www.hpc.cineca.it/hardware/marconi}{Marconi Tier-0 system} from September 2019 to January 2020. The post-processing activity was in part conducted on the \href{https://hpc.polito.it/hactar_cluster.php}{HACTAR cluster}, and the computational resources were provided by HPC@POLITO, a project of Academic Computing within the Department of Control and Computer Engineering at the Politecnico di Torino (\href{http://www.hpc.polito.it}{http://www.hpc.polito.it}). %The authors would also like to thank prof. J\"{o}rg Schumacher of Technische Universit\"{a}t Ilmenau and dr. Federico Fraternale for all their support and scientific discussion.
	
	\nocite{*}
	\bibliography{references}% Produces the bibliography via BibTeX.
	
\end{document}